\documentclass[twocolumn,onecolappendix]{aastex701}

\usepackage{lipsum}
\usepackage{amsmath}
\usepackage{fontenc}

\defcitealias{monteiro-oliveira_unveiling_2022}{MO}

\begin{document}

\title{LoVoCCS. III. Third Generation Pipeline \& The Hercules Supercluster}

\correspondingauthor{Anthony M. Englert}
\email{astroenglert@Gmail.com}

\author[orcid=0000-0003-2314-5336]{Anthony M. Englert}
\affiliation{Department of Physics, Brown University, 182 Hope Street, Box 1843, Providence, RI 02912, USA}
\affiliation{Department of Physics and Astronomy, University of California, Davis, Davis CA, USA}
\email[hide]{astroenglert@Gmail.com}

\author[orcid=0000-0001-5422-1958]{Shenming Fu}
\affiliation{NSF-DOE Vera C. Rubin Observatory / SLAC National Accelerator Laboratory, 2575 Sand Hill Road, Menlo Park, CA 94025, USA}
\affiliation{Kavli Institute for Particle Astrophysics and Cosmology, Stanford University, Stanford, CA 94305, USA}
\email[hide]{shenming.fu.astro@gmail.com}

\author[orcid=0000-0003-0751-7312]{Ian Dell'Antonio}
\affiliation{Department of Physics, Brown University, 182 Hope Street, Box 1843, Providence, RI 02912, USA}
\email[hide]{ian_dellantonio@brown.edu}

\author[orcid=0000-0003-3595-7147]{Mohamed H. Abdullah}
\affiliation{Department of Physics, University of California Merced, 5200 Lake Road, Merced, CA 95343, USA}
\affiliation{Department of Astronomy, National Research Institute of Astronomy and Geophysics, Cairo, 11421, Egypt}
\email[hide]{mohamedelhashash@ucmerced.edu}
\author[orcid=0009-0008-3612-8942]{Shrouk Abdulshafy}
\affiliation{Department of Physics, University of California Merced, 5200 Lake Road, Merced, CA 95343, USA}
\affiliation{Department of Astronomy, Cairo University, 1 Gamaa Street, Giza, 12613, Egypt}
\email[hide]{shroukabdulshafy@ucmerced.edu}
\author[orcid=0009-0002-2253-4583]{Eddie Aljamal}
\affiliation{Department of Physics and Astronomy, Michigan State University, East Lansing, MI 48824, USA}
\email[hide]{ealjamal@umich.edu}
\author[orcid=0000-0003-4811-7913]{William K. Black}
\affiliation{Department of Physics and Astronomy, Brigham Young University, N283 ESC, Provo, UT 84602, USA}
\email[hide]{wkblack@umich.edu}
\author[orcid=0009-0002-8065-1454]{Nicole Chidester}
\affiliation{Department of Physics, Northeastern University, 110 Forsyth St, Boston, MA 02115}
\email[hide]{chidester.n@northeastern.edu}
\author[orcid=0000-0003-2416-1557]{Douglas Clowe}
\affiliation{Department of Physics and Astronomy, Ohio University, 1 Ohio University, Athens, OH 45701, USA}
\email[hide]{clowe@ohio.edu}
\author[orcid=0000-0003-1371-6019]{M. C. Cooper}
\affiliation{Department of Physics and Astronomy, University of California, Irvine, Irvine CA, USA}
\email[hide]{cooper@uci.edu}

\author[orcid=0000-0002-2808-0853]{Megan Donahue}
\affiliation{Department of Physics and Astronomy, Michigan State University, East Lansing, MI 48824, USA}
\email[hide]{donahu42@msu.edu}

\author[orcid=0000-0003-0936-7223]{Zacharias Escalante}
\affiliation{Department of Physics, Brown University, 182 Hope Street, Box 1843, Providence, RI 02912, USA}
\email[hide]{zacharias_escalante@brown.edu}
\author[orcid=0000-0002-4876-956X]{August Evrard}
\affiliation{Department of Physics, University of Michigan, Ann Arbor, MI 48109, USA}
\affiliation{Leinweber Institute for Theoretical Physics, University of Michigan, Ann Arbor, MI 48109, USA}
\email[hide]{evrard@umich.edu}

\author[orcid=0009-0003-9547-0952]{Eric Habjan}
\affiliation{Department of Physics, Northeastern University, 110 Forsyth St, Boston, MA 02115}
\email[hide]{erichabjan@gmail.com}
\author[orcid=0000-0003-4774-4288]{Soren Helhoski}
\affiliation{Department of Physics, Brown University, 182 Hope Street, Box 1843, Providence, RI 02912, USA}
\email[hide]{soren_helhoski@brown.edu}

\author[orcid=0009-0004-6900-6707]{Ruoning Lan}
\affiliation{Department of Physics, Brown University, 182 Hope Street, Box 1843, Providence, RI 02912, USA}
\email[hide]{ruoning_lan@brown.edu}

\author[orcid=0000-0002-0561-7937]{Binyang Liu}
\affiliation{Purple Mountain Observatory, Chinese Academy of Sciences, Nanjing 210023, China}
\email[hide]{binyang_liu@alumni.brown.edu}
\author[orcid=0000-0002-9883-7460]{Jacqueline McCleary}
\affiliation{Department of Physics, Northeastern University, 110 Forsyth St, Boston, MA 02115}
\email[hide]{j.mccleary@northeastern.edu}
\author[orcid=0000-0001-7964-9766]{Hironao Miyatake}
\affiliation{Kobayashi-Maskawa Institute for the Origin of Particles and the Universe (KMI), Nagoya University, Nagoya, 464-8602, Japan}
\affiliation{Institute for Advanced Research, Nagoya University, Nagoya 464-8601, Japan}
\affiliation{Kavli Institute for the Physics and Mathematics of the Universe (WPI), The University of Tokyo Institutes for Advanced Study (UTIAS), The University of Tokyo, Chiba 277-8583, Japan}
\email[hide]{hironao.miyatake@nagoya-u.jp}
\author[orcid=0000-0001-7847-0393]{Mireia Montes}
\affiliation{Institute of Space Sciences (ICE, CSIC), Campus UAB, Carrer de Can Magrans, s/n, 08193 Barcelona, Spain.}
\email[hide]{mireia.montes.quiles@gmail.com}
\author[orcid=0000-0002-5554-8896]{Priyamvada Natarajan}
\affiliation{Department of Astronomy, Yale University, 266 Whitney Avenue, New Haven, CT 06511, USA}
\affiliation{Department of Physics, Yale University, 217 Prospect Street, New Haven, CT 06520, USA}
\email[hide]{priyamvada.natarajan@yale.edu}
\author[orcid=0000-0003-3397-6838]{Jessica Nelson}
\affiliation{Department of Physics, Brown University, 182 Hope Street, Box 1843, Providence, RI 02912, USA}
\email[hide]{jessica_n_nelson@brown.edu}
\author[orcid=0000-0002-0144-387X]{Michelle Ntampaka}
\affiliation{Space Telescope Science Institute, Baltimore, MD 21218, USA}
\affiliation{Department of Physics and Astronomy, Johns Hopkins University, Baltimore, MD 21218, USA}
\email[hide]{mntampaka@stsci.edu}
\author[orcid=0000-0002-7957-8993]{Elena Pierpaoli}
\affiliation{University of Southern California, Los Angeles, CA 90089, USA}
\email[hide]{pierpaol@usc.edu}
\author[orcid=0000-0002-9365-7989]{Marc Postman}
\affiliation{Space Telescope Science Institute, 3700 San Martin Drive, Baltimore, MD 21218, USA}
\email[hide]{postman@stsci.edu}
\author[orcid=0000-0002-7342-3229]{Rahul Shinde}
\affiliation{Department of Physics, Brown University, 182 Hope Street, Box 1843, Providence, RI 02912, USA}
\email[hide]{rahul_Shinde@brown.edu}
\author[orcid=0000-0002-9254-144X]{Jubee Sohn}
\affiliation{Astronomy Program, Department of Physics and Astronomy, Seoul National University, 1 Gwanak-ro, Gwanak-gu, Seoul 08826, Republic of Korea}
\affiliation{SNU Astronomy Research Center, Seoul National University, 1 Gwanak-ro, Gwanak-gu, Seoul 08826, Republic of Korea}
\email[hide]{jbsohn@astro.snu.ac.kr}
\author[orcid=0009-0001-5246-9826]{Fuyuko Tanaka}
\affiliation{Division of Physics and Astrophysical Science, Graduate School of Science, Nagoya University, Nagoya 464-8602, Japan}
\email[hide]{tanaka.fuyuko.r0@s.mail.nagoya-u.ac.jp}
\author[orcid=0000-0001-9658-1396]{David Turner}
\affiliation{Department of Physics and Astronomy, Michigan State University, East Lansing, MI 48824, USA}
\email[hide]{djturner@umbc.edu}
\author[orcid=0000-0002-7196-4822]{Keiichi Umetsu}
\affiliation{Academia Sinica Institute of Astronomy and Astrophysics (ASIAA), No. 1, Section 4, Roosevelt Road, Taipei 106319, Taiwan}
\email[hide]{keiichi@asiaa.sinica.edu.tw}
\author[orcid=0000-0001-6161-8988]{Yousuke Utsumi}
\affiliation{Kavli Institute for Particle Astrophysics and Cosmology (KIPAC), SLAC National Accelerator Laboratory, Stanford University, 2575 Sand Hill Road, Menlo Park, CA 94025, USA}
\email[hide]{yousuke.utsumi@nao.ac.jp}
\author[orcid=0000-0003-2102-8646]{Ray Wang}
\affiliation{Department of Physics and Astronomy, Michigan State University, East Lansing, MI 48824, USA}
\email[hide]{wangru46@msu.edu}
\author[orcid=0000-0002-6572-7089]{Gillian Wilson}
\affiliation{Department of Physics, University of California Merced, 5200 Lake Road, Merced, CA 95343, USA}
\email[hide]{gwilson@ucmerced.edu}

\collaboration{all}{LoVoCCS Collaboration}

\begin{abstract}

The Local Volume Complete Cluster Survey (LoVoCCS) is a volume-complete survey of over one-hundred nearby ($0.03 < z < 0.12$), X-ray luminous ($L_{500} > 10^{44} \text{ erg s}^{-1}$) galaxy clusters in the southern sky. Observations for the survey concluded in December 2025, reaching Vera C. Rubin Observatory's Legacy Survey of Space and Time (LSST) Year 1-2 depth in each field and providing observations with $\lesssim 1"$ seeing for weak lensing science. In this paper, we present the latest pipeline for reducing observations using the third-generation of the LSST Science Pipelines. We use recent observations of the Hercules Supercluster to validate the pipeline's data-products and conduct an extensive multi-plane weak-lensing analysis of a $\sim 16 \text{ deg}^2$ complex covering Abell 2147, 2151, 2152, and several additional structures. We confirm that the dynamical mass of the complex is biased due to the dynamical state of Abell 2147, which is consistent with being $\sim 0.2-0.4 \text{ Gyr}$ out-of periapsis, and estimate that the total mass of the supercluster is $8.9^{+1.7}_{-1.4} \times 10^{14}~M_{\odot}$.

\end{abstract}




\section{Introduction} 

Under the $\Lambda$CDM cosmology, structure formation in the early universe was ``seeded'' by nearly gaussian fluctuations. Cold dark matter accreted into these wells, forming the initial large scale structure of the universe. As the universe evolved, these regions collapsed under gravity, accreted additional matter, and merged together, leading to the formation of regions with overdensities of galaxies and hot gas. These regions are known as galaxy clusters and are composed of three distinct majority components: a host dark matter halo, a galaxy population, and the intracluster medium (ICM) composed of hot, ionized plasma \citep{rosati_evolution_2002,kravtsov_formation_2012}.

The properties of clusters have changed over their evolution with redshift. Early clusters/proto-clusters likely lack a built-up ICM and have yet to accrete as much mass as the nearest ($z\lesssim 0.5$) clusters \citep{rohr_cooler_2025,mcdonald_evolution_2016,ohara_evolution_2007}. Due to their large dynamical timescale (typically between 1--10 Gyr), $z\lesssim1$ clusters have recently virialized and their three components are nearly in equilibrium. Of course, the nearest \emph{merging} clusters, which exhibit rich out-of-equilibrium physics at high energies \citep[e.g.][]{clowe_direct_2006,omiya_indications_2024,finner_weak-lensing_2025} are not in equilibrium.

The number density of clusters, their mass distribution, and the evolution of these parameters with redshift are established cosmological probes. The direct observable, the cluster mass function, is particularly sensitive to $S_8$ and is complementary to $3\times2\text{pt}$ methods \citep{allen_cosmological_2011}. Cluster cosmology, however, requires a selection of clusters with a high sample purity and established selection criteria, along with a method to estimate the mass of a cluster; cluster cosmology thus requires building a complete understanding of cluster astrophysical properties and how they evolve with redshift.

Building this understanding requires multi-wavelength observations: the state of the intracluster-medium can be inferred from its bremsstrahlung and line emission in X-rays \citep{cavaliere_x-rays_1976}, the effect of merger shocks from the existence of radio relics through synchrotron emission \citep{van_weeren_diffuse_2019}, and the integrated ICM pressure along the line of sight from the Sunyaev–Zeldovich (SZ) effect at sub-millimeter wavelengths \citep{sunyaev_microwave_1980}. Additionally, cluster galaxy contents can be studied through their optical emission with broadband imaging or through spectroscopy \citep{fasano_wings_2006,cariddi_characterization_2018}, and the underlying dark-matter halo can be studied via gravitational lensing. In principle, a scaling-relation for the masses of clusters can be built by combining multiple methods of estimating the mass of a halo. In practice, these relations can be subject to large biases and an intrinsic scatter \citep{evrard_model_2014}. Estimates of the cluster mass from the SZ effect, the X-ray luminosity ($L_X$), and even dynamical masses inferred from cluster kinematics, can be biased by the dynamical state of a cluster \citep[e.g.][]{aguado-barahona_velocity_2022}. Gravitational lensing may be the only nearly unbiased estimator for cluster mass; unfortunately, lensing masses are particularly sensitive to systematics and themselves have a large scatter due to shot noise.

Several surveys, motivated by the prospects of cluster cosmology, have worked to establish scaling relations across different subsets of clusters. Programs such as the Massive Cluster Survey \citep{ebeling_macs_2001}, CLASH \citep{postman_cluster_2012}, and the Hubble Frontier Fields \citep{lotz_frontier_2017} have targeted some of the most massive galaxy clusters at $z \gtrsim 0.2$ that can be observed at a high spatial resolution with the Hubble Space Telescope. Nearby ($z \lesssim 0.2$) clusters can be observed at similar resolution with ground-based observatories, which has motivated programs such as the Red-sequence Cluster Survey \citep{gladders_red-sequence_2005,gilbank_red-sequence_2011}, the Canadian Cluster Comparison Project \citep{hoekstra_canadian_2012}, the Multi-Epoch Nearby Cluster Survey \citep{sand_intracluster_2011}, the Local Cluster Substructure Survey \citep{zhang_locuss_2008}, Weighing the Giants \citep{von_der_linden_weighing_2014}, and numerous case-studies of nearby clusters \citep[e.g.][]{2015ApJ...805...40M}. 

Existing X-ray surveys are mass-complete to $M_{200} \gtrsim 2.5\times 10^{14} M_{\odot}$\footnote{200 here refers to the mass overdensity factor, such that $M_{200} = 200 \rho_c R_{200}^3(4\pi/3)$ and the critical density is $8\pi H(z)^2/3G$, and $H(z)=H_0(\Omega_m (1+z)^3 + \Omega_\Lambda)^{1/2}$.} in the local volume ($z \lesssim 0.15$) \citep{ebeling_rosat_2000}, but fully characterizing their properties requires deep optical observations along with supporting X-ray and sub-millimeter observations. The Local Volume Complete Cluster Survey, LoVoCCS, was created to address this need and cover a volume-complete sample of massive, X-ray selected galaxy clusters \citep{fu_lovoccs_2022}. Since it is volume complete, the full LoVoCCS sample enables several auxiliary science cases; this list includes studying the cores of galaxy ``superclusters''.

The large-scale structure of the universe is dominated by the cosmic-web, a massive, ``foamy'' structure composed of voids deprived of matter, dense knots containing many clusters, and thin filaments connecting everything together \citep{bahcall_large-scale_1988,einasto_structure_1994}. Structures on scales this large are not necessarily gravitationally bound, but they can be defined using clusters as a tracer with linking-lengths on the order of $\sim$ 10--50 Mpc depending on their redshift; these structures are galaxy superclusters \citep{oort_superclusters_1983,einasto_optical_2001}. The dense cores of superclusters, containing multiple rich clusters within $\lesssim 5\text{ Mpc}$, can be gravitationally bound, making them compelling to study as they host complex interactions between galaxy clusters in the early stages of a merger \citep[e.g.][]{higuchi_shapley_2020,monteiro-oliveira_unveiling_2022}. Additionally, with dynamical timescales on the order of $\sim 10\text{ Gyr}$ and smaller over-densities ($\frac{\bar{\rho}}{\rho_c} \lesssim 10$), these structures have yet to virialize and still contain traces of their formation history \citep{einasto_evolution_2021}



Superclusters have masses on the order of $10^{15}$--$10^{16} M_{\odot}$, with their gravitationally bound cores typically having $M \sim 10^{15} M_{\odot}$. Much of the matter in superclusters is in the clusters themselves: rich clusters contribute $\sim 75\%$, supercluster gas contributes $\sim 10\%$, and poor groups/isolated-galaxies contribute the remaining $\sim15\%$ of the total mass \citep{einasto_galaxy_2025}. 
{ $\Lambda$CDM places strict upper-bounds on the sizes and masses of structures in the Universe; therefore, weighing superclusters serves as a test of the cosmology \citep{einasto_galaxy_2025}. }
Beyond this, the cores of superclusters are among the most rich environments in the universe, with excesses of gravitationally-bound X-ray luminous clusters; these regions are excellent environments for studying merging clusters in a variety of stages \citep{bohringer_cosmic_2021-2,bohringer_cosmic_2021-1}. 


Recent studies have begun examining the distribution of dark matter in superclusters, their total mass, and have predicted which regions of superclusters consist of gravitationally bound structures \citep[e.g.][]{heymans_dark_2008,higuchi_shapley_2020,monteiro-oliveira_unveiling_2022,mondelin_dissecting_2025,sifon_chances_2025}.
We collected observations of three superclusters whose members were partially covered by LoVoCCS: the Hercules Supercluster (MSCC-474) and the Abell 2029 Superclusters (MSCC-457 \& MSCC-458) \citep{chow-martinez_two_2014}.
The goal of our survey is to estimate their masses via weak gravitational lensing to confirm existing dynamical and X-ray masses, and infer which clusters are currently out of equilibrium (interacting, merging) by looking for discrepancies between these mass estimates.

In this paper, we introduce the latest version of the data reduction pipeline for LoVoCCS and demonstrate its fidelity by carrying out a weak lensing analysis of the Hercules Supercluster. First, we provide background on LoVoCCS (Section \ref{sec:lv}) and the Hercules Supercluster (Section \ref{sec:scl}). Then, the fundamentals of weak gravitational lensing are reviewed, and problems unique to estimating the mass of a supercluster via weak lensing are discussed (Section \ref{sec:wl}). We then review the new, ``Third Generation Pipeline'' for LoVoCCS (Section \ref{sec:gen3}) with an emphasis on the changes that have taken place since the original data reduction. Extensive validation tests are carried out in Section \ref{sec:valid} and the lensing analysis of Hercules is summarized in Section \ref{sec:wl_herc}. We assume a concordance flat-$\Lambda$CDM cosmology ($h = 0.7$, $\Omega_m$ = 0.3) and assume that structures in Hercules are located at $z=0.04$\footnote{Separate clusters in Hercules are located at slightly different redshifts, ranging from $z\sim 0.03 - 0.05$; uncertainties introduced by placing them at the median redshift will be swamped by weak-lensing shape noise.}, with a distance scale of $\sim 0.79 \text{ kpc} \text{ arcsec}^{-1}$.

\section{Background}

\subsection{LoVoCCS}\label{sec:lv}

A core goal of LoVoCCS is to use weak gravitational lensing to estimate the mass of nearby clusters using on deep optical observations which cover \emph{at least} the virial radius of a cluster. Supporting X-ray observations come from archival data taken with the Chandra X-ray Observatory \citep{2000SPIE.4012....2W} and XMM-Newton \citep{jansen_xmm-newton_2001}. The initial optical component of the project was supported by a NOIRLab Survey Program (PI: Ian Dell'Antonio) carried out with the Dark Energy Camera (DECam) on the 4m Victor M. Blanco telescope \citep{fu_lovoccs_2022,fu_lovoccs_2024}.

Clusters were selected from the Meta-Catalog of X-Ray Detected Clusters of Galaxies \citep[MCXC-I;][]{piffaretti_mcxc_2011} with X-ray luminosity $L_{X500} > 10^{44} \text{ erg s}^{-1}$. We required that the Galactic extinction be $<0.5\text{ mag}$ in $r$-band at the cluster center and that no-more than 30 stars with $G \le 13$ are within $15'$ of the peak X-ray luminosity \citep{gaia_collaboration_gaia_2023}. Additionally, clusters had to be observable with an airmass $\lesssim1.5$, which limits the survey's coverage to declinations $\delta < 20^{\circ}$. The survey covers clusters in the redshift range from $0.03 < z < 0.12$; the lower bound is set by the requirement that $r_{200}$ be covered by DECam's $\sim 3\text{ deg}^2$ field-of-view and the upper bound is set by the mass-completeness of all-sky X-ray surveys \citep{ebeling_rosat_2000}. This resulted in an initial sample of $107$ galaxy clusters.

\begin{figure}
    \centering
    \includegraphics[width=0.98\linewidth]{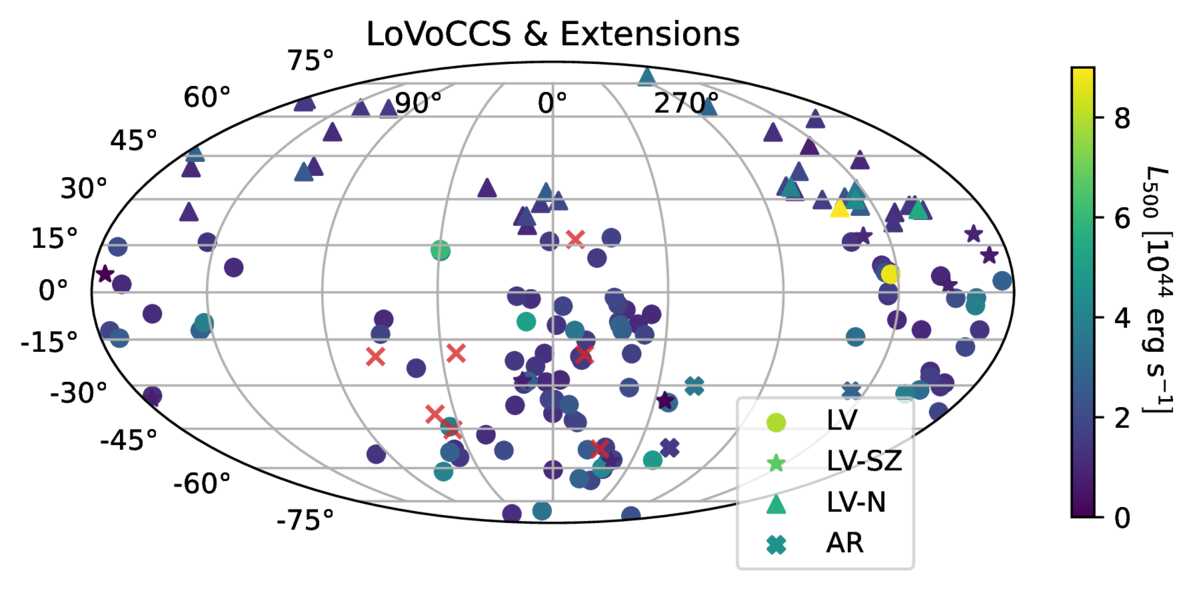}
    \caption{The complete sample of LoVoCCS clusters, colored by their X-ray luminosity, and drawn over a Mollweide projection. The markers denote the origin of a cluster from our sample, including the original LoVoCCS catalog (LV; circle), the LoVoCCS-SZ extension (LV-SZ; star), the LoVoCCS-North extension (LV-N; triangle), and the archival extension (AR; cross). Clusters with incomplete observations are drawn with red X's.}
    \label{fig:lv_plus}
\end{figure}

A major justification for LoVoCCS is to serve as a precursor dataset to the Vera C. Rubin Observatory's upcoming Legacy Survey of Space and Time (LSST) \citep{ivezic_lsst_2019}. This means that the observations must be comparable to early LSST data in their depth and photometric/astrometric precision. To reach this goal, the LoVoCCS dataset has been reduced entirely within the LSST Science Pipelines (LSP) \citep{10.71929/rubin/2570545}; this ensures that the observations have been reduced with the latest algorithms developed for use on LSST, keeping them as similar as possible. This also places the data products from LoVoCCS in a format compatible with future LSST science, serving as deep Y0 static-sky templates and supporting early-science validation in cluster fields \citep[e.g.][]{fu_decam_2024}.

The survey's initial goal was to reach LSST-Y1 depth across five bands ($ugriz$); strict seeing requirements were placed on the lensing band, $r$, and supporting bands reached a comparable depth to enable robust photometric redshifts \citep{lsst_science_collaboration_lsst_2009} and to support detailed studies of the galaxy population. The primary observing runs were carried out from 2019-2022, however poor weather and interruptions due to COVID-19 left the survey nearly complete across $griz$, but incomplete in $u$; we carried out several extensions taking place from 2023-2025 to complete the observations.

Throughout these extensions, in addition to completing the initial sample of clusters, several lower-priority auxiliary programs were proposed to improve the SZ-completeness. When applying the same mass threshold to masses inferred via the SZ-effect (selected from \citet{planck_collaboration_planck_2016}), there are an additional $17$ clusters absent from the X-ray sample. These objects are particularly interesting, as they can indicate what underlying mechanisms are responsible for incompleteness in samples of clusters used for cosmology.

With the second release of the Meta-Catalogue of X-ray detected Clusters of galaxies (MCXC-II; \citet{sadibekova_mcxc-ii_2024}), an additional $28$ clusters pass the selection criteria for LoVoCCS. $17$ of these meet the requirements for LoVoCCS's depth in at least two bands from archival data alone, and have been added to the queue for data reduction; they will be included in future analyses. A northern extension to LoVoCCS is also being carried out with the Hyper Suprime Cam (HSC) on the Subaru Telescope \citep{2012SPIE.8446E..0ZM} which, when complete, will add an additional $37$ clusters.

In total, the LoVoCCS plus extensions catalog contains $138$ galaxy clusters in the southern sky (Fig.~\ref{fig:lv_plus}). Of the initial $107$ clusters, $100$ were completed with the target LoVoCCS depth, $4$ were completed with partial $u$-band coverage, and three clusters (A3104, A514, A2589) have been left incomplete. The SZ-extension provides an additional $14$ targets, with $10$ having complete observations and $4$ being incomplete. The archival observations adds $17$ partially-observed clusters, bringing the total number of clusters in the current queue for data reduction to $104 + 10 + 17 = 131$. The southern portion of LoVoCCS plus extensions is presented in Table \ref{tab:lv}.

The datasets presented in LoVoCCS I \citep[LVI]{fu_lovoccs_2022} and LoVoCCS II \citep[LVII]{fu_lovoccs_2024}, henceforth referred to as the ``Gen2'' data, cover a total of $\sim 60$ nearby clusters and have led to several additional projects. This includes a search for intracluster filaments \citep{shinde_weak-lensing_2025}, efforts to classify the dynamical state of a cluster from optical and lensing data alone \citep{kong_merger-induced_2026}, and correcting for non-linearity in cluster lensing shear \citep{liu_measurement_2024}. Since LVII, the data reduction pipeline has undergone a major overhaul, which is discussed in Section \ref{sec:gen3}; this pipeline and its associated data products will be referred to as the ``Gen3'' versions. Partial reductions of this data have already spurred several additional projects, including an initial assessment of the intracluster light of Abell 3667 \citep{englert_intracluster_2025} and an analysis of the extended profile of the Tucana Dwarf galaxy \citep{huang_measurement_2025}.

\begin{figure}
    \centering
    \includegraphics[width=0.98\linewidth]{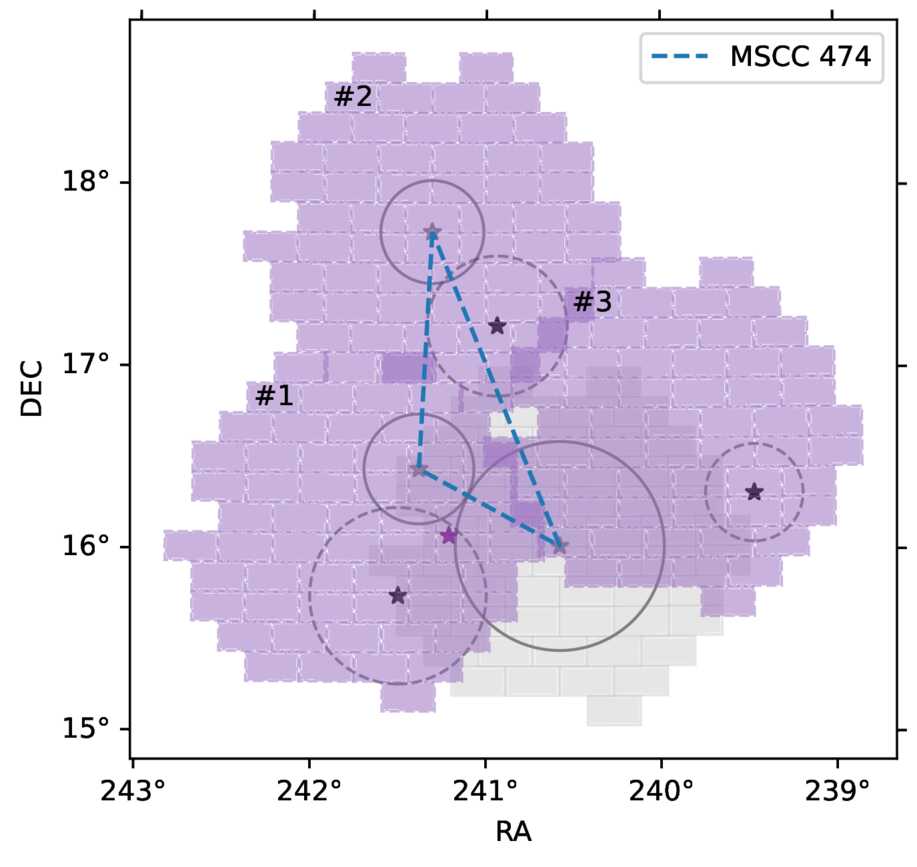}
    \caption{A sketch of the Hercules Supercluster with the positions of major clumps starred; circles extend to their respective $r_{200}$. Numbered magenta regions are the observations carried out to survey the supercluster, and the unlabeled gray region is the original pointing from LoVoCCS.}
    \label{fig:hercules_pointing}
\end{figure}

\subsection{The Hercules Supercluster}\label{sec:scl}

\begin{figure*}
    \centering
    \includegraphics[width=0.98\linewidth]{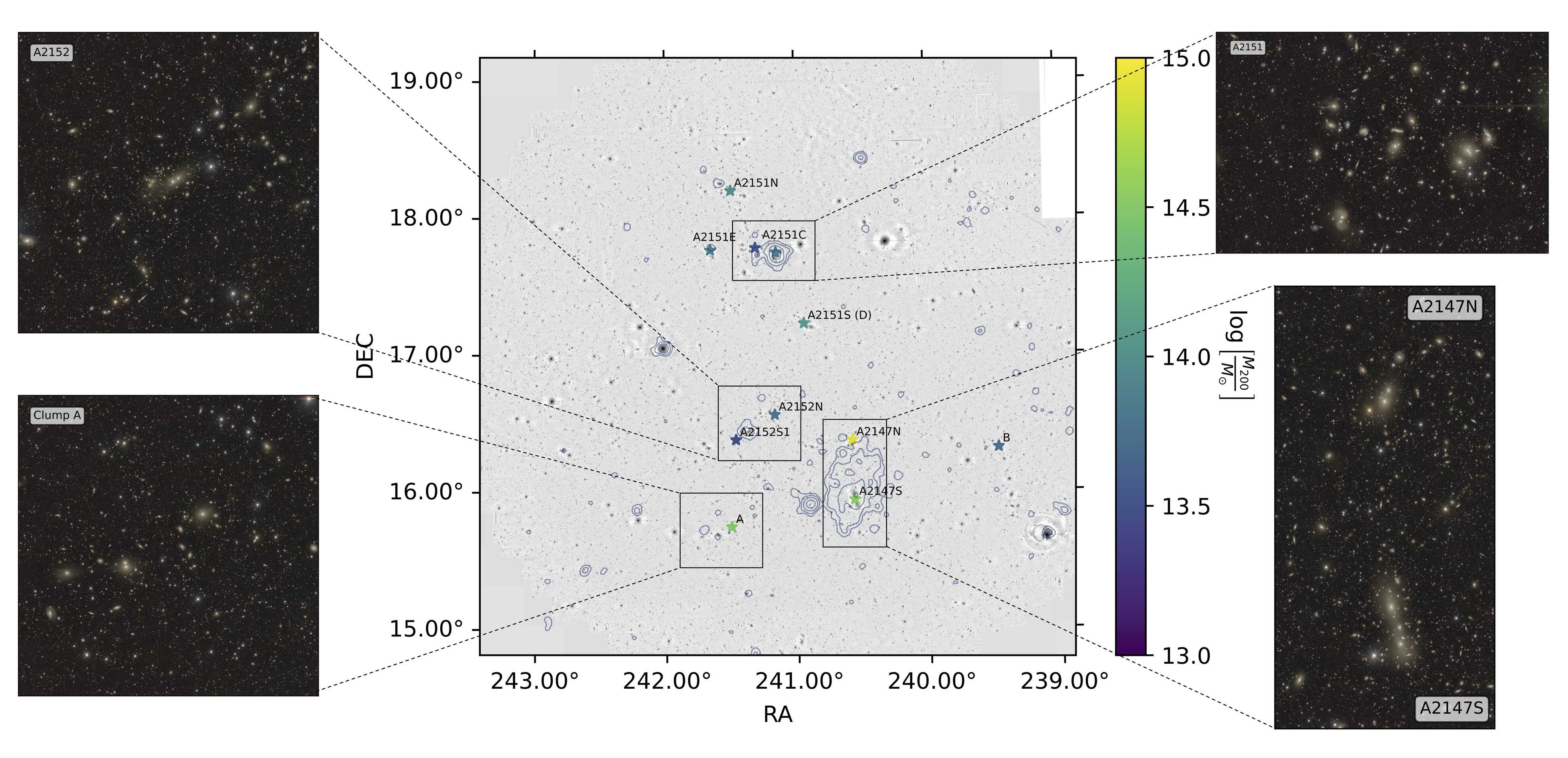}
    \caption{A $\sim 16\text{ deg}^2$ field covering the densest complex of the Hercules Supercluster, composed of A2147, A2152, A2151, and two additional ``clumps''. The central panel contains the stretched, inverted, stacked DECam observations (\ref{sec:gen3}) with contours from the ROSAT All-Sky Survey overlaid \citep{voges_rosat_1993}. Individual RGB cutouts showcase the $irg$ sky-corrected stacks.}
    \label{fig:hercules_full}
\end{figure*}

The dense core of the Hercules supercluster, containing Abell 2147, 2151, and 2152, was first identified in the late 1930s \citep{shapley_structural_1934}.
In 1979, Tarenghi obtained $> 150$ redshifts over a 28 deg$^2$ field and confirmed that these clusters were located at similar redshifts \citep{tarenghi_hercules_1979,tarenghi_hercules_1980}. A followup kinematic study by \citet{barmby_kinematics_1998} included $468$ redshifts and confirmed that the supercluster was gravitationally bound with a total mass of $\sim (7.6 \pm 2) \times 10^{15}~M_{\odot}$. 

The most recent analysis of the supercluster as a whole was carried out by \citet[][hencforth MO]{monteiro-oliveira_unveiling_2022}, who analyzed the kinematics of the complex using a magnitude-complete sample of SDSS spectra across a $\sim 16\text{ deg}^2$ region. \citetalias{monteiro-oliveira_unveiling_2022} identified several additional structures, ``Clump A'' and ``Clump B'', which were gravitationally bound to the primary clusters and contribute significantly to the total mass of the structure. Masses are also estimated for the complex, though \citetalias{monteiro-oliveira_unveiling_2022} cautions that the dynamical mass may be biased as the clusters in the complex are not in equilibrium and demonstrates this using simulated observations. 

It is worthwhile to summarize the established properties of each complex in the supercluster:

\emph{Abell 2147:} Abell 2147 lacks a single paper analyzing its extended X-ray morphology (Fig.~\ref{fig:hercules_full}). Nonetheless, several authors have noted that the structure is bi-modal and distinguished between the northern (A2147N) and southern (A2147S) clumps. X-ray and dynamical mass estimates tend to disagree, with the former assigning it a mass $M_{200} \sim 3\times10^{14}~M_{\odot}$ and the latter assigning this cluster a mass of $M_{200} \sim 1\times10^{15}~M_{\odot}$ \citep{barmby_kinematics_1998,monteiro-oliveira_unveiling_2022,sadibekova_mcxc-ii_2024}; based on the extended X-ray morphology, \citetalias{monteiro-oliveira_unveiling_2022} concluded that the complexes are most likely outgoing following a collision $\sim 0.8\text{ Gyr}$ ago.

\emph{Abell 2152:} Abell 2152 (A2152) is a ``double cluster'' that features multiple BCG's from two distinct clusters overlapping along the line-of-sight. The reported centroid of A2152 has varied by several arcminutes across literature due to the presence of large groups north and south of the X-ray morphology, which itself is blended with a background cluster \citep{blakeslee_lensing_2001}. \citetalias{monteiro-oliveira_unveiling_2022} most recently decomposed the complex into a northern and southern group, but noted that tweaking the smoothing scale of their density map leads to a single peak located between the two structures; this could make the centroid more consistent with the observed X-ray source, but its exact position is not discussed.

\emph{Abell 2151:} Abell 2151, otherwise known as the ``Hercules Cluster'', is composed of five separate ``clumps'' that, unless otherwise noted below, are all gravitationally bound \citep{barmby_kinematics_1998,monteiro-oliveira_unveiling_2022}. The northern clump, A2151N, does not have any clear X-ray morphology and is consistent with a group currently in-falling and merging with the cluster core. A2151E, east of the cluster's centroid, has a weak, but non-zero, X-ray emission and is most likely gravitationally bound to A2151N but not the central clumps. The components nearest to the centroid, A2151C-F and A2151C-B, are associated with the faint and bright peaks respectively in the X-ray emissions and are actively merging \citep{tiwari_hercules_2021}. A2151S was located during followup observations with the ROSAT High Resolution Imager \citep{huang_high-resolution_1996} and is likely infalling towards the center of the cluster. Overall, this cluster is unique due to its excess of blue and spiral galaxies, which indicates that the system is in an early dynamical state where star formation has yet to be fully quenched \citep{agulli_built-up_2016}.

\emph{Clump A \& B:} ``Clump A'' was first noted by \citetalias{monteiro-oliveira_unveiling_2022}; it is a massive structure without any clear X-ray morphology in ROSAT. It is gravitationally bound to A2152 and A2147 and is less rich compared to other structures in the field. Similarly, ``Clump B'' was detected by \citetalias{monteiro-oliveira_unveiling_2022}, lacks any X-ray emission, and is gravitationally bound to A2147. 

Three additional pointings covering Hercules were requested to study the primary structures identified in \citetalias{monteiro-oliveira_unveiling_2022} and estimate their weak lensing mass (Fig.~\ref{fig:hercules_pointing}). The total on-sky area covered is $\sim 10$\text{ deg}$^2$, encompassing A2147, A2151, A2152, Clump A, and Clump B (Fig.~\ref{fig:hercules_full}). The depth is matched to LoVoCCS and the total exposure times per field and per band are listed in Table \ref{tab:point}. In principle, the observations are sensitive to structures with projected density contrasts $\Delta \Sigma \sim 10^{14}~M_{\odot}~\text{ Mpc}^{-2}$ with $2\sigma$-confidence (assuming background sources at $z_s\sim1$ and a lens at $z_l\sim0.04$), which is sufficient to detect the most massive structures in the complex: A2147, A2151, A2152, and Clump A.

\begin{deluxetable*}{c|c|c|c|c|c|c|c}[t]
    \centering
    \tablehead{
    \colhead{Field} & \colhead{RA (deg)} & \colhead{DEC (deg)}& \colhead{\emph{u}} & \colhead{\emph{g}} & \colhead{\emph{r}} & \colhead{\emph{i}} & \colhead{\emph{z}}
    }
    \startdata
    1 & 241.75144 & 16.06458 & 5100 & 2114 & 4088 & 3310 & 1724 \\
    2 & 241.31246 & 17.74852 & 5100 & 2135 & 4086 & 3310 & 2026 \\
    3 & 239.92792 & 16.62 & 5100 & 2034 & 4112 & 3310 & 2080 \\
    LV & 240.54984 & 15.91987 & 5100 & 2300 & 4200 & 3400 & 2300 \\
    \enddata
    \caption{Table of pointings and exposure times (in seconds) for the Hercules supercluster, including exposure times from the original observations of A2147 for LoVoCCS.}
    \label{tab:point}
\end{deluxetable*}

\section{Gravitational Lensing}\label{sec:wl}

Estimating the mass of low-redshift clusters spanning a wide field of view is a non-trivial process. Outside of reducing the observations to produce images and catalogs suitable for weak lensing science, there are two problems that must be addressed:

\begin{enumerate}
    \item How to build a robust estimator of the mass in foreground structures in the presence of massive background deflectors.
    \item How to map out the cluster weak lensing signal and select statistically significant structures.
\end{enumerate}

In this section, we review the formalism for cluster weak lensing. The problem of selecting statistically significant structures from a mass reconstruction will be addressed in Section \ref{sec:wl_herc}.

\subsection{Single Lens Plane}

Assuming that there is a source located behind some thin massive deflector, the observed position of the source ($\theta$) and its true position ($\beta$) are linked through the lens equation

\begin{equation}\label{eq:lens}
    \beta = \theta - {\alpha}(\theta) \frac{D_{LS}^{(A)}}{D_S^{(A)}},
\end{equation}

where $\hat{\alpha}$ is the deflection angle in the lens plane, $D_{LS}^{(A)}$ is the angular diameter distance between the lens and source, and $D_S^{(A)}$ is the angular-diameter distance to the source \citep{blandford_cosmological_1992}. The angle vectors are measured in a tangent plane with some orthogonal set of axes $(\theta_1,\theta_2)$.

The deflection angle is linked to the projected gravitational potential $\psi$:

\begin{equation}\label{eq:psi_sig}
    {\alpha} = \nabla \psi, \quad \frac{1}{2} \nabla^2 \psi = \frac{\Sigma}{\Sigma_c},
\end{equation}

where $\Sigma$ is a projected mass density in the lens plane, and $\Sigma_c$ is the critical projected mass density:

\begin{equation}
    \Sigma_c = \frac{c^2 D_S^{(A)}}{4\pi G D_{LS}^{(A)} D_L^{(A)}}.
\end{equation}

In the case where multiple deflectors exist in the lens plane and individual deflections are small, Equation \ref{eq:psi_sig} is linear in the projected density, so the deflection angles add together, e.g. the net deflection in the presence of multiple deflectors at the same redshift is

\begin{equation}
    {\alpha} = \sum_{i=1}^{N_D} {\alpha}_i,
\end{equation}

where there are $N_D$ deflectors in the lens plane. Assuming the deflectors are spherically symmetric, this can be reduced to the magnitude of a deflection and a unit-vector specifying the direction of the deflection

\begin{equation}\label{eq:sph_defl}
    \hat{\alpha}(\theta) = \sum_{i=1}^{N_D} \left| \alpha(\theta_i - \theta)\right| \times \frac{\theta_i - \theta}{\left| \theta_i - \theta \right|},
\end{equation}

where $\theta_i$ is the center of each deflector. Images of background sources are transformed from the source plane to the lens plane; the Jacobian of this transformation is called the magnification tensor:

\begin{equation}\label{eq:mag}
    A_{ij} = \frac{\partial \beta_i}{\partial \theta_j} = 1 - \frac{\partial \alpha_i}{\partial \theta_j} \frac{D_{LS}^{(A)}}{{D_S}^{(A)}}.
\end{equation}

This can be decomposed into a traced and trace-free term, often parameterized by the convergence ($\kappa$) and shear ($\gamma_1,\gamma_2$):

\begin{equation}\label{eq:amp_conv_sh}
    A_{ij} = (I - \kappa) - \begin{bmatrix}
        \gamma_1 & \gamma_2 \\ \gamma_2 & - \gamma_1 \\
    \end{bmatrix}.
\end{equation}

In the weak lensing limit ($\kappa, \, \gamma \ll 1$), both of these quantities are linear in the gravitational potential:

\begin{equation}\label{eq:psi_shear}
    \kappa = \frac{1}{2} \nabla^2 \psi = \frac{\Sigma}{\Sigma_c}, \quad \gamma_1 = \frac{1}{2} \left( \partial_1^2 \psi - \partial_2^2 \psi \right), \quad \gamma_2 = \partial_1\partial_2 \psi.
\end{equation}
    
For the special case of spherically symmetric deflectors, it is convenient to parameterize the projected mass density in the form of a dimensionless function

\begin{equation}
    \Sigma(r) = \Sigma_s \times f\left(\frac{r}{r_s}\right),
\end{equation}

where $\Sigma_s$ is some scale density, $r_s$ is some scale length, and $r = D_L^{(A)} \theta$ is the distance between the deflector and a point in the lens plane; it is also convenient to define $F(x) = \int_0^x t f(t) dt$. For a coordinate system centered on the symmetric deflector, the deflection angle and its derivative can be written as

\begin{equation}\label{eq:sph_dimn}
    \hat{\alpha}(\theta) = \left[ \frac{8\pi G}{D_L^{(A)} c^2} \Sigma_s r_s^2 \right] F\left(x\right) \frac{\theta}{|\theta|},
\end{equation}

\begin{equation}\label{eq:sph_jac}
    \frac{\partial \hat{\alpha}}{\partial \theta} = \alpha(\theta) \left[ x \frac{f(x)}{F(x)} \frac{D_L^{(A)}}{r_s} - \frac{2}{\theta} \right] R_{\phi} + \frac{\alpha(\theta)}{\theta},
\end{equation}

where $x = \frac{D_L^{(A)} \theta}{r_s}$ and

\begin{equation}
    R_{\phi} = \begin{bmatrix}
        \cos^2(\phi) & \cos(\phi) \sin(\phi) \\ \cos(\phi) \sin(\phi) & \sin^2(\phi)
    \end{bmatrix},
\end{equation}

with $\phi$ some position-angle in the tangent plane.

\subsection{Multiple Lens Planes}\label{sec:multiplane}

\begin{figure}
    \centering
    \includegraphics[width=0.98\linewidth]{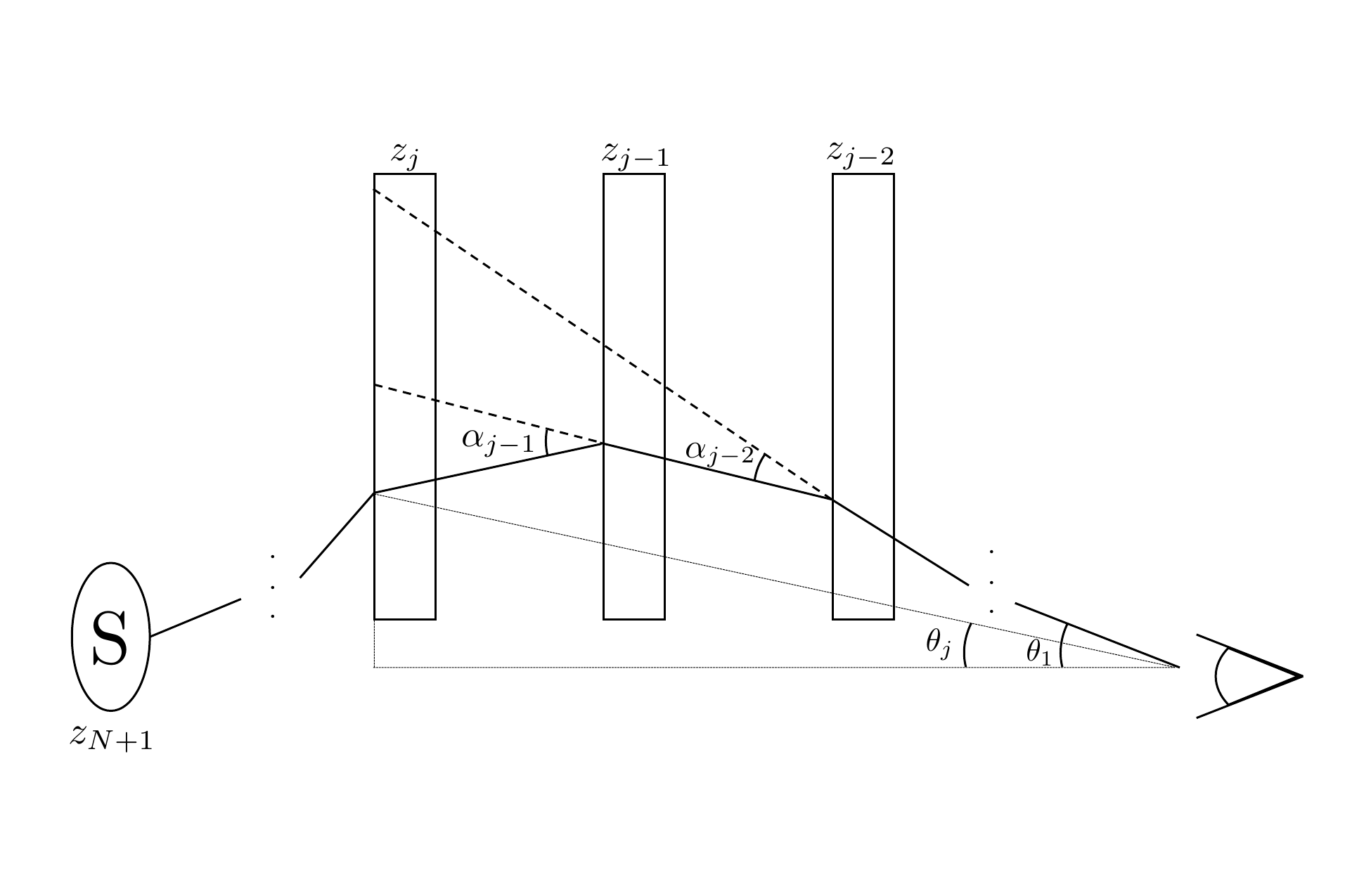}
    \caption{A diagram of the multi-plane geometry; each plane is labeled by their redshift. The observed position of an object is $\theta_1$, while the un-deflected position is $\beta = \theta_{N+1}$ and can be recovered by using Equation \ref{eq:rec_alpha}. { In principle, there are an arbitrary number of deflectors prior to plane $j$ and following plane $j-2$.}}
    \label{fig:multiplane_lensing}
\end{figure}

In practice, there are often multiple deflectors along the line-of-sight at different redshifts, which can also bend light from background sources. In the limit where these deflectors are thin relative to their separations\footnote{Clusters have characteristic sizes $r_{200} \sim 1~\text{Mpc}$. However, along the line-of-sight they are typically separated by distances on the order of $\sim 10^2\text{ Mpc}$, making this a reasonable approximation for our science case.}, the multi-plane lens equation must be used: 

\begin{equation}\label{eq:multi_lens}
    \theta_j = \theta_1 - \sum_{i=1}^{j-1} \alpha_i \frac{D_{ij}^{(A)}}{D_j^{(A)}},
\end{equation}

where $\theta_j$ is the position of an incident ray in plane $j$, $D_{ij}^{(A)}$ is the angular diameter distance between planes $i$ and $j$, and $D_j^{(A)}$ is the angular diameter distance to plane $j$ from the observer; the geometry is shown in Figure \ref{fig:multiplane_lensing}. Over a collection of $N$ lens planes, we can interpret $\theta_{N+1}$ as the un-deflected position of an object and define $z_{N+1} \equiv z_s$. Assuming a flat cosmology, Equation \ref{eq:multi_lens} and its derivative can be written recursively as 

\begin{widetext}
\begin{eqnarray}\label{eq:rec_alpha}
    \theta_{j+1} = \theta_j \frac{D_j}{D_{j+1}} \frac{\Delta_{j+1,j}}{\Delta^*} - \theta_{j-1} \frac{D_{j-1}}{D_{j+1}} \frac{\Delta_{j+1,j}}{\Delta_{j,j-1}} - \alpha_j \frac{\Delta_{j+1,j}}{D_{j+1}},
\end{eqnarray}

\begin{eqnarray}\label{eq:rec_jac}
    \frac{\partial \theta_{j+1}}{\partial \theta_1} = \frac{\partial \theta_j}{\partial \theta_1} \frac{D_j}{D_{j+1}} \frac{\Delta_{j+1,j}}{\Delta^*} - \frac{\partial \theta_{j-1}}{\partial \theta_1} \frac{D_{j-1}}{D_{j+1}} \frac{\Delta_{j+1,j}}{\Delta_{j,j-1}} - \frac{ \partial\alpha_j}{\partial \theta_j} \frac{\partial \theta_j}{\partial \theta_1} \frac{\Delta_{j+1,j}}{D_{j+1}},
\end{eqnarray}

\end{widetext}

where the $\Delta$'s are defined as

\begin{equation}
    \Delta_{lm} = D_l - D_m, \quad \frac{1}{\Delta^*} = \frac{1}{\Delta_{j+1,j}} + \frac{1}{\Delta_{j,j-1}}.
\end{equation}

$D_j$ denotes the co-moving distance to plane $j$ and $D_{ij}$ denotes the co-moving distance between planes $i$ and $j$. By letting $j=N$, Equation \ref{eq:rec_jac} provides an explicit formula for the magnification tensor.

\subsection{Weak Lensing}\label{ch:wl}

A core concept in weak gravitational lensing is that the signal due to gravitational lensing can be detected by studying the average shape of background objects. Consider a single object whose image is lensed in the weak limit; the second moments of the image transform as a rank-2 tensor

\begin{equation}
    I_{ij} \simeq \frac{\partial \theta_j}{\partial \beta_{j'}}\frac{\partial \theta_i}{\partial \beta_{i'}} I_{i'j'},
\end{equation}

where $I_{i'j'}$ are the original moments of the source \citep{bernstein_shapes_2002}. There exists a combination of the moments which, due to the random orientation of galaxies along the line of sight, vanishes on average; these are the ellipticities

\begin{equation}\label{eq:ellip}
    e_1 \equiv \frac{I_{11} - I_{22}}{I_{11} + I_{22}}, \quad e_2 \equiv \frac{2 I_{12}}{I_{11} + I_{22}}.
\end{equation}

These are zero on average, \emph{except in the presence of a massive lens in the foreground}; in that case, the mean ellipticity of a background source is directly proportional to the reduced shear: $g_i = \frac{\gamma_i}{1-\kappa}$.





The ``tangential shear'' (and its corresponding tangential reduced shear) can be defined with respect to the center of a lens

\begin{equation}\label{eq:tanshear}
    \gamma^{(+)} \equiv -\gamma_1 \cos(2\phi) - \gamma_2 \sin(2\phi),
\end{equation}

which is orthogonal to the ``cross shear''

\begin{equation}
    \gamma^{(\times)} \equiv \gamma_1 \sin(2\phi) - \gamma_2 \cos(2\phi).
\end{equation}

For a spherically symmetric deflector, the tangential shear is a radial function and the cross shear vanishes. The ellipticity and shear are spinors\footnote{Rotating an image by $\phi$ rotates the shears by $2\phi$.}, e.g. the complex-valued ellipticity can be written

\begin{equation}
    e = e_1 + i e_2.
\end{equation}

Thus, the complex shear due to lensing by a spherical source is

\begin{equation}
    \gamma = \gamma(\theta) e^{2i\phi}.
\end{equation}

Before discussing the systematics involved in precision shear estimation, it is helpful to review how we have implemented single- and multi-plane lens models to compute $g(\theta)$. First, for a collection of lenses at the same redshift, the procedure is:

\begin{enumerate}
    \item Define a collection of deflectors (masses and density profiles) and center positions.
    \item Assign each deflector a scale density, scale radius, then define some $f(x)$ and $F(x)$.
    \item Use Equation \ref{eq:sph_dimn} and \ref{eq:sph_jac} to compute $\hat{\alpha}$ and $\frac{\partial \alpha}{\partial \theta}$.
    \item Add together each $\frac{\partial \alpha}{\partial \theta}$ and compute the magnification tensor (Equation \ref{eq:mag}).
    \item Compute the convergence by taking the trace of Equation \ref{eq:amp_conv_sh} and the shear by computing the traceless component of $A_{ij}$.
    \item Compute the reduced shear.
\end{enumerate}

This procedure has been wrapped in a \texttt{SymmetricDeflectorSky} base class and implemented through \texttt{SISDeflector} and \texttt{NFWDeflector} classes; this lets us model lenses as a singular-isothermal sphere (SIS) or a Navarro-Frenk-White \citep[NFW;][]{navarro_universal_1997} deflector. A collection of \texttt{SymmetricDeflectorSky} objects can be wrapped into a \texttt{CompositeDeflector} object.

For the multiplane case, the following procedure is implemented:

\begin{enumerate}
    \item Starting from $j=0$, compute the position on plane $j=1$ using Equation \ref{eq:rec_alpha}, and the Jacobian of the transformation from the observed position with Equation \ref{eq:rec_jac} (let $\theta_{-1} = 0$ during this step).
    \item Repeat this up to $j=N$, the source plane; this provides the full magnification tensor and the original position of a source $\theta_{N+1}$.
    \item Compute the convergence by taking the trace of Equation \ref{eq:amp_conv_sh}, and the shear by computing the traceless component of $A_{ij}$.
    \item Compute the reduced shear.
\end{enumerate}

\begin{figure}
    \centering
    \includegraphics[width=0.98\linewidth]{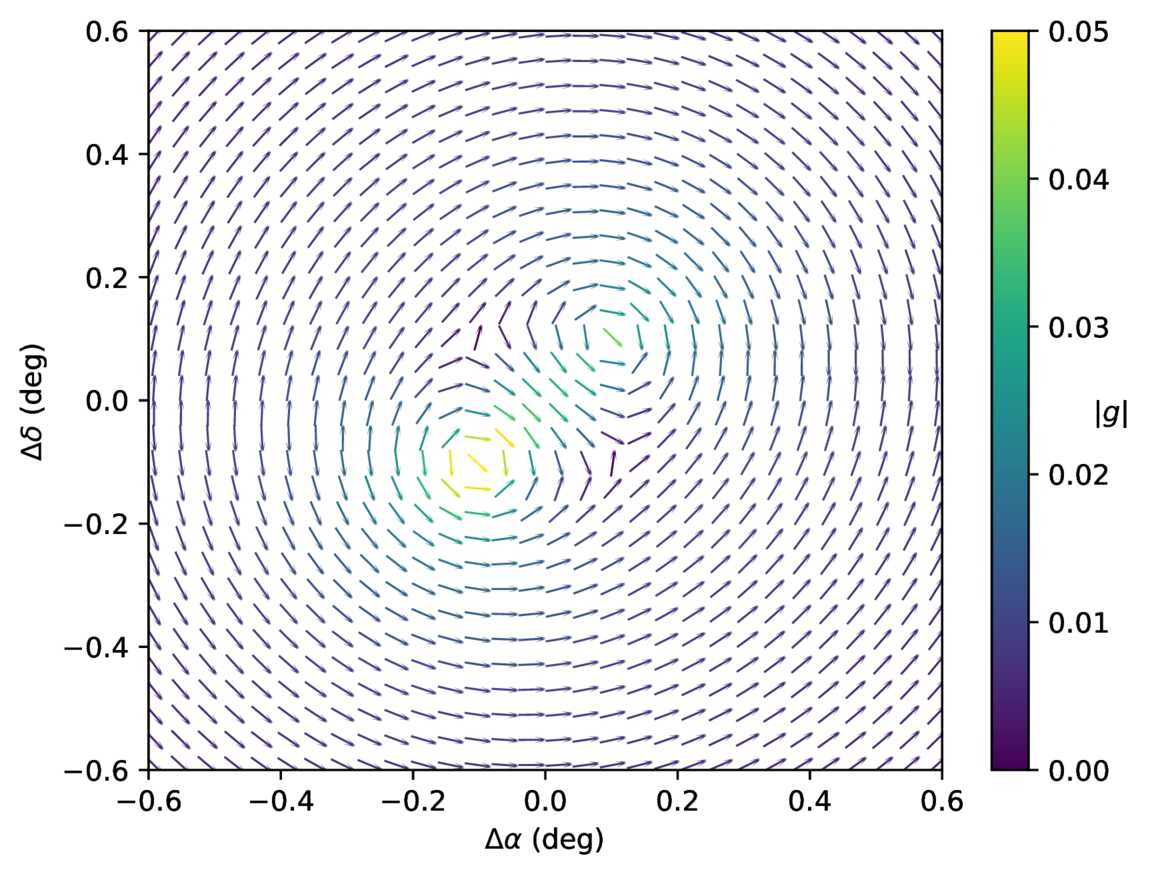}
    \caption{The shear-field for a source at $z=1$ due to a ``double cluster'' with two deflectors of equal mass ($M_{200} = 5\times10^{14} M_{\odot}$) that are placed at different redshifts: $z=0.13$ at $(-0.1,-0.1)$ and $z=0.04$ at $(0.1,0.1)$.}
    \label{fig:mock_dcl}
\end{figure}

In practice, when this is applied to a catalog of galaxies, checks must be made to confirm whether a source is in front of the current lens plane and raise a flag to stop updating the tensor when this occurs. To higher orders, shearing by successive lens planes can introduce physical rotations in the moments of order $\gamma^2$; these are negligible in the weak lensing limit and our approximation is correct to first order in $\kappa$, $\gamma$, and $\kappa\gamma$ \citep{petkova_glamer_2014}. An example multi-plane lens is presented in Fig.~\ref{fig:mock_dcl}.

\subsection{Shear Calibration}\label{ch:shearcal}

The shear due to lensing, or simply the lensing signal, is typically $\sim10 \times$ smaller than the intrinsic ellipticity of a galaxy. Therefore, systematics that can bias the ellipticity must be corrected to the percent level or better; applying these corrections is called  ``shear calibration'' \citep{hirata_shear_2003,heymans_shear_2006}. The goal of weak lensing is to estimate the average shear of a population of sources with some distribution function $P(e)$; on average, the ellipticity of this population is zero, but that distribution is augmented by a foreground lens which shifts the distribution $P(e) \to P_g(e)$. Consider an estimator for the average ellipticity of some galaxy population; given some choice of weights $w(e)$ which minimizes the variance, the estimator is

\begin{equation}
    \hat{\gamma} \equiv \sum_i w_i \hat{e}_i,
\end{equation}

where the weights are normalized ($\sum_i w_i = 1$). For a population of galaxies with random orientation, the expected value of this estimator vanishes:

\begin{equation}
    \langle \hat{\gamma} \rangle = \iint d^2e~w(e) P(e) e = 0.
\end{equation}

When the population is sheared by some small $\delta g$, exchange $P(e) \to P_{\delta g}(e) \simeq P(e) + \frac{\partial P}{\partial g} \delta g$; to first order in shear, the estimator becomes

\begin{equation}
    \langle \hat{\gamma} \rangle  = \delta g \times \iint d^2e~w(e) \frac{\partial P}{\partial g} e \equiv \delta g \times R.
\end{equation}

$R$ is called the \emph{shear responsivity}, and it must be modeled during shear calibration \citep{bernstein_shapes_2002}. More explicitly, the response is a matrix defined by taking the derivative of a shear-estimator

\begin{equation}\label{eq:res_deriv}
    R_{ij} \equiv\frac{\partial \langle \hat{\gamma_i} \rangle}{\partial g_j}.
\end{equation}

Shear calibration is made challenging by are several other complications as well (see \citet{mandelbaum_weak_2018} for a complete review of the problem):

\begin{enumerate}
    \item Objects have been convolved with a point spread function (PSF) that varies across the focal plane, ``smearing'' their shape \citep{kaiser_method_1995,bernstein_shapes_2002}. Additionally, the PSF can be discontinuous on small scales\footnote{Strictly-speaking, the PSF is a smooth function across the focal plane \emph{in a given visit}. However, when individual sources are aligned and stacked to form a deep-coadded image, sources on the boundary between individual detectors and sources with some fraction of masked-pixels are convolved with a dis-continuous PSF.}, contributing to systematic uncertainties when modeling the PSF with a static kernel.
    \item Ellipticities of overlapping galaxies, particularly blends, are often biased; recognized blends can be culled from a catalog to mitigate this, but un-recognized blends contribute to systematic uncertainty in these analyses \citep{maccrann_dark_2022,liang_catalog-based_2026}.
    \item The noise present in a stacked image is correlated; accurate shape measurements must be optimally weighted to maximize the signal \citep{kacprzak_measurement_2012}.
\end{enumerate}

Initially, shear calibration was carried out by first measuring ellipticities with an explicit correction for the PSF \citep{kaiser_method_1995}. Given an understanding of the underlying galaxy population, the impact of other sources of uncertainty can be quantified by injecting galaxies with known shears into simulated observations with a realistic PSF. This is the approach taken by the HSC-Y1 shear calibration algorithm previously used during Gen2 data reductions, where calibrated shears were estimated via

\begin{equation}
    \hat{g}_i = \frac{1}{1 + m_i} \left( \frac{e_i}{2R_i} - c_i \right).
\end{equation}

Here, $m_i$, $c_i$, and $R_i$ are estimators for the multiplicative bias, the additive bias, and the response, respectively \citep{mandelbaum_weak_2018}. However, this calibration was built for a different telescope with different observing conditions.\footnote{Notably, at Mauna Kea HSC typically delivers a $\sim 0.6^"$ or better seeing, compared to the $\sim 1^"$ seeing delivered by DECam in $r$-band.} Moreover, several adjustments have been made since the Gen2 data reductions, including overhauling the PSF modeling, deblending, and detection algorithms; combined, these can bias the response $R$ through implicit selection criteria, causing $P(e)$ to deviate from its assumed distribution in the simulations. In principle, this can be corrected by building a ``global mass-calibration'' by injecting lensed galaxies into LoVoCCS observations (Helhoski et al., in-prep).

Over the past decade, an alternate approach in the form of ``self-calibration'' algorithms have emerged. Rather than calibrate with simulated data, self-calibration methods explicitly correct for the PSF by de-convolving the observations and estimating the response by either shearing the data \citep{huff_metacalibration_2017} or analytically estimating the response \citep{li_analytical_2023}. The resulting catalogs can be further processed to produce optimal weights for statistical averages from the data itself. The benefit of these schemes is that they are agnostic to one's choice of instrument or data reduction pipeline. To address the sources of systematic uncertainty introduced while applying the HSC shear calibration to LoVoCCS observations, we have built an implementation of the self-calibration algorithm \texttt{metadetect} \citep{sheldon_mitigating_2020}; this will be discussed in Section \ref{sec:meta}.

\subsection{Mass Estimation \& Reconstruction}\label{ch:mass_map}

The challenges for carrying out weak lensing are two-fold: first there is the problem of shear-calibration discussed above, and second the estimator produced---following shear calibration---has a large, intrinsic noise. This problem can be modeled rather simply: for a shear signal $g(\theta)$ that has been discretely sampled at some random set of on-sky galaxy positions $\{\theta_i\}$ and where each sample has some shot-noise $\hat{\sigma}_{e,i}$, the point estimator for the shear is

\begin{equation}
    \hat{g}_i = \hat{\sigma}_{e,i} + g(\theta_i).
\end{equation}

Once a collection of weights $w(e)$ has been built, an observed shear field can be explicitly fitted by minimizing a weighted chi-squared. Considering some collection of $N$ deflectors with masses $\{M_j\}$ and a series of background galaxies with calibrated shears $g_i$ each at redshift $z_i$, following Section \ref{sec:multiplane} the shear field can be computed explicitly and each galaxy position can be fit individually:

\begin{equation}
    \chi^2(\{M_j\}) \equiv \sum_i w_{e,i} \left| \left( g_{i} - g(\theta_i|\{M_j\},z_i) \right) \right|^2.
\end{equation}

The shear is estimated per object since, in the case of multiple deflectors, there is no general symmetry that can be used to bin the observations.

To reconstruct the projected mass density, the shears of background galaxies need to be averaged together with the correct kernel. In this work, we use the aperture mass statistic \citep{schneider_detection_1996}, defined as

\begin{equation}
    M_{ap}^{(+)}(\theta) =  \iint d^2\theta' ~Q(\left| \theta - \theta' \right|)~g^{(+)}(\theta'),
\end{equation}

where $g^{(+)}(\theta_i)$ is the tangential shear at $\theta'$ measured relative to some point on-sky $\theta$, and the integral is evaluated over some aperture of radius $\theta_{ap}$ centered on $\theta$. The aperture mass is also equal to a smoothed convergence field

\begin{equation}
    M_{ap}^{(+)}(\theta) =  \iint d^2\theta' ~U(\left| \theta - \theta' \right|)~ \kappa (\theta'),
\end{equation}

where the filters over shear ($Q$) and convergence ($U$) are related by

\begin{equation}\label{eq:q_to_u}
    Q(x) = - U(x) + \frac{2}{x^2} \int_0^x dx' x' U(x'). 
\end{equation}

This also ensures that the filter over convergence is compensated, e.g. $\int_0^x dx'~x'U(x') = 0$. We use the Schirmer filter,

\begin{equation}
    Q(x) = \left[ 1 + e^{6 - 150x} + e^{50x - 47} \right]^{-1} \times \frac{\tanh(x/0.15)}{x/0.15},
\end{equation}

which is optimized for the detection of NFW-like structures embedded in large-scale structure \citep{maturi_searching_2007,schirmer_gabods_2007}. The aperture mass can be sampled over a grid on-sky with an arbitrary resolution, but the effective resolution of the map itself can be characterized by its full-width at half-maximum (FWHM) in convergence space\footnote{This can be computed directly from $U(x)$ by inverting Equation \ref{eq:q_to_u}.}, $\sim 0.133 \times \theta_{ap}$. Larger apertures have an improved signal, but come at the expense of resolution in the reconstruction; in subsequent sections we will use a $20'$ aperture, resulting in a $\sim 2.7'$ resolution. To compute the aperture mass, a weighted estimator can be built:

\begin{equation}\label{eq:map}
    \hat{M}_{ap}^{(+)} (\theta) = \frac{1}{\pi \theta_{ap}^2 \sum_i w_i} \times \sum_i Q(|\theta - \theta_i|) w_i \hat{g}_i^{(+)}.
\end{equation}

Here, the sum is evaluated within all galaxies in the aperture, and their tangential shear is computed relative to $\theta$ following Equation \ref{eq:tanshear}. The variance in the aperture mass can also be estimated by

\begin{equation}
    \hat{V} (\theta) = \left( \frac{1}{\pi \theta_{ap}^2 \sum_i w_i} \right)^2 \times \sum_i Q(|\theta - \theta_i|)^2 w_i^2 \frac{ \hat{g}_i^2  }{2},
\end{equation}

where $\hat{g}^2 = \hat{g}_1^2 + \hat{g}_2^2 $ is the magnitude of the calibrated shear per-object. A signal-to-noise map can then be constructed:

\begin{equation}
    S(\theta) = \frac{M_{ap}^{(+)}(\theta)}{\sqrt{V(\theta)}}.
\end{equation}

It is sometimes helpful to construct a map from the cross-component of shear rather than the tangential component:

\begin{equation}\label{eq:map}
    \hat{M}_{ap}^{(\times)} (\theta) = \frac{1}{\pi \theta_{ap}^2 \sum_i w_i} \times \sum_i Q(|\theta - \theta_i|) w_i \hat{g}_i^{(\times)}.
\end{equation}

The cross-mode of the map is zero for isolated sources, making it a useful null-test for systematics. The problem of detecting sources on one of these mass reconstructions is addressed in Section \ref{sec:detect_maps}.

Mass-maps reconstructed from the shear are proportional to the projected \emph{overdensity} from some mean projected mass, but suffer from the mass-sheet degeneracy as the reduced shear is invariant under the transformation $\Sigma(\theta) \to \Sigma(\theta) + \Sigma_0$ \citep{kaiser_method_1995}. Aperture-mass maps are compelling since the filters in convergence space are compensated, leaving the reconstruction immune to the mass-sheet degeneracy and directly proportional to a convolution of $\frac{\Delta \Sigma}{\Sigma_c}$ \citep{schneider_detection_1996}. The mass-sheet degeneracy can be broken given an additional observable from the magnification of background sources; efforts to implement this for LoVoCCS observations are currently being explored (Helhoski et al., in-prep).

\section{The Third Generation Pipeline}\label{sec:gen3}

The Gen2 data reduction for LoVoCCS was supported by \texttt{v19\_0\_0} of the LSP, but the LSP has undergone several major updates since that time. The structure of data repositories has been entirely overhauled and processing observations now requires users to explicitly use the \texttt{Butler} to ingest and manipulate data stored in a repository \citep{jenness_vera_2022}. Several key tasks in the LSP have been updated, including support for multiband deblending through \texttt{scarlet\_lite} \citep{melchior_scarlet_2018} and sophisticated PSF modeling with \texttt{piff} \citep{jarvis_piff_2021}. The third generation pipeline now relies on \texttt{v26\_0\_0} of the LSP, though a small number of bugfixes from later versions have been applied.

The Gen3 pipeline is stored in a public repository called \texttt{lovoccs\_pipe}. Each step in the pipeline, representing an underlying \texttt{Pipetask} or collection of \texttt{Pipetask}s to be executed via the LSP's Batch Processing Service, is referred to as a ``processing step''. It takes on the order of one week to completely run a cluster through the pipeline depending on the when allocated $20$-cores and $500$GB of RAM. The updated pipeline is described below; we exclude processing steps which involve downloading data from the NOIRLab Astro Data Archive, the creation or ingestion of data into the repository, and disk space management strategies. Additional documentation on these steps and a detailed walkthrough of the workflow can be found on the pipeline's \href{https://github.com/astroenglert/lovoccs_pipe/tree/main}{GitHub repository}. The LSP portion of the pipeline derives heavily from the \href{https://github.com/lsst/drp\_pipe/blob/main/pipelines/DECam/DRP-Merian.yaml}{Merian Data Release Pipeline} (DRP) \citep{danieli_first_2025}, but with several modifications that are summarized below. Throughout this section, the terminology of visit/exposure are used interchangeably in reference to individual images captured by DECam.

\subsection{Calibrations \& Reference Catalogs}

\begin{figure}
    \centering
    \includegraphics[width=0.98\linewidth]{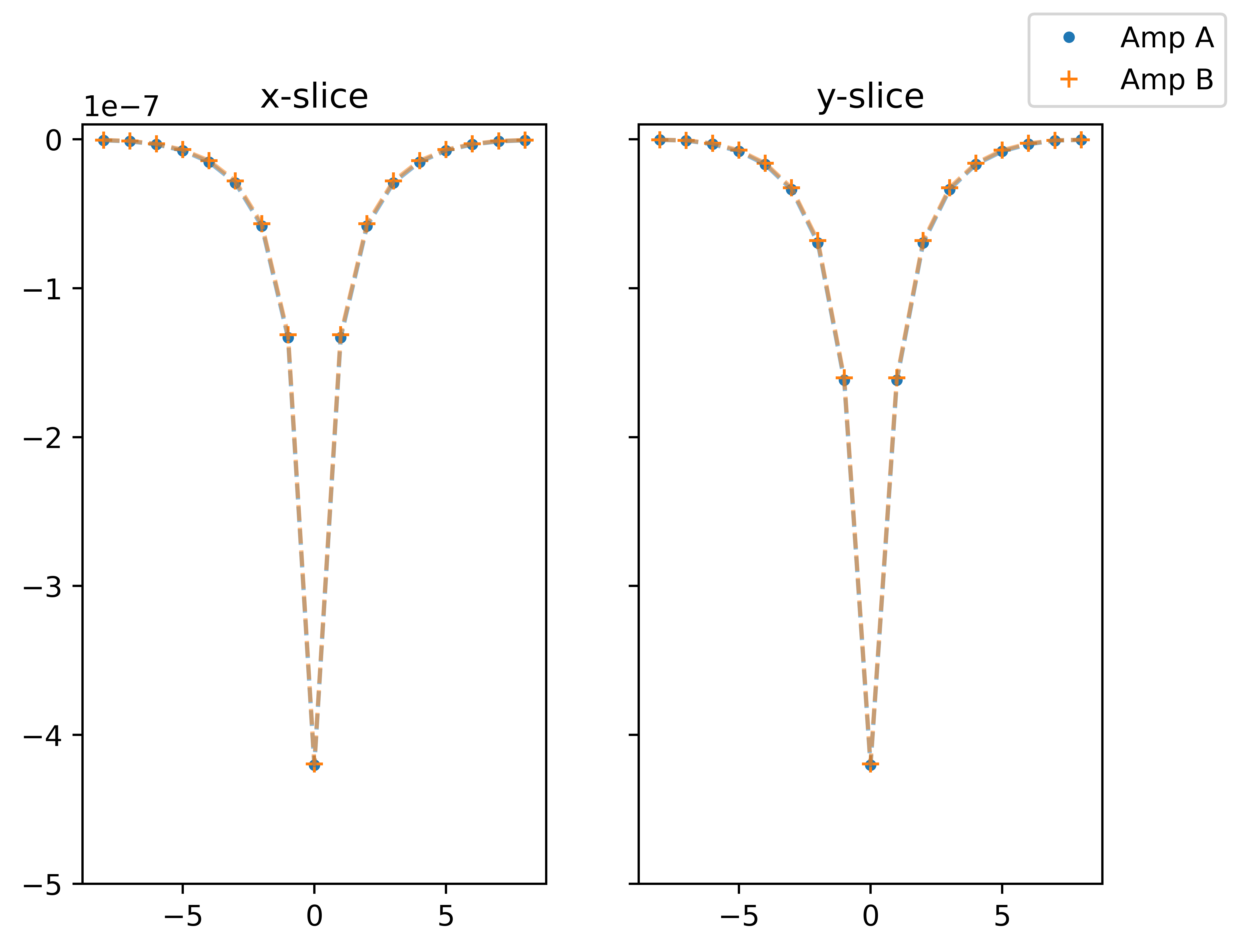}
    \caption{Cross-sections of the kernel used to correct the brighter-fatter effect; horizontal axes are pixel-displacements. The slope of the kernel is proportional to the strength of the field displacing incident charges; effectively, the correction redistributes flux deflected into adjacent pixels. }
    \label{fig:calib_bfk}
\end{figure}

Recent updates to the LSP have removed native support for master calibration frames generated by NOIRLab with the DECam Community Pipeline \citep{valdes_decam_2014}, which were used throughout the Gen2 reduction. New master-bias, master-flat, and fringe frames starting from raws were constructed using the \href{https://github.com/lsst/cp_pipe}{Calibration-products Production Package} (\texttt{cp\_pipe}). The reference frames gradually vary over time; we sampled this with a monthly cadence, starting from September, 2012, which was sufficient to capture their evolution. 

During the Gen2 reduction, the brighter-fatter effect (BFE) was not corrected due to a lack of supporting calibrations compatible with the LSP. For Gen3 processing, \texttt{cp\_pipe} was used to build kernels to correct for this effect within the formalism created by \citet{coulton_exploring_2018}. To generate the kernels (Fig.~\ref{fig:calib_bfk}), flats from 2017-2019 were downloaded, and the \texttt{cpPtc} pipeline was used to create a photon transfer curve (PTC).
The PTC and flats were passed to \texttt{cpBfk}, which explicitly computed additional covariances and solved for a per-amplifier kernel. Since LoVoCCS uses observations from a single band to estimate the shape of an object, kernels from the lensing band ($r$) are used to correct all visits for the BFE. Nonetheless, the per-band kernels were constructed and showed a weak color dependence, with bluer bands requiring a more aggressive correction. This is expected; redder photons penetrate deeper into a CCD before scattering an electron, which leaves the charge closer to the base of a pixel's potential well. The charge has a smaller distance over which it can be displaced by the accumulated charges, resulting in less displacement on average \citep{astier_correction_2023}.

\begin{figure}
    \centering
    \includegraphics[width=0.98\linewidth]{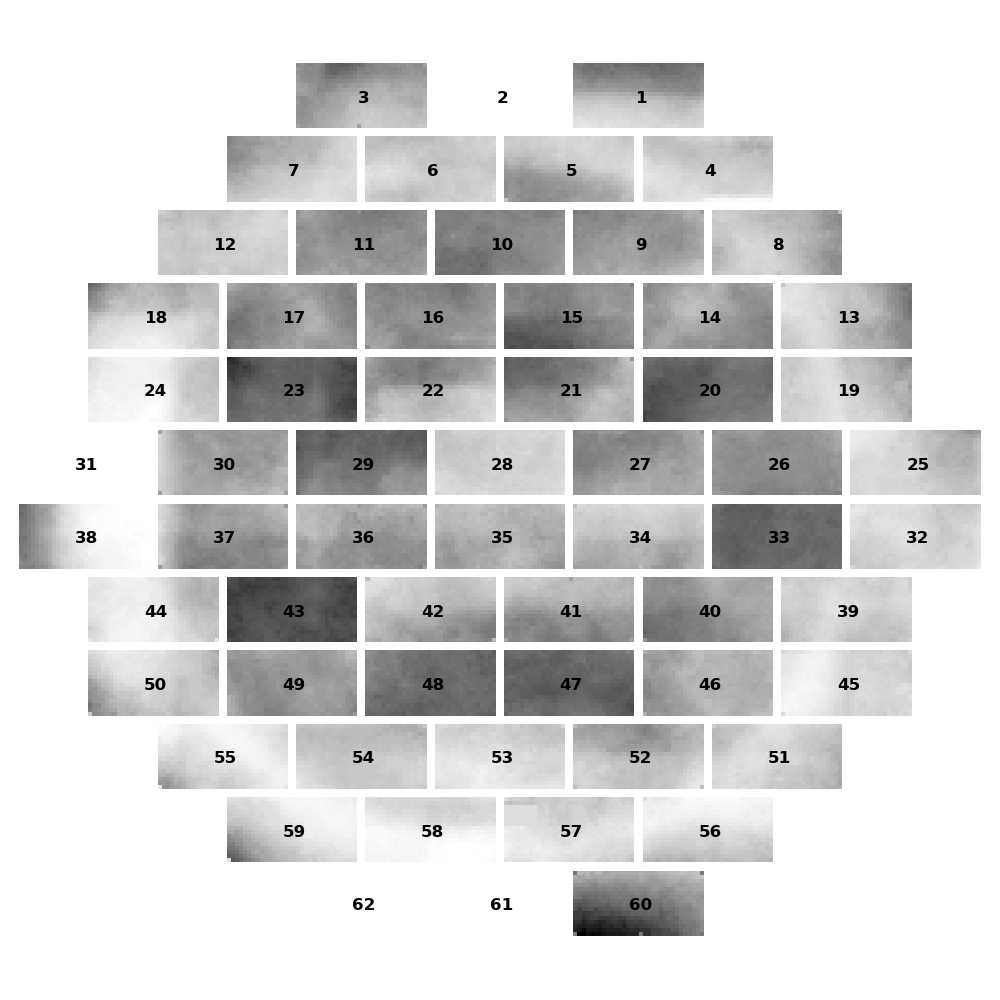}
    \caption{A stretched, inverted image of the skyframe for $r$-band. The ``ghost pupil''---an image of DECam and its supporting struts---is clearly visible.}
    \label{fig:calib_sky}
\end{figure}

The DRP creates coadded images suitable for weak lensing science at the expense of extended low surface brightness features, which are removed by the default chip-level background model \citep{watkins_strategies_2024}. To address this, \texttt{skycorr} was developed and successfully applied to data from HSC to better preserve the extended profiles of large galaxies \citep{aihara_second_2019,aihara_third_2022}. To support future low surface brightness (LSB) science with LoVoCCS observations \citep[Escalante et al., in-prep, ][]{englert_intracluster_2025} the Gen3 pipeline now uses \texttt{skycorr} to produce sky-corrected coadds in addition to the default \texttt{deepCoadd}'s used for lensing-science.

An essential ingredient in \texttt{skycorr} is the creation of a sky-frame, which captures the response of the telescope to the sky. The NOIRLab Astro Data Archive was queried, and a subset of exposures whose distribution of airmasses, moon phases, and moon separations resembling LoVoCCS were selected for stacking. The exposures were separated by a minimum of $ 1^{\circ}$ from one another and from any LoVoCCS clusters. All exposures within $15^{\circ}$ of the galactic plane were also pruned, leaving $\sim 200$ exposures per band. \texttt{cpSky} was then used to correct for instrumental noise, mask detections, and stack the exposures to create the sky-frame (Fig.~\ref{fig:calib_sky}).


The reference catalogs which support LoVoCCS data reduction have been updated to reflect recent data releases since Gen2. Gaia DR3 \citep{gaia_collaboration_gaia_2016,gaia_collaboration_gaia_2023} is used for astrometric calibration and, depending on a cluster's declination, Pan-STARRS DR2 ($\delta \gtrsim -15^{\circ}$) \citep{chambers_pan-starrs1_2016} or SkyMapper DR4 ($\delta \lesssim -15^{\circ}$) \citep{onken_skymapper_2024} is used for photometric calibration; this is supplemented by the Sloan Digital Sky Survey (SDSS) as necessary to cover $u$-band \citep{adelman-mccarthy_sixth_2008}. Stacks for LoVoCCS utilize all existing archival data, making it inappropriate to calibrate using surveys which implicitly include LoVoCCS data in their stacks; instead, photometric catalogs from other DECam surveys, such as the Dark Energy Survey (DES) \citep{abbott_dark_2018} and the DESI Legacy Imaging Survey \citep{dey_overview_2019}, are downloaded and reserved for validating LoVoCCS photometry.

\subsection{Processing Steps}

\begin{figure}
    \centering
    \includegraphics[width=0.98\linewidth]{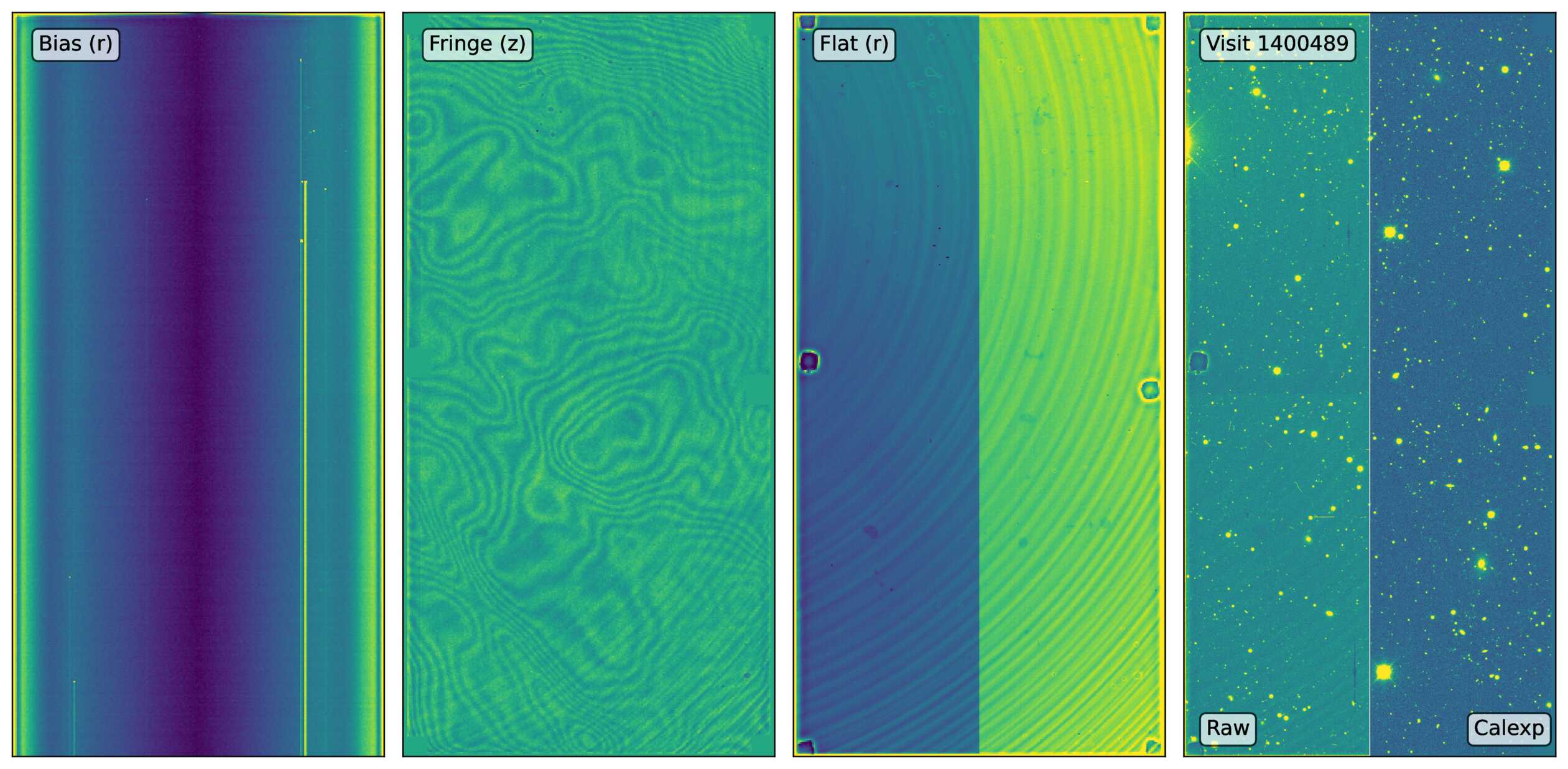}
    \caption{A bias ($r$-band), fringe ($z$-band), and flat ($r$-band) from chip 10. The right-most cutout shows the same chip from visit 1400489 ($r$-band) before and after applying \texttt{process\_ccd}.}
    \label{fig:calib_comp}
\end{figure}

\emph{process\_ccd:} This removes the instrumental signal, including detecting streaks (satellites crossing a chip during an exposure) and cosmic-rays (which can also scatter electrons during an exposure), then correcting for bias, overscan, crosstalk, linearity, BFE, fringing ($z$ only), and flat-fielding. A chip-level background subtraction is carried out along with initial detections and measurements. Following this, preliminary photometric and astrometric calibrations are performed with user-specified reference catalogs. Lastly, a sample of PSF stars is selected and a model is built with PSFEx \citep{bertin_psfex_2013}; everything is then collected into a chip-level catalog. At this stage, the calibrated frames are referred to as \texttt{calexp}'s.

\emph{check\_visit \& select\_visit:} The seeing conditions are checked per-chip by fitting a Moffat profile \citep{moffat_theoretical_1969} to PSF-stars and their ellipticity is computed (Equation \ref{eq:ellip}). Chips whose stars have a PSF with FWHM $> 1\arcsec$ ($> 1.7\arcsec$) or an ellipticity $|e| > 0.13$ ($ > 0.33$) in $r$ ($ugiz$) are rejected. To prevent failures downstream, chips with too few stars in a given visit (with a $2\sigma$ threshold, computed from all chips in a visit) and visits with fewer than 10 chips matching these conditions are rejected; visits with too few stars relative to all exposures of a cluster are also rejected (with a $3\sigma$-threshold).

\emph{visit\_summary:} Measurements from \texttt{process\_ccd} are collected into per-visit summaries.

\emph{jointcal:} A joint astrometric and photometric solution is solved per-visit. Joint-calibration assumes that variations in the photometric zero-point and the position of reference stars can be modeled with a polynomial whose coefficients vary between exposures, and depends on an object's position in the focal plane. This polynomial can be solved for across all visits simultaneously by minimizing a joint $\chi^2$ \citep{2019AAS...23336324P}. Fixing the models to a second order polynomial minimized the residuals with negligible improvements at higher-orders; therefore, a second order model is used by default to avoid over-fitting the data.

\emph{final\_visit\_summary:} The final astrometric and photometric solutions from jointcal are applied to the catalogs, and a final PSF model is built using \texttt{piff}.

\emph{coadd\_3a:} First, a PSF-matched stack is built and used as a static-sky template for flagging candidate transients\footnote{Detecting transients requires studying the difference between exposures, but different exposures have different seeing conditions and PSFs; naively subtracting them leads to spurious false detections and sub-optimal signal-to-noise. Instead, difference imaging typically relies on exposures whose PSFs have been matched by first selecting a reference-image, then matching all exposures to the reference's PSF by convolving with an appropriate kernel \citep{alard_method_1998}.}. Transients are selected based on their persistence in differences between the static-sky and the PSF-matched visits; their associated pixels are masked and interpolated over \citep{bosch_hyper_2018}. Following this, the visits are warped onto a tangent plane and stacked. For Hercules, the RINGS tesselation of the sphere is used due to the large area covered by our visits \citep{waters_pan-starrs_2020}; for individual clusters in LoVoCCS, a single tangent plane is centered on each cluster's X-ray peak.

\emph{coadd\_3b:} Detection and deblending is run on the coadd. The deblender, \texttt{scarlet\_lite}, has been re-configured so it runs on nearly all sources\footnote{The default settings have been tweaked so that galaxies with a large angular size ($\gtrsim 1 \text{ arcmin}^2$) are fully deblended; in practice, background sources behind these galaxies are masked during the lensing analysis.}. This leads to large memory usage, but ensures that even the brightest cluster galaxies are deblended and do not drop-out of the final catalogs. Measurements are then run on the detected footprints.

\emph{coadd\_3c:} ``Forced'' measurements are made across multiple bands using a source's footprint from a reference band; the reference is chosen from either $r$ or $i$ depending on which has the largest signal. Final object catalogs are then written.

\emph{coadd\_3d:} Sky-corrected visits are created using \texttt{skycorr} and stacked following the same rejection scheme as in \texttt{coadd\_3a}.

\begin{figure}
    \centering
    \includegraphics[width=0.98\linewidth]{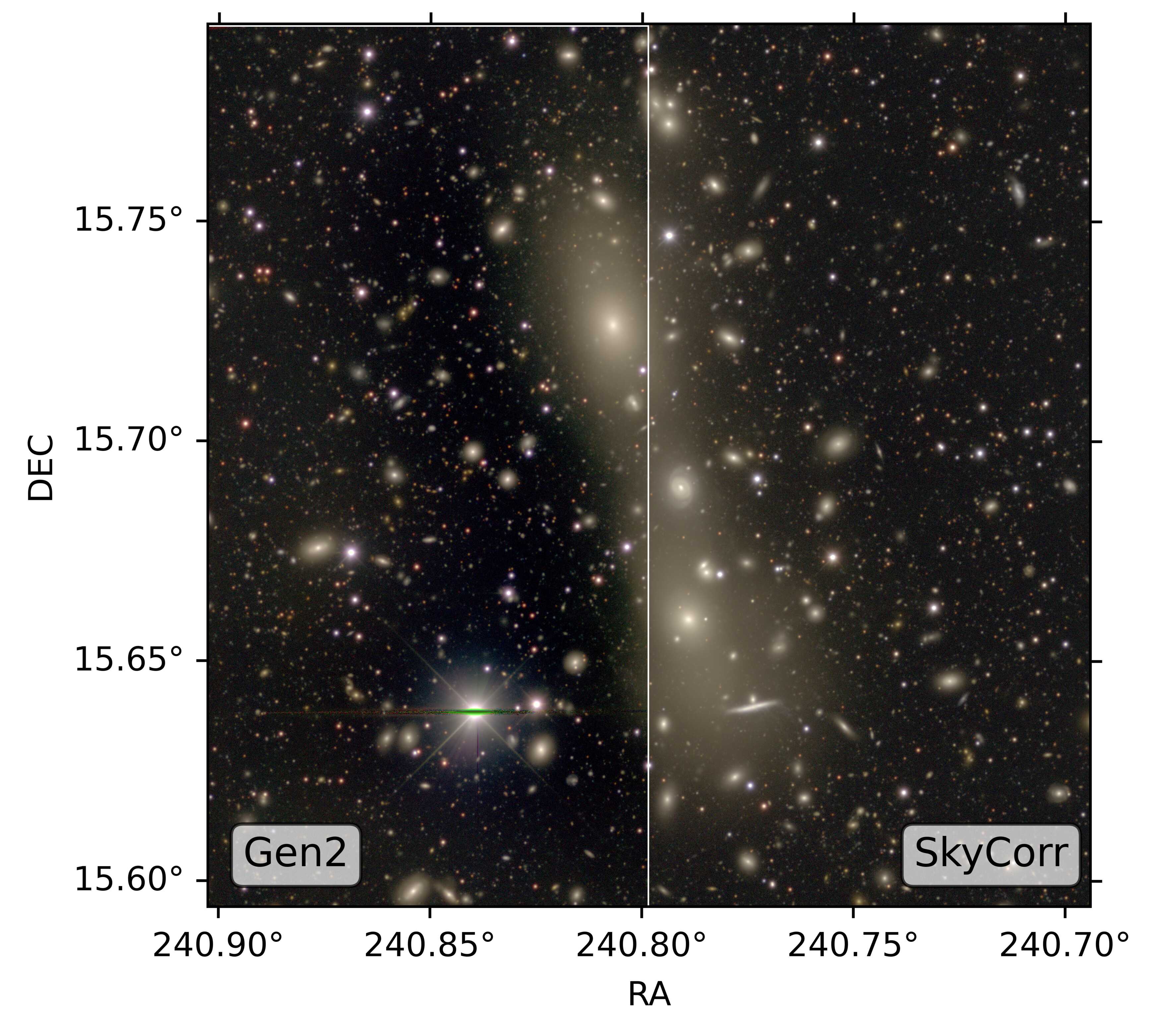}
    \caption{A stretched $irg$ cutout showing the Gen2 (left) and sky-corrected Gen3 (right) coadds.}
    \label{fig:run_skycorr}
\end{figure}

\subsection{Post-LSP Steps}

\emph{export\_data:} The coadds, sky-corrected coadds, and object catalogs are exported from the repository for ease of access. Crucially, this step exports CModel and PSF photometry \citep{abazajian_second_2004} along with adaptive second moments of sources and the PSF \citep{hirata_shear_2003,mandelbaum_systematic_2005}.

\emph{photometric\_correction:} In Gen2, extinction was carried out from low-resolution IRSA maps which were assigned a fiducial $10\%$ error \citep{irsa_galactic_2022}. This has been changed and the Planck 2013 extinction maps are now used and their uncertainties are explicitly propagated into the noise per-object \citep{planck_collaboration_planck_2014}. Following this, a photometric zero-point is solved for by matching the $griz$ catalogs to either Pan-STARRS or SkyMapper and identifying the stellar-locus in $u$.

\emph{photo\_z:} The photometric redshift code has been overhauled so that objects missing photometry in a single band are included and so that the algorithm can be applied to other instruments (transmissions from DECam, HSC, and SDSS are included). Priors from the Next Generation Virgo Cluster Survey, which better match the $ugriz$ filter-set and depth of LoVoCCS \citep{raichoor_next_2014}, are used instead of the traditional Hubble Deep-Field North priors \citep{benitez_bayesian_2000,coe_galaxies_2006}. A followup paper discussing improvements to photometric redshifts for LoVoCCS is currently in-preparation (Lan et al., in-prep).

\emph{shear\_calibration:} A fiducial shear calibration is carried out using HSM shapes \citep{hirata_shear_2003,mandelbaum_systematic_2005} by applying the HSC-Y1 shear calibration \citep{mandelbaum_weak_2018}. For Hercules, a new shear calibration scheme is adopted (Sec. \ref{sec:meta}).

\emph{mass\_map:} The mass map is computed using the aperture mass as outlined in Chapter \ref{ch:mass_map}. By default, the statistic is sampled across a tangent plane centered on the cluster over a grid whose pixels are separated by $\sim 0.3\arcmin$, with apertures varying between $3000-29000\text{ px}$ ($0.2^{\circ} - 2.1^{\circ}$). For Hercules, the statistic was sampled on a higher resolution grid, with samples spaced by $\sim 0.13\arcmin$ with a single $20\arcmin$ aperture.

\emph{mass\_fit:} The map with the largest signal is used to center an NFW profile \citep{navarro_structure_1996,navarro_universal_1997} that is fit to the calibrated shears. For our analysis of Hercules, we adopt the multi-plane lens model outlined in Section \ref{ch:wl}; this will be generalized to all LoVoCCS clusters in the future.

\emph{quality\_check:} Plots to check the coadd PSF, depth, photometry, and astrometry are created.

\emph{red\_sequence:} Following the algorithm from LVII, a map of the red sequence galaxies across the cluster is created. For Hercules, the galaxy population will be studied in-depth in a future paper.

\subsection{Metadetection}\label{sec:meta}

In the newest reductions, we have begun implementing the \texttt{metadetect} algorithm for precise shear-calibration. The core idea behind \texttt{metadetect} is that the average shear can be calibrated by creating sheared realizations of the data itself and re-running the detection and measurement steps to estimate the response. Explicitly, the shear due to lensing can be estimated as

\begin{equation}\label{eq:calib_shear}
    \langle g_i \rangle \simeq \langle R_{ij} \rangle^{-1} \langle e_j \rangle,
\end{equation}

where the response itself is estimated using finite differences

\begin{equation}\label{eq:resp}
    \langle R_{ij} \rangle \simeq \frac{ \langle e_{j} \rangle^{i+} - \langle e_{j} \rangle^{i-} }{2 \gamma}.
\end{equation}

Each average is computed over catalogs ran through an identical pipeline for carrying out measurements, but with a coadd that has been explicitly sheared by some small $\pm\gamma$. 

\begin{figure}
    \centering
    \includegraphics[width=0.98\linewidth]{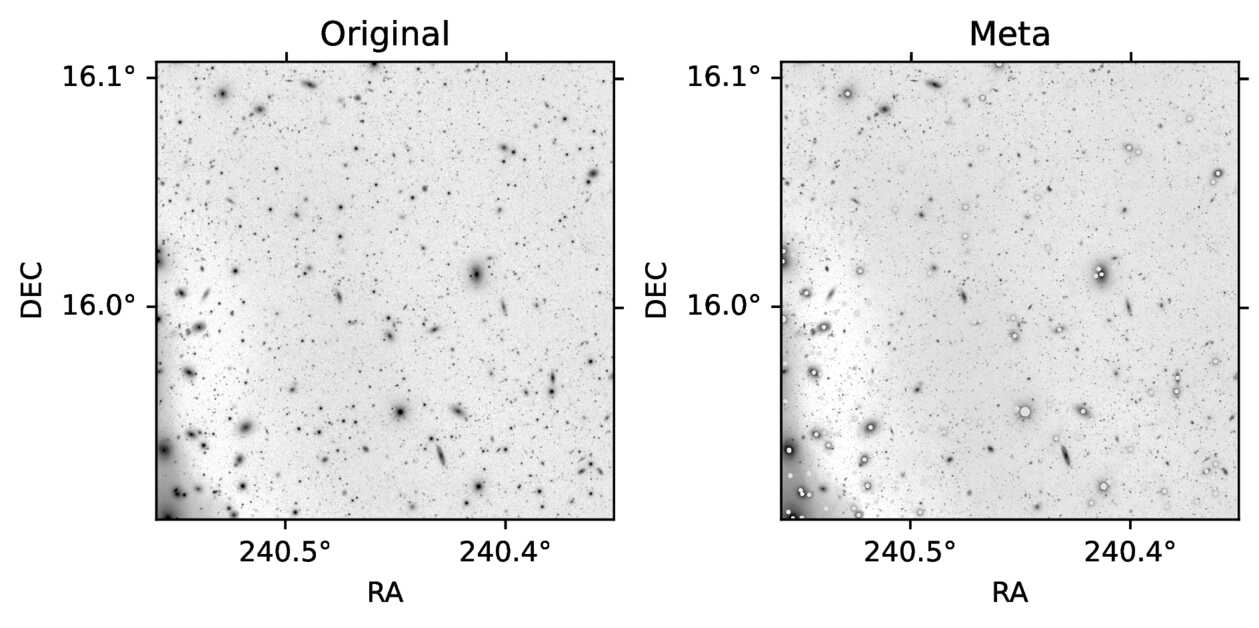}
    \caption{A single patch, before (left) and after (right) using the \texttt{ShearCoaddTask} to generate the \texttt{noshear} realization of the coadd.}
    \label{fig:meta_comp}
\end{figure}

Formally, the shear we aim to measure is applied prior to light passing through the atmosphere and the telescope's optical path; this can be approximated by explicitly deconvolving each coadd using its model PSF, applying a shear, then reconvolving with a slightly larger PSF. In practice, this is carried out over small regions where the PSF is nearly uniform and deconvolution can be carried out with a static kernel. The convolutions and shears are applied in Fourier space, so apodized masks are used to cover bright stars and the edges of a patch to prevent the formation of high-frequency artifacts. Lastly, a noise image whose covariance matches the underlying noise in the coadd is also needed to mitigate the creation of spurious noise sources. This is repeated five times, producing four synthetically sheared images and one ``control'' with zero-applied shear that shapes are measured on.

Our implementation of \texttt{metadetect} follows this procedure with some deviations from its most recent application to the DES Y6 shear catalog \citep{yamamoto_dark_2025}. First, per-object response is assumed to be a function of the signal-to-noise ($S$) and resolvedness ($\text{res}$) of a given population of galaxies:

\begin{equation}
    S = \frac{2.5}{\sigma_r \ln(10)}, \quad \text{res} = 1 - \frac{T_{\text{PSF}}}{T_{\text{gal}}},
\end{equation}

where $T_{\text{PSF}}$ and $T_{\text{gal}}$ are traces of the second moments from the PSF model and the galaxy respectively; $S$ is estimated from the uncertainty in $r$-band magnitude, $\sigma_r$. This ensures that the shear is calibrated over ensemble averages---though any individual point estimate may have a small systematic scatter. Second, this procedure is carried out over rather large $\sim 20\arcmin \times 20\arcmin$ ``patches'' on-sky. Fortunately, owing to the strict observing conditions used for LoVoCCS, the coadd-level PSF is nearly uniform over these regions, making this a reasonable approximation. To generate a noise image for the patch, all detections are masked, then the covariances and corresponding power spectrum of the un-masked pixels are estimated; this is then used to generate a Gaussian random field with an identical power spectrum. This procedure is remarkably effective at reconstructing the correlated noise and becomes an excellent approximation in the limit where a large number of exposures are stacked with random dithers.

Measurements are carried out with the same processing pipeline used for generating the HSM shape catalog. The only difference is that \texttt{metadetect} explicitly corrects for the PSF during deconvolution, so the chosen shape measurement does not need to be PSF-corrected; due to their signal-to-noise advantage, we have chosen to use weighted second moments measured with the adaptive-Gaussian scheme from SDSS \citep{abazajian_second_2004}. The details of how the catalogs are combined to build a model of the response and optimal weights is discussed in Section \ref{sec:shear_calib}.

Within \texttt{lovoccs\_pipe}, this implementation of \texttt{metadetect} is carried out via the following steps:

\emph{meta\_4a:} The \texttt{ShearCoaddTask} generates a synthetic noise image and uses the \texttt{metadetect} library\footnote{\url{https://github.com/esheldon/metadetect}} to generate the four sheared and single un-sheared realizations of each patch. These are then stored in the repository. 

\emph{meta\_4b:} Each realization is ran through a streamlined version of \texttt{coadd\_3b} and \texttt{coadd\_3c} which only measures the magnitudes and shapes needed for weak lensing.

\emph{meta\_export:} Catalogs from each realization are exported from the LSP and saved on-disk.

\emph{meta\_process:} Each \texttt{metadetect} catalog is passed through \texttt{photometric\_correction} and \texttt{photo\_z}.

\emph{meta\_lensing:} For each LoVoCCS cluster, a fiducial model for the response is built with a coarse adaptive grid covering the $S-\text{res}$ parameter space; this model is used to calibrate the shapes which are passed to \texttt{mass\_map} and \texttt{mass\_fit} as before. The calibration carried out for Hercules, which will soon be applied to all LoVoCCS clusters, is discussed in Section \ref{sec:shear_calib}.

\subsection{Summary}

\begin{figure*}
    \centering
    \includegraphics[width=0.98\linewidth]{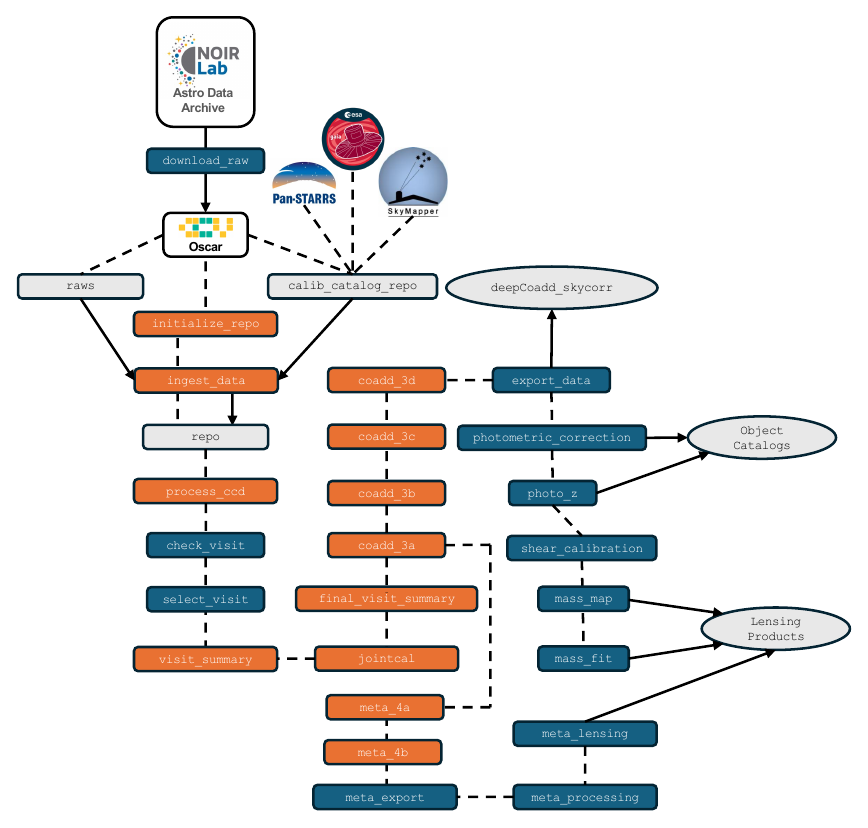}
    \caption{A sketch of the complete Gen3 pipeline. The data is primarily reduced on OSCAR, Brown University's high performance computing cluster. Transparent boxes are folders saved on OSCAR, orange colored boxes are processing steps that rely on the LSP, blue boxes are post-LSP steps, and planned data products are enclosed in elliptical boxes.}
    \label{fig:pipeline}
\end{figure*}

An outline of the complete ``Gen3'' pipeline is presented in Figure \ref{fig:pipeline}. At present, we anticipate releasing mass estimates and mass maps derived from both the HSM and \texttt{metadetect} shape catalogs as lensing products; the calibrated shape catalogs themselves will also be released. The sky-corrected coadds and star/galaxy object catalogs, including photometric redshifts, will also be released. At the time of writing, nearly $50\%$ of the original LoVoCCS clusters have been fully reduced with this pipeline.

\section{Science Validation}\label{sec:valid}

LoVoCCS weak lensing science goals require that shear estimates must be accurate down to the percent-level ($\delta e \lesssim 0.05$), astrometry should be accurate to within one pixel ($\delta \theta < 0.263\arcsec$), and photometry should be correct to $\delta m \lesssim 50 \text{ mmag}$ so that the photometric redshifts are sufficiently accurate. The stretch goal of the Gen3 reduction is to reach the standards outlined by the LSST Dark Energy Science Collaboration (DESC) Science Requirements Document and the LSST System Science Requirements \citep{2018arXiv180901669T,ivezic_lsst_2019}; this includes reaching $\delta e \lesssim 0.01$, $\delta m \sim 10 \text{ mmag}$, and $\delta \theta \sim 10~\text{mas}$ (Table \ref{tab:lsst_standards}). In this section, observations from the Hercules Supercluster will be used to validate \texttt{lovoccs\_pipe} and test whether it meets these goals.

Unless stated otherwise, scatter in the photometry, astrometry, and photometric redshifts is estimated using the normalized median absolute deviation

\begin{equation}
    \sigma_{\text{NMAD}} = 1.48 \times \text{median}\left[ \left| x - \text{median}(x) \right| \right].
\end{equation}

We prefer this estimator over the sample standard deviation since it is less sensitive to outliers. The factor of $1.48$ normalizes the statistic such that, when $x$ is a normal random variable, the NMAD is equal to the standard deviation.

\subsection{Masking}

\begin{figure}
    \centering
    \includegraphics[width=0.98\linewidth]{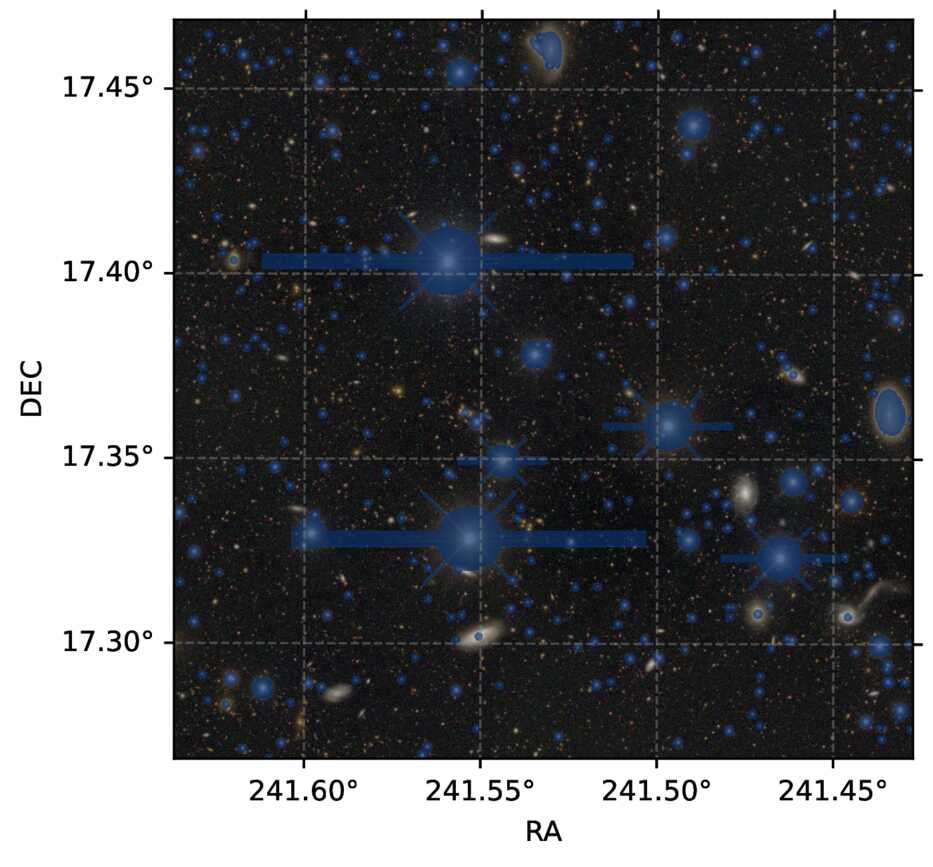}
    \caption{An $irg$ cutout south of A2151 with the star+galaxy mask overlaid in blue.}
    \label{fig:valid_mask}
\end{figure}

Prior to assessing other statistics, regions where known systematics can bias the shapes or photometry must be masked. First, the wings of bright stars, their diffraction spikes, and bleed rows are masked automatically; this is done by masking all objects present in the Gaia catalog\footnote{Objects included in Gaia are almost entirely stars, $\lesssim1\%$ of sources are extragalactic \citep{gaia_collaboration_gaia_2023}.} with a circular mask whose radius is set by a star's $G$-band magnitude. For reference, it is helpful to define the mask radius used by DES \citep{yamamoto_dark_2025}:

\begin{equation}
    \log_{10} \left[ \frac{R_{\text{DES}}(G)}{1''} \right] = 0.004432 G^2 - 0.2257 G + 2.996.
\end{equation}

LoVoCCS observations are more than a magnitude deeper than DES\footnote{DES Y6 reached $i\sim23.4$ at $S\sim10$ \citep{bechtol_dark_2026}, but LoVoCCS reaches $i\sim24.8$ at $S\sim10$.}; as a result, the extended profiles of stars are more pronounced and require more aggressive masking. The masks are built to cover regions where the source density of objects systematically drops\footnote{The extended profiles of stars are brighter than most sources used for lensing science; background sources obscured by these profiles can have biased shapes and photometry. Additionally, the number density of sources systematically drops due to the added Poisson noise.}. The star-masks were assigned a radius $R_{\text{LV}}$ according to

\begin{equation}
    R_{\text{LV}} = \begin{cases}
        1.85 \times R_{\text{DES}}(G) & G < 14 \\
        R_{\text{DES}}(G) + 10 & G \geq 14 
    \end{cases}
\end{equation}

The brightest stars also have prominent bleed rows, depending on the brightness of the star; these were masked with lengths

\begin{equation}
    L = \begin{cases}
        0 & G \geq 12 \\
        10 R_{\text{DES}}(G) & 10 \le G < 12 \\
        20 R_{\text{DES}}(G) & 8 \le G < 10 \\
        25 R_{\text{DES}}(G) & 7 \le G < 8  \\
        35 R_{\text{DES}}(G) & G < 7 \\
    \end{cases}
\end{equation}

Bleed rows, at their largest, can occur when a star is located on the border of adjacent chips; therefore we truncated the length at $8192\text{ px}$. The widths of bleed rows were also scaled with magnitude

\begin{equation}
    W = \begin{cases}
        0 & G \geq 12 \\
        30 & 10 \le G < 12 \\
        60 & 8 \le G < 10 \\
        70 & 7 \le G < 8  \\
        100 & G < 7 \\
    \end{cases}
\end{equation}

The diffraction spikes of stars brighter than $G < 12$ were masked with rectangles of dimensions $4R_{\text{LV}}\times12\text{ px}^2$, which were tilted by $45^{\circ}$ and centered on the star.

The deblender's updated settings ensure that large foreground galaxies will be deblended, but this can lead to spurious sources created by the ``shredding'' of galaxies with well-defined knots in their disk. Galaxies susceptible to this were masked by passing each coadd through SExtractor \citep{bertin_sextractor_1996} with conservative settings so that only large, bright galaxies ($r < 19$ and angular size $\gtrsim 1 \text{ arcmin}^2$) are detected. The corresponding segmentation image was turned into a binary mask, which was then grown by $\sim5^"$.

A handful of artifacts were identified by eye and manually masked. This includes the DECam-exclusive edge bleeds, which are not automatically detected and masked by the LSP. Spurious collections of artifacts exclusive to $z$-band surrounded a subset of the bright stars; these were masked. A small fraction of bright transients can be flagged as persistent and left un-masked during visit-level processing, producing a short, bright streak on the coadd surrounded by a monochrome ``halo''; these were identified and masked. Lastly, sources beyond $1^{\circ}$ from the center of each pointing (Table \ref{tab:point}) were dropped.

\subsection{Astrometry \& Photometry}

\begin{figure}
    \centering
    \includegraphics[width=0.98\linewidth]{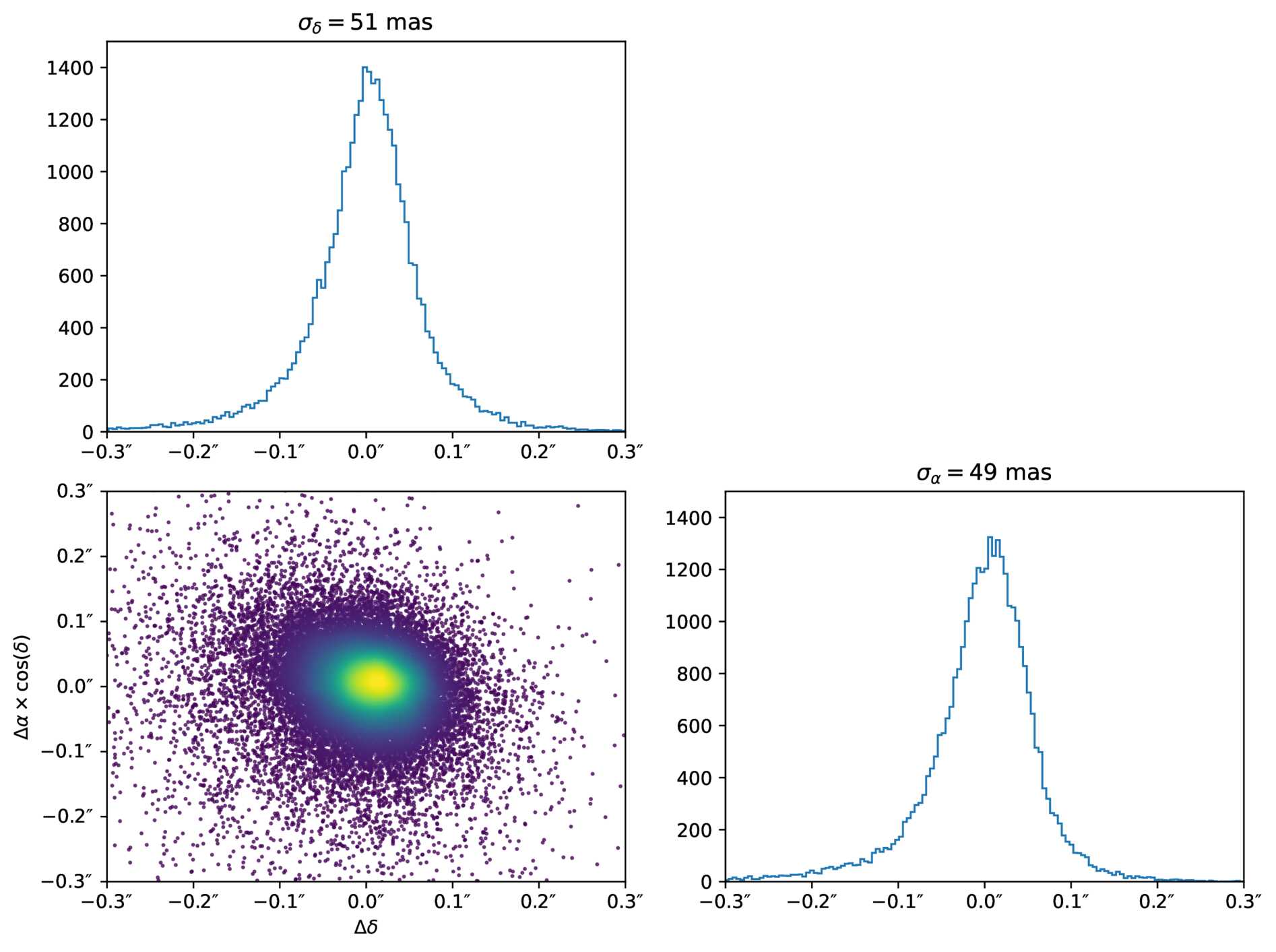}
    \includegraphics[width=0.98\linewidth]{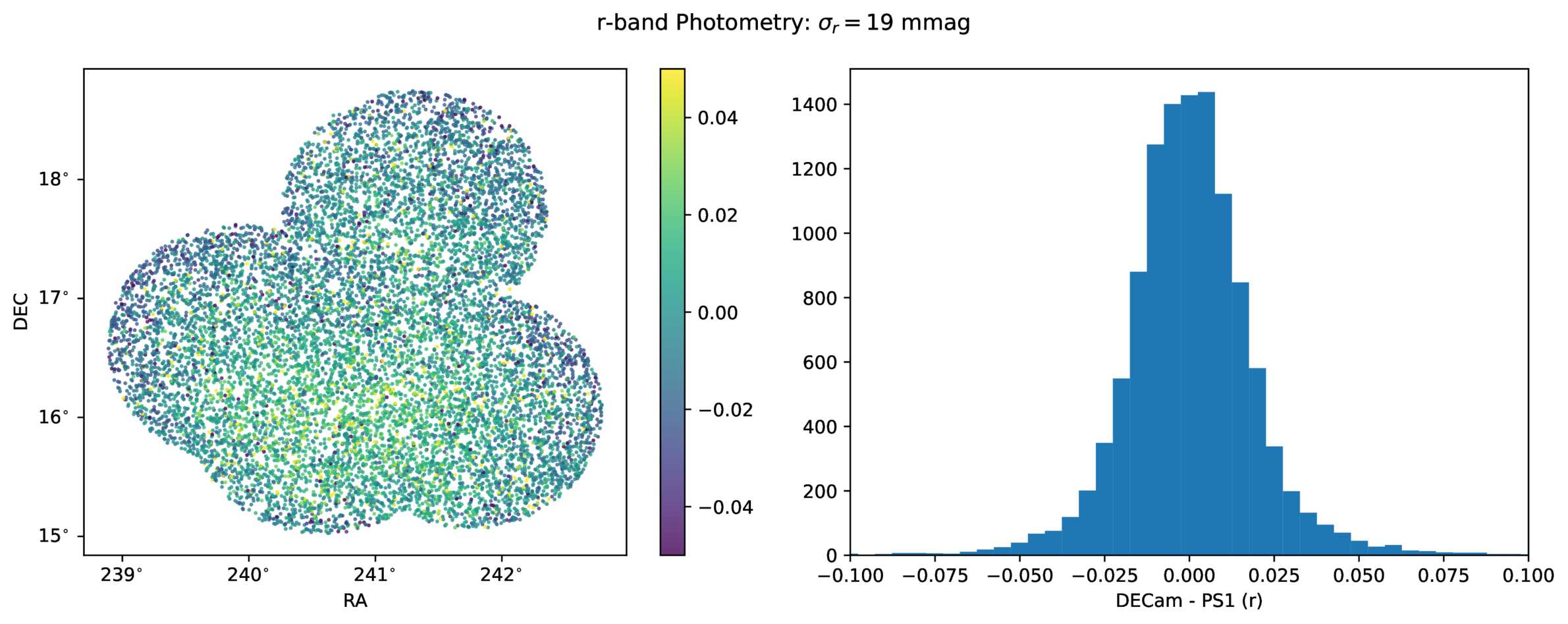}
    \caption{\emph{Top:} A corner plot showing scatter between the measured and Gaia centroids of stars selected for validating the astrometry. \emph{Bottom:} A plot showing the scatter in photometry relative to Pan-STARRS across our dataset (left), and a histogram showing the overall photometric scatter (right). We see residual variations in the photometry across the field, but they are generally within $\sim 2\sigma_r$ of zero.}
    \label{fig:valid_phot}
\end{figure}

To quantify the photometry uniformity, zero-point corrected Hercules catalog sources were matched to Pan-STARRS sources with $S > 10$, $15 < g < 21$, and color $g - i < 0.7$, where the color-terms built in LoVoCCS I are valid\footnote{The photometric precision of these stars are set by PanSTARR's systematic floor of $\sim 10\text{ mmag}$ \citep{magnier_pan-starrs_2020}.}. An example of the variation in photometry across Hercules is provided in Fig.~\ref{fig:valid_phot} and the photometric scatter per-band is provided in Table \ref{tab:lsst_standards}; across $griz$ the photometric scatter $\sigma \sim 10 - 20 \text{ mmag}$. The photometric uniformity in $u$ was quantified by applying the same set of cuts to Hercules's SDSS catalog and matching it to the zero-point-corrected Hercules catalog; the observations contain a large ($\sigma \sim 60 \text{ mmag}$) systematic scatter in $u$. This is due to a combination of poor ultraviolet transmission at Cerro Tololo---leading to shallow individual visits---and the lack of a deep photometric reference catalog covering this region in $u$.

To validate the astrometry, Gaia stars with $15 < G < 20$ and $S>10$ were selected\footnote{For these stars, Gaia's astrometric uniformity is $\sigma_{\theta} \lesssim 1\text{ mas}$.}. This was matched to the Hercules star catalog after selecting stars with $15 < r < 21$ and a signal $S > 10$. The differences in RA and DEC for this sample is shown in Fig.~\ref{fig:valid_phot}; the Hercules field reaches a competitive scatter of $\lesssim 50 \text{ mas}$ in the RA and DEC, while the overall scatter in angular separation is $\lesssim 40\text{ mas}$.

Scatter for both the photometry and astrometry reach the target precision for LoVoCCS science goals but fall short of the LSST-Y1 requirements. This is primarily due to joint calibration's limitations, as it is fundamentally an empirical model for the impact of the atmosphere and optical path on the visit-level astrometric and photometric solutions. Reaching the hefty $5-10\text{ mmag}$ photometric and the $\sim 10\text{ mas}$ astrometric uniformity required of LSST with such a method is unlikely.

Reaching the photometric requirements requires a full model of the atmosphere during observations; this can be built from observations of standard stars from either the data being reduced, or auxiliary observations from a different telescope on the same night. Matching the astrometric scatter requires explicit modeling which includes the drift of charges within a CCD, turbulence in the atmosphere, small shifts in the CCDs over time, and other systematics. The calibration methods developed by the Dark Energy Survey, Forward Global Calibration Method \citep{burke_forward_2018} and its astrometric model \citep{bernstein_astrometric_2017}, are being used as the base for matching these requirements with LSST.

\subsection{Photometric Redshifts}

\begin{figure}
    \centering
    \includegraphics[width=0.98\linewidth]{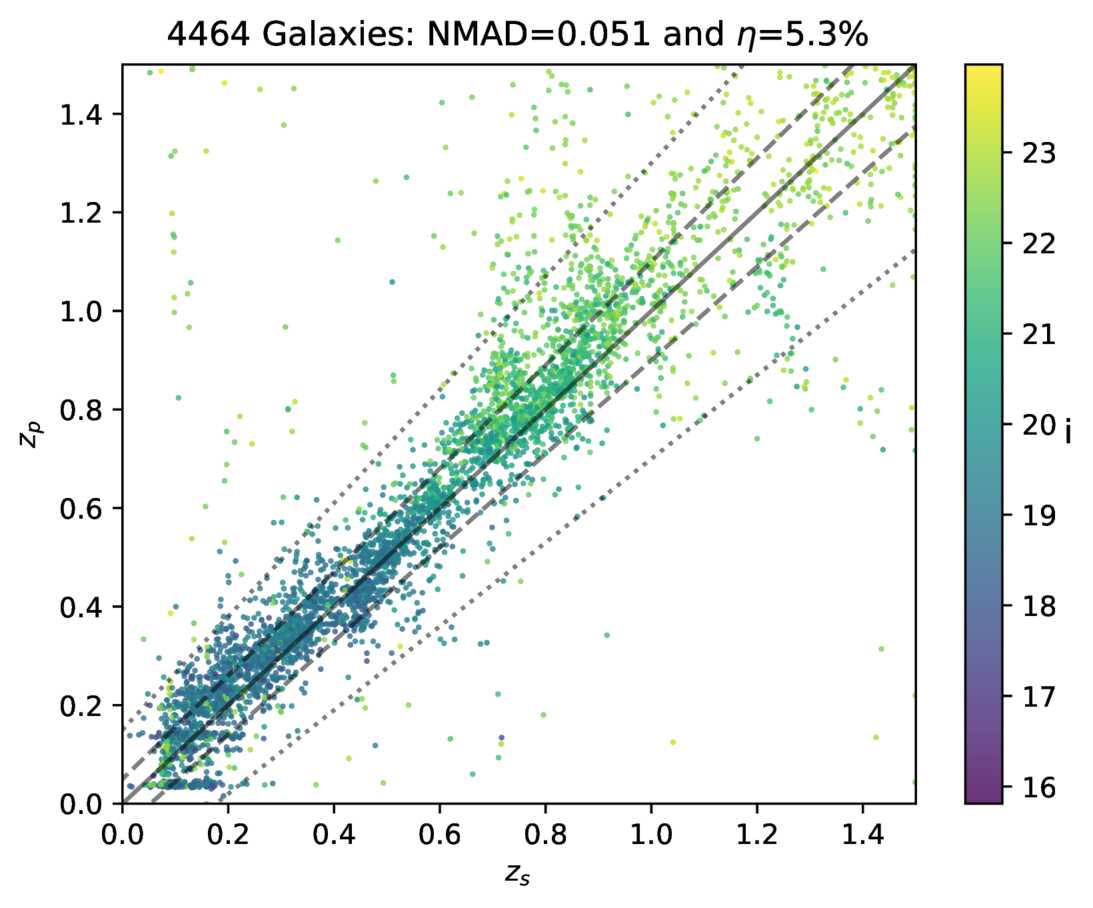}
    \caption{A comparison of the spectroscopic and photometric redshifts, colored by their $i$-band magnitude. The solid line is $\frac{\Delta z}{1+z_s} = 0$; the dashed and dotted lines are $\frac{\Delta z}{1+z_s} = 0.05$ and $\frac{\Delta z}{1+z_s} = 0.15$ respectively.}
    \label{fig:pz_comp}
\end{figure}

There are several competing algorithms for estimating photometric redshifts, including a variety of template-fitting and machine-learning codes. These are currently being implemented in the Redshift Assessment Infrastructure Layers (RAIL) library to support extragalactic science with LSST \citep{rail_team_redshift_2026}. For LoVoCCS science, a photometric redshift algorithm which has a low outlier rate and scatter for \emph{bright} objects (such as cluster members) was needed; additionally the special case of cluster weak lensing can tolerate larger scatter at high redshifts\footnote{For mass estimation, the source redshift is used to estimate the distance-ratio $\frac{D_{LS}}{D_S}$ which, for our low-redshift clusters, is nearly unity for redshifts $z > 0.5$. At redshifts beyond this, the photometric redshifts only need to be \emph{good enough} to keep the distance-ratio near unity.}. Achieving a small scatter at low redshifts is challenging without added flexibility in the set of templates and choice of priors; it is for these reasons that BPZ was chosen for estimating photometric redshifts.

\begin{figure}
    \centering
    \includegraphics[width=0.98\linewidth]{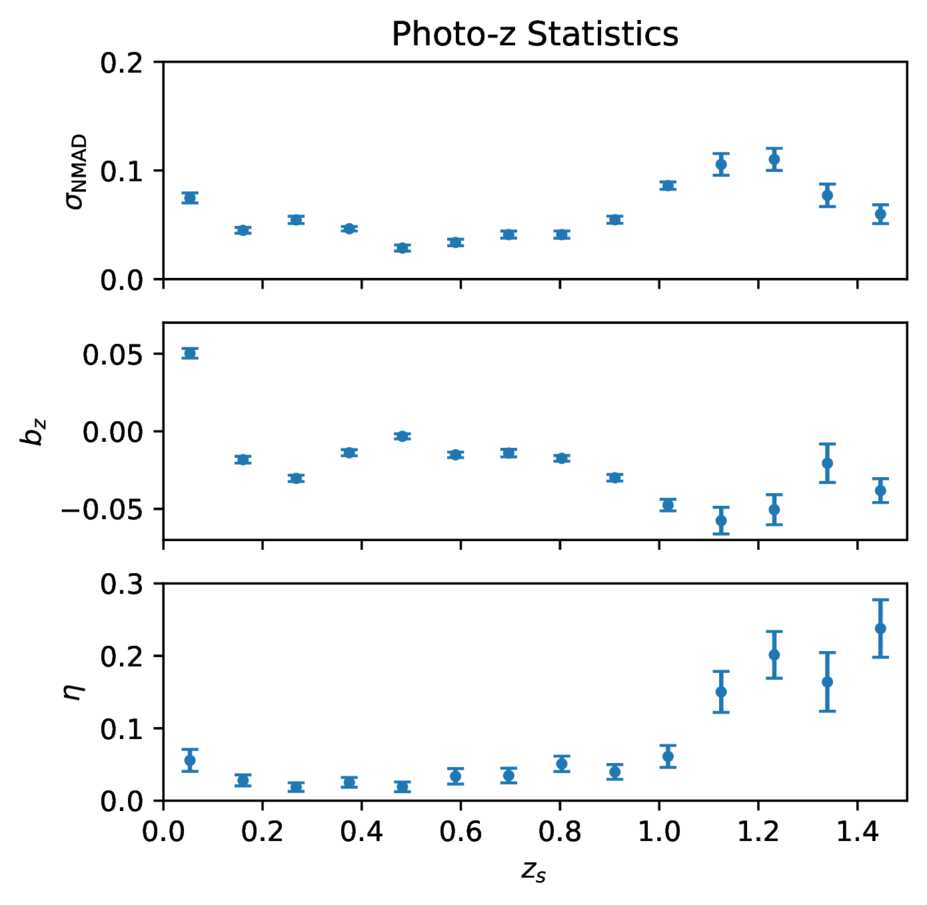}
    \caption{From top to bottom: plots showing the scatter, bias, and outlier rates in our photometric redshifts as a function of their spectroscopic redshift. Generally, we see larger statistics at higher redshifts where most of the emission has been redshifted out of $ugri$ and into $z$.}
    \label{fig:pz_stats}
\end{figure}

Through DESI DR1 \citep{desi_collaboration_data_2025}, a total of $10,988$ galaxies covering Hercules are available for validating the photometric redshifts. First, the Hercules galaxy catalog was truncated following the cuts outlined in Table \ref{tab:meta_cuts} to select galaxies suitable for lensing science. The DESI and Hercules catalogs were then matched with a minimum separation of $0.2\arcsec$, and the statistics shown in Figs. \ref{fig:pz_comp}, \ref{fig:pz_stats} were computed. The overall scatter is $\frac{ \sigma_z }{1+z} \sim 0.051$ with an outlier rate $\eta \sim 5.3\%$. This matches the most generous precision required of photometric redshifts from LSST ($\frac{\sigma_z}{1+z} < 0.1$), but does not pass the requirements for $3\times2\text{ pt}$ nor cluster-cosmology. In principle, a smaller scatter can be achieved by implementing cuts on BPZ's internal quality-check statistics, such as the \texttt{odds} or the modified-$\chi^2$, but this comes at the expense of a large fraction (often $> 50\%$) of sources. 

For a typical source with $z_s\sim0.7$ and foreground cluster at $z\sim0.06$, a $5\%$ scatter in the photometric redshift contributes to only a $\sim 1-3\%$ scatter in the distance ratio, which directly propagates to an identical scatter in mass estimates (Equations \ref{eq:psi_sig}, \ref{eq:psi_shear}). Assuming the shape noise adds an intrinsic scatter of $\sim 30\%$ to each cluster mass estimate \citep{fu_lovoccs_2022}, lensing science using the total sample of $\sim 100$ clusters will be shape-noise limited down to the $\sim 3\%$ level. This implies that the final science results from LoVoCCS will generally be limited by its sample size and the accuracy of its photometric redshifts; using quality cuts to improve the photometric redshifts will be explored in future papers, but it will come at the cost of reducing the source density.

\begin{deluxetable*}{c|c|c}
    \centering
    \tablehead{
    \colhead{Quality Cut} & \colhead{Number of Objects Removed} & \colhead{Percent Pruned}
    }
    \startdata
    $ \text{res} > 0.3 $ \& $ \text{res} < 0.9 $ & $247798$ & $35.6\%$ \\
    $z_p > 0.2$ & $61475$ & $8.8\%$ \\
    $\frac{\sigma_e}{T} < 0.02$ & $59475$ & $8.5\%$ \\
    $T_{\text{gal}} > 1$ \& $T_{\text{gal}} < 40$ & $21563$ & $3.1\%$ \\
    blendedness $< 0.4 $ & $12706$ & $1.8\%$ \\
    $ | r - i| < 10 $ & $2672$ & $0.4\%$ \\
    \hline
    Total (from $S>10$) & $301998$ & $43.4\%$ \\
    \enddata
    \caption{Quality cuts applied to each masked \texttt{metadetect} catalog prior to calibrating; percent prune are computed \emph{after} applying a cut to the signal-to-noise ratio $S > 10$. }
    \label{tab:meta_cuts}
\end{deluxetable*}

Most photometric redshift codes struggle with low redshift sources \citep{newman_photometric_2022}; this is relatively mild in LoVoCCS observations thanks to the $u$-band observations, which ensures the filter set covers the $4000$\r{A} break. Nonetheless,  to avoid the large scatter at low redshifts and select background sources, the final shear catalog has a cut of $z_p > 0.2$. There is an increased scatter, bias, and outlier rate at high redshifts $z > 1$. This occurs as key emission lines (most notably H$\alpha$ and [OIII]) and the $4000$\r{A}-break are redshifted out of the filter-set; however, since the distance ratio rapidly approaches unity for high-redshift sources this does not have a large impact on our mass estimates.

\subsection{Shear Calibration}\label{sec:shear_calib}

\begin{figure*}
    \centering
    \includegraphics[width=0.98\linewidth]{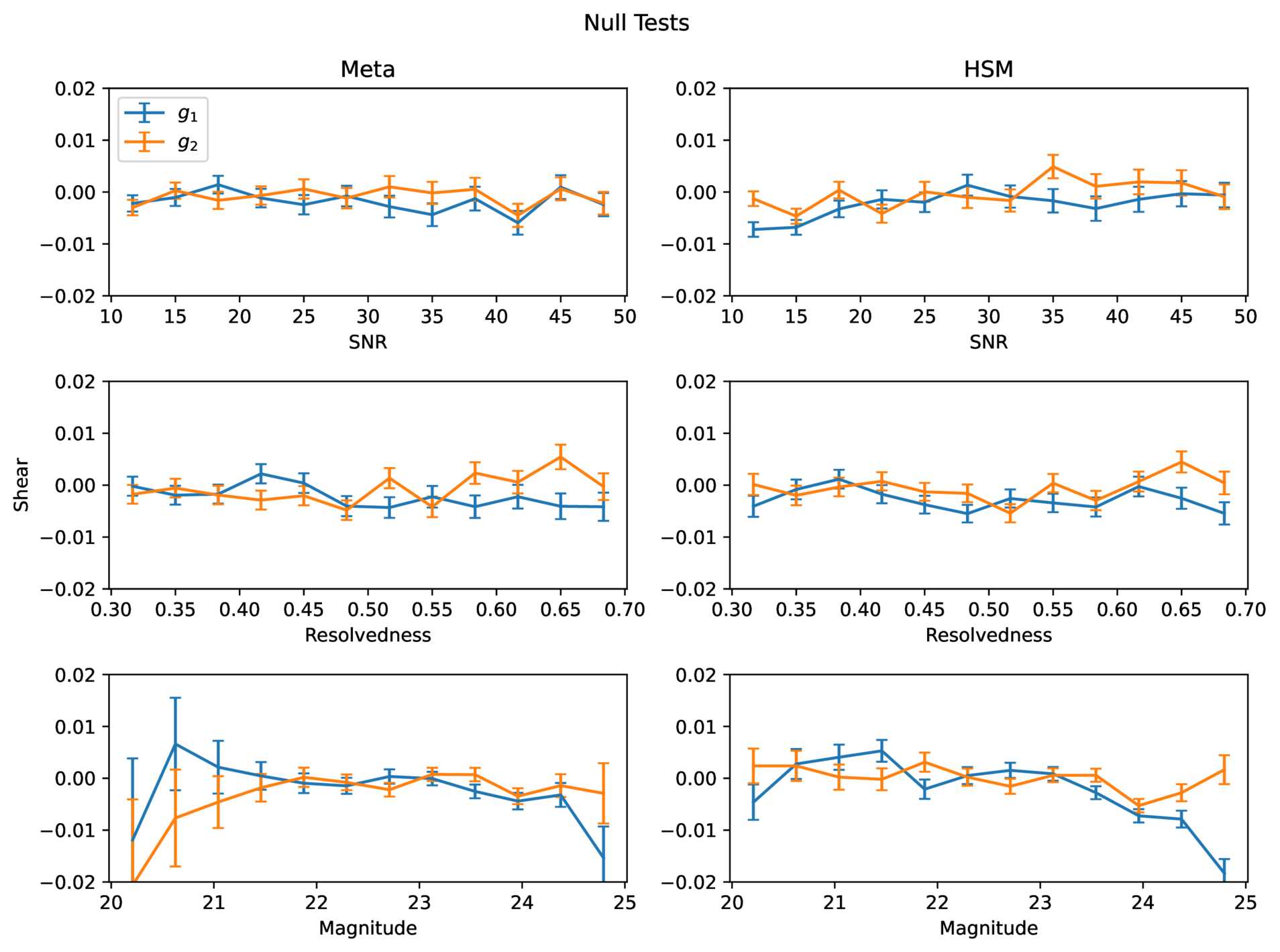}
    \caption{A set of null tests applied to both the \texttt{metadetect} (left) and HSM (right) shape catalogs. From top-to-bottom, these check that the mean shear as a function of $S$, $\text{res}$, and $r$-band magnitude is consistent with zero.}
    \label{fig:shear_null}
\end{figure*}

Prior to calibrating the shears, a series of quality cuts is applied to both the \texttt{metadetect} and HSM shape catalogs. The cuts applied to the HSM catalog are consistent with those applied in \citet{mandelbaum_weak_2018}. The cuts applied to the \texttt{metadetect} catalog are specified as follows. Objects must have a minimum signal-to-noise ratio $S>10$, a resolvedness $\text{res} > 0.3$, and a blendedness $< 0.4$. To prune noise detections and unmasked transients which printed into the final coadds, a color cut ($| r - i | < 10$)  is applied. Most sources with $T_{\text{gal}} < 1$, $T_{\text{gal}} > 40$, or $\text{res} > 0.9$ were artifacts or shredded-galaxies that were not fully masked; these are also dropped from the catalog. When computing shapes from the SDSS moments, errors from the second moments are explicitly propagated to estimate the uncertainty in each component of ellipticity; this is known to underestimate the true uncertainty \citep{mandelbaum_systematic_2005}. However, a cut of $\frac{\sigma_e}{T} < 0.02$ was found to mimic the more accurate cuts on shape-uncertainty used by HSC \citep{li_three-year_2022}; therefore, this was applied to the \texttt{metadetect} catalog. A summary of these cuts and the fraction of objects pruned is provided in Table \ref{tab:meta_cuts}.

Since Hercules covers a large area on sky, there are enough objects to build a robust model for the response and statistical weights. The data was collected into seven bins along $S \in [10,50]$ and $\text{res} \in [0.3,0.9]$, for a total of $49$ bins across the entire parameter space. First, the variance per-bin was estimated with

\begin{equation}
    \hat{V}(S,\text{res}) = \sum_{i \in \text{bin}} \frac{e_{1,i}^2 + e_{2,i}^2}{2}.
\end{equation}

A grid of weights was built by inverting this and interpolating the result to produce a smooth model. These weights were used to compute $\langle e_i \rangle^{j\pm}$ for the diagonal components of the response, $R_{11}$ and $R_{22}$, per-bin according to Equation \ref{eq:resp}. The response was smoothed by a gaussian kernel to reduce the shot noise and interpolated to produce a smooth model. The off-diagonal elements of the response were consistent with zero, reducing Equation \ref{eq:calib_shear} to element-wise division.

With the response model, calibrated shears per-galaxy $i$ were estimated with

\begin{equation}
    g_i^{(j)} = e_{i}^{(j)} \times R_{jj}^{-1}(S_i,\text{res}_i),
\end{equation}

and the per-galaxy weight was estimated with

\begin{equation}
    w_i^{(g)} = w_i^{(e)} \times \left[ \frac{ R_{11}(S_i,\text{res}_i) + R_{22}(S_i,\text{res}_i) }{2} \right]^2.
\end{equation} 

Objects with shears outside of $g_1^2 + g_2^2 < 4$ were dropped; this preserves a small-fraction of non-physical shears ($g_1^2 + g_2^2 > 1$) to avoid truncating their distribution, which can produce an additive bias \citep{mandelbaum_systematic_2005}.

\subsection{Shear Null Tests}

A series of null tests is carried out to assess which of the two shape catalogs, HSM and \texttt{metadetect}, offer better control of systematics. First, on average the shear as a function of $S$, res, and $r$-band magnitude should vanish; this is true for the \texttt{metadetect} catalog, but the HSM catalog fails this check at the $\sim 2\sigma$ level for $S < 15$ and $r > 24$ (Fig.~\ref{fig:shear_null}). The quality of each catalog's PSF correction can be checked by computing the correlation between the shapes of PSF stars and galaxies post-calibration; this can be summarized with the correlation function

\begin{equation}
    \zeta_{\text{sys}} \equiv \frac{ \langle g_{i} , e_{j}^{*} \rangle^2 }{ \langle e_{i}^{*} , e_{j}^{*} \rangle }.
\end{equation}

\begin{figure}
    \centering
    \includegraphics[width=0.98\linewidth]{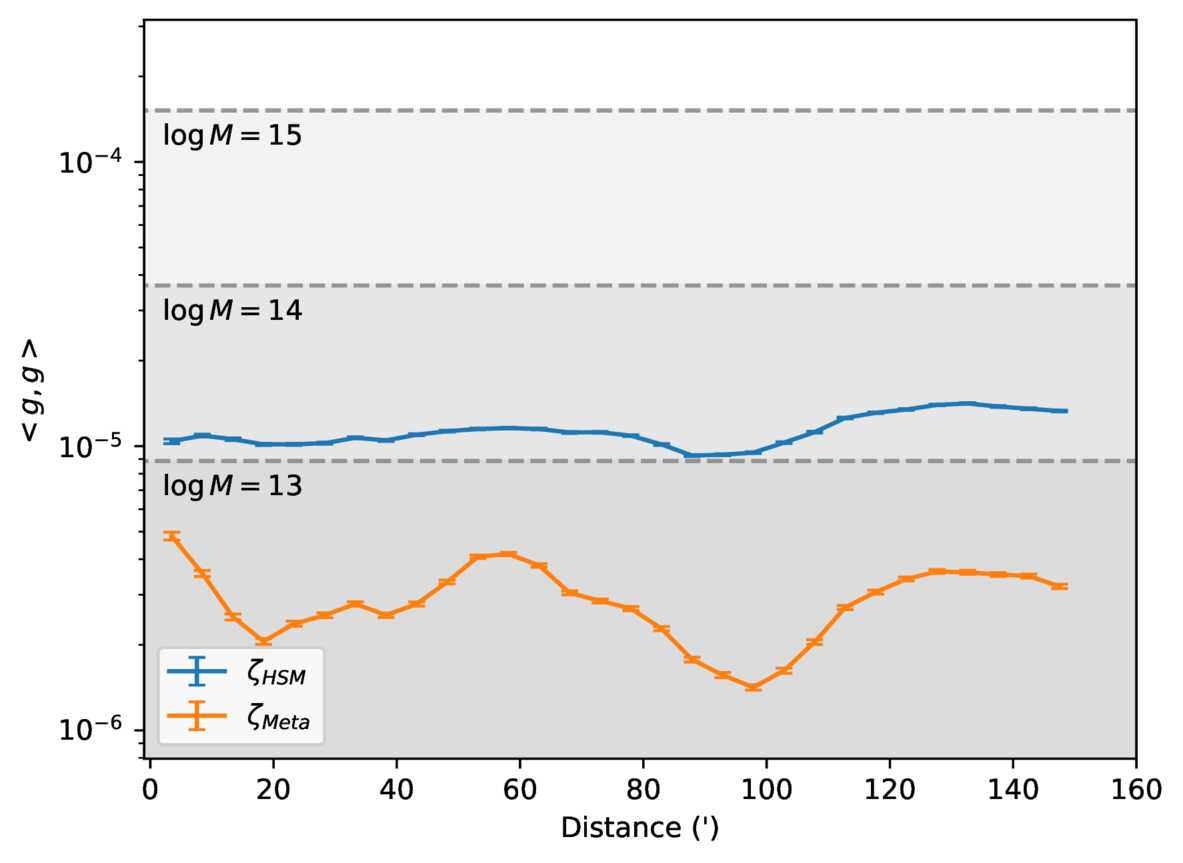}
    \caption{Correlation between the shapes of PSF stars and galaxies; horizontal lines show the root-mean-square shear due to lensing over by a circle of radius $r_{200}$ for halos of various masses assuming a lens at $z=0.05$ and a source at $z=1$.}
    \label{fig:shear_corr}
\end{figure}

$\sqrt{\zeta_{\text{sys}}(\theta)}$ serves as an estimate of root mean-square fluctuations in the shear estimator $g$ on a scale $\theta$ due to the PSF \citep{mandelbaum_first-year_2018}. \texttt{metadetect} provides a strong correction for the PSF, as $\zeta_{\text{sys}} < 10^{-5}$ implies that fluctuations in shear due to the PSF will be on the order of $\delta g \lesssim 0.003$, matching LoVoCCS requirements. This requirement is also fulfilled by the HSM catalog, but with a tighter margin as $10^{-4} < \zeta_{\text{sys}} \lesssim 10^{-5}$ implying that $\delta g \lesssim 0.01$ (Fig.~\ref{fig:shear_corr}).

An additional test is passing star shapes through Equation \ref{eq:map} to generate a ``star-map'' and verifying that both it and the galaxy map are uncorrelated \citep{oguri_two-_2018,mandelbaum_first-year_2018}. This can also test for leakage between the tangential and cross modes of the PSF and the corresponding modes in the mass map. This can be done by computing the Pearson correlation coefficient between each set of maps; correlations between these maps is weak ($ \lesssim 0.1$), so the test is considered passed for both catalogs. Unfortunately, this is not a true null test as the tangential mode of the mass map has a non-zero signal across the field of view. This test is also flawed as the correlation between the mass map and star map is a spatially varying field that cannot be summarized in a single statistic. 

\begin{figure}
    \centering
    \includegraphics[width=0.98\linewidth]{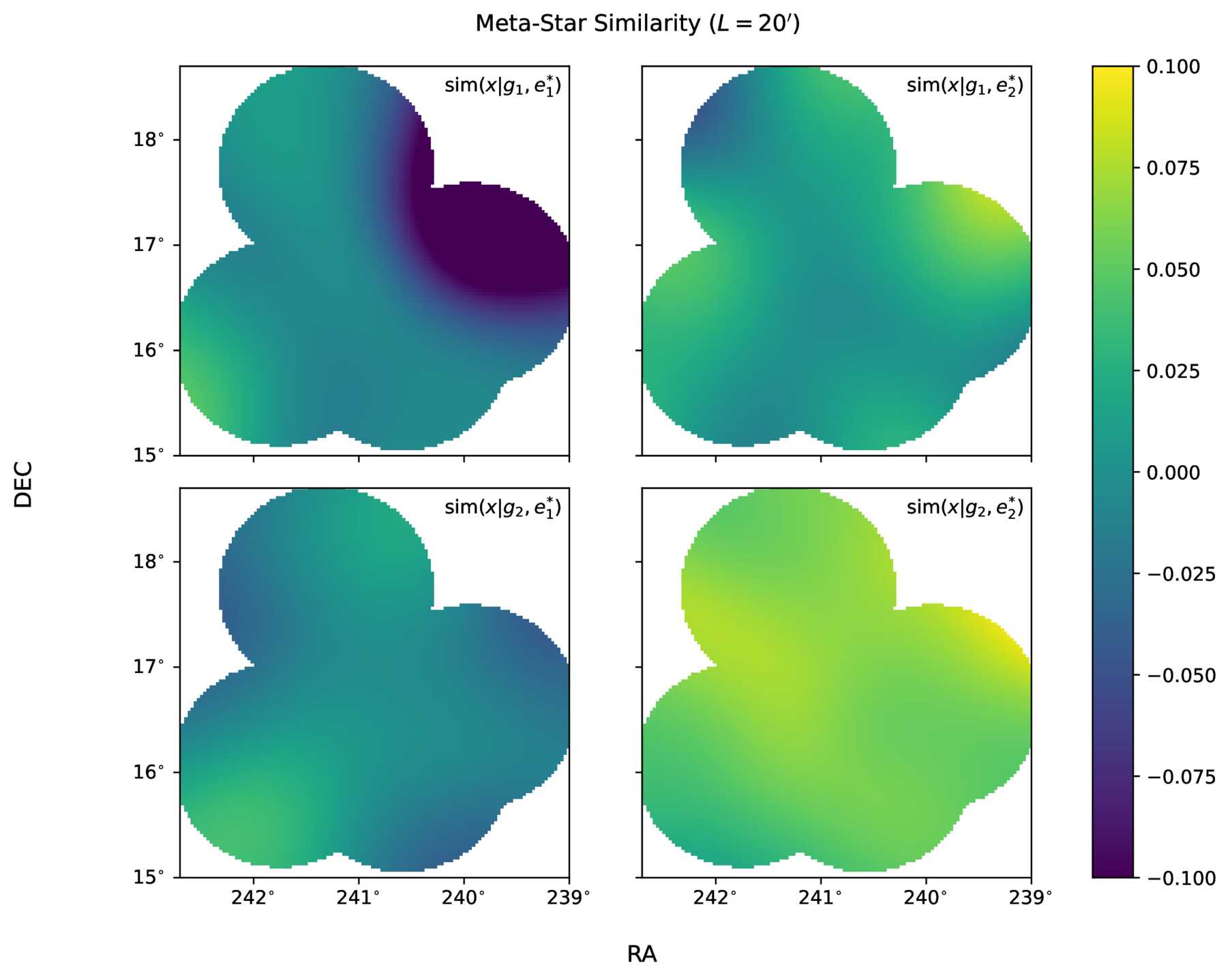}
    \caption{The similarity score, defined in Equation \ref{eq:simscore}, evaluated across the \texttt{metadetect} catalog. There is a region where the correlation is particularly strong, fortunately it does not cover any gravitationally bound members of Hercules; sources in this region will be masked during the lensing analysis.}
    \label{fig:shear_simscore}
\end{figure}

A modified version of this test can be carried out by computing a spatially varying ``similarity score.'' first define a function

\begin{equation}
    f(x;L,M_1,M_2) = \int~dA'~T(|x - x'|; L)~\tilde{M}_1(x')~\tilde{M}_2(x') ,
\end{equation}

where $\tilde{M}_i$ is

\begin{equation}
    \tilde{M}_i(x) = M_i(x) - \int~dA'~T(|x - x'|; L)~M_1(x').
\end{equation}

When $T$ is a top-hat filter, $f(x)$ is directly proportional to the Pearson correlation coefficient between the fields $M_1$ and $M_2$ in a region of length $L$ centered on $x$. We use this function to define a normalized similarity score which ranges from $(-1,1)$:

\begin{equation}\label{eq:simscore}
    \text{sim}(x|M_1,M_2) = \frac{f(x;L,M_1,M_2)}{ \sqrt{ f(x;L,M_1,M_1)~f(x;L,M_2,M_2)} }.
\end{equation}

In practice, computing this with a top-hat kernel leads to artifacts, so a Gaussian filter whose FWHM is set to $L$ is used instead. The galaxy- and star-shapes are binned across a $150\times150$-pixel grid spanning $\alpha \in (238.5^{\circ},243^{\circ})$ and $\delta \in (14.5^{\circ},19^{\circ})$. The similarity score from the \texttt{metadetect} catalog is shown in Fig.~\ref{fig:shear_simscore}, which confirms that some PSF leakage has occurred leading to a strong correlation between $g_1$ and $e_1^*$. Fortunately, this region did not cover any massive gravitationally bound members of the supercluster; to avoid this region, we mask all objects where $|\text{sim}(x|g_i,e_j^*)| < 0.1$.

\begin{figure}
    \centering
    \includegraphics[width=0.95\linewidth]{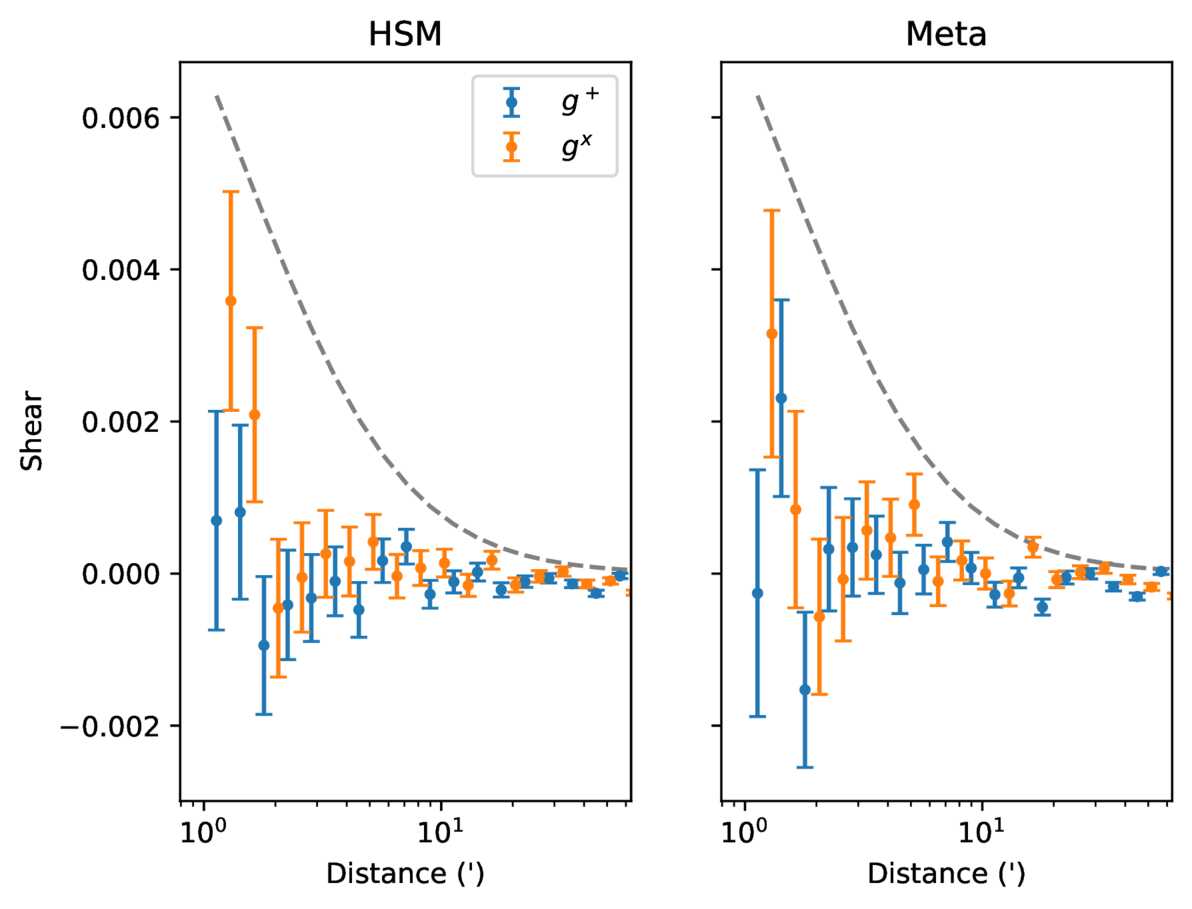}
    \caption{The stacked tangential ($g^+$) and cross ($g^{\times}$) radial shear signal surrounding bright stars with $13 < G < 15$; the radial binning used for each shear-mode is identical and $g^{\times}$ has been shifted to the right for visual clarity. The dashed-curves show the expected shear profile for a $10^{13} M_{\odot}$ lens at $z\sim0.05$. }
    \label{fig:star_valid}
\end{figure}

A final null test can be carried out by estimating the stacked tangential- and cross-mode signals in annuli surrounding bright stars. Bright stars provide a pseudo-random set of points in the sky uncorrelated with any structure outside the Milky Way, so the resulting shear should be zero for both modes. The stacked profile surrounding bright stars also confirms that the extended wings of the PSF, which should be subtracted during the visit-level background model and at least partially masked, are not biasing the shapes of nearby galaxies. Both the HSM and \text{metadetect} catalogs pass this test (Figure \ref{fig:star_valid}) and have radial signals consistent with zero.

\subsection{Additional Considerations}

The \texttt{metadetect} and HSM shape catalogs are only partially overlapping; the noise-image added during \texttt{metadetect} reduces the limiting magnitude of the catalog, cutting its length by $\sim 30\%$ relative to HSM. Fortunately, most objects dropped from the \texttt{metadetect} catalog have a low-signal and are otherwise down-weighted in the HSM analysis. There is a small number of objects which are detected in the \texttt{metadetect} lensing catalog, but absent from the HSM catalog. A fraction of these objects have their signal enhanced by the added noise, while the majority become detectable due to the re-convolution step which uses a slightly larger PSF than the original observations.

The statistical power of the catalogs can be compared by computing the effective number density,

\begin{equation}
    n_{\text{eff}} = \frac{1}{\Omega} \frac{ (\sum w_i)^2 }{\sum w_i^2 },
\end{equation}

where $w_i$ is the shear weight per-galaxy and $\Omega$ is the area on-sky over which these sums are evaluated \citep{heymans_cfhtlens_2012}. The effective source densities for the \texttt{metadetect} and HSM catalogs is provided in Table \ref{tab:lsst_standards}. HSM has a larger effective source density; however, the weighting scheme used in the HSC-Y1 shape calibration was optimized for observations with better seeing ($\sim0.6"$). Comparing the source density from Hercules against HSC fields with $\sim 1\arcsec$ seeing \citep{li_three-year_2022}, the effective source density is reduced to levels comparable to our \texttt{metadetect} catalog ($n_{\text{eff}} \sim 10\text{ gal arcmin}^{-2}$).

\begin{figure}
    \centering
    \includegraphics[width=0.95\linewidth]{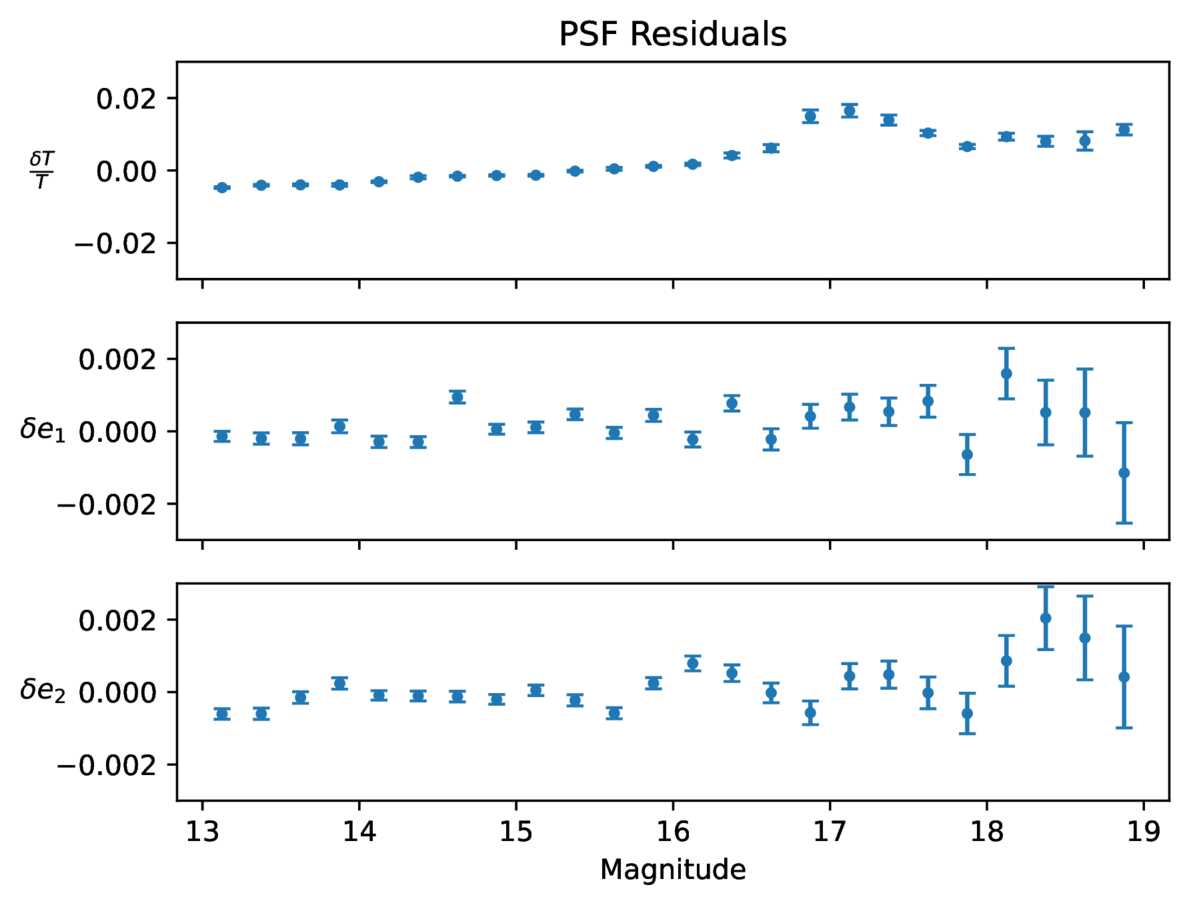}
    \caption{Residuals in the size of the PSF and its ellipticity as a function of instrumental magnitude. Percent-level biases in $\delta T$ are still present, indicating that the BFE has only been partially corrected.}
    \label{fig:valid_psf}
\end{figure}

The BFE systematically biases the shape of bright stars, contributing to a $\sim 5-10\%$ bias in the size of the PSF. If the BFE is not fully corrected, the PSF will over-estimate the size of faint stars and under-estimate the size of bright stars; moreover, because the effect is asymmetric with a preferential increase in size along the read-direction, an uncorrected effect can bias the ellipticity of the PSF-model. For Hercules, the effect is corrected down to the percent-level in the size of the PSF, with no statistically significant bias in the ellipticity of the PSF (Fig.~\ref{fig:valid_psf}).

%

\subsection{Comparison to LSST Y1}

The Hercules stack matches several of the expectations outlined in the DESC Science Requirements Document: this includes matching the expected source density in $r$-band by Y1 after accounting for blends, hitting the photometric redshift requirements for Large Scale Structure, and matching the expected depth. The stack falls short in the photometric and astrometric scatter, but this is not surprising given the empirical nature of joint-calibration. An explicit comparison between the Hercules fields and the LSST-Y1 requirements is presented in Table \ref{tab:lsst_standards}.

\begin{deluxetable*}{c|c|c}
    \centering
    \tablehead{
    \colhead{Name} & \colhead{LSST Target/Expectation} & \colhead{LoVoCCS Result}
    }
    \startdata
    LSST Y1 Depth ($ugriz$, $5\sigma$ point-sources) & $24.1, 25.6, 25.8, 25.1, 24.1$ & $24.8, 25.6, 25.5, 25.3, 24.5$ \\
    LSS1 Redshift Scatter $\frac{\sigma_z}{1+z}$ & $ < 0.1$ & $ < 0.05 $ \\
    CL2 Redshift Scatter $\frac{\sigma_z}{1+z}$ & $ < 0.006$ & $ < 0.05 $ \\
    WL2 Redshift Scatter $\frac{\sigma_z}{1+z}$ & $ < 0.002$ & $ < 0.05 $ \\
    Photometric Uniformity ($ugriz$, mmag) & $15,10,10,10,15$ & $65,19,16,22,19$ \\
    Astrometric Uniformity ($\sigma_{\alpha},\sigma_{\delta},\sigma_{\theta}$, mas) & $10,10,10$ & $51,49,40$ \\ 
    LSST Y1 $n$ (gals arcmin$^{-2}$) & 10.4 & 12.0 (\texttt{metadetect}), 16.5 (HSM) \\
    LSST Y1 $n_{\text{eff}}$ (gals arcmin$^{-2}$) & 8.2 & 10.2 (\texttt{metadetect}), 15.5 (HSM) \\
    \enddata
    \caption{A collection of LSST target goals and science requirements. The Y1 depth, LSS1, CL2, WL2 requirements and the anticipated source densities are pulled from \citet{2018arXiv180901669T}; the restrictions on photometric and astrometric uniformity are from \citet{ivezic_lsst_2018}.}
    \label{tab:lsst_standards}
\end{deluxetable*}

\section{Weak Lensing Analysis of Hercules}\label{sec:wl_herc}

Based on the null tests, the \texttt{metadetect} shape catalog offers better control over systematics than the HSM catalog; therefore, for the science presented in this section, we use the \texttt{metadetect} shape catalog.

\subsection{Mass Maps}\label{sec:detect_maps}

A separate problem from constructing maps is detecting sources on them and assigning them accurate statistical confidences. Techniques adopted in literature for carrying this out are varied, with the primary methods being selecting peaks manually or using a peak finder such as SExtractor to detect them. Once sources are detected, the statistical confidence is typically assumed to be the peak signal within a given footprint.

In the case where sources are separated by large distances ($\theta > \theta_{ap}$), the noise can be approximated as a Gaussian random variable\footnote{This is true regardless of the underlying shot noise as, according to the central limit theorem, the weighted average carried out by Equation \ref{eq:map} converges to a normal variable in the limit of $N\to\infty $.}, and pixels with $S > S_0$ will have an $S_0\times\sigma$ confidence level. However, any two pixels separated by less than $\theta_{ap}$ will share a fraction of the galaxies used to compute the aperture mass; therefore, nearby pixels and sources are \emph{correlated}. Since the collection of peak pixels are correlated, they do not constitute an independent sampling and the peak-$S$ of a given source is an over-estimate of the statistical confidence of some source.

This problem can be overcome by explicitly computing a relation between the peak signal and a true statistical confidence. To carry this out, $\sim 900$ noise maps were generated from the data itself by rotating each galaxy's shear by some random phase $e^{i\phi}$; this dissolves the tangential shear signal and leaves a collection of sources with similar noise properties to the original catalog. SExtractor was used to detect sources on each noise map and catalog their peak signal; then, the distribution of peak-signal due to the noise itself is inferred by collecting the detections together into a single catalog and generating a kernel density estimate to reconstruct $p(S)$. The probability of observing a source with some signal $S > S_0$ is then computed and presented as a Gaussianized statistical confidence.
From this, a rule-of-thumb for this dataset is that a source with signal $S > S_0$ is detected with an $(S_0-1)\times\sigma$ statistical confidence.

The true mass map was processed with SExtractor using the same settings as the noise maps, creating a final lensing catalog. These are each assigned a statistical confidence;
a handful of other statistics are also recorded, including each source's bounding-box, centroids, and an integrated flux.\footnote{Strictly speaking, the ``flux'' here is the total aperture mass across a source ($\sum_i M_{ap}^{(+)}(\theta_i)$).}


Before confirming that each peak is associated with a galaxy cluster or structure in Hercules, a catalog of background clusters in the field was constructed. Clusters were pulled from optically selected catalogs including the SDSS redMaPPer (RM) catalog \citep{rykoff_redmapper_2016}, the WHL galaxy cluster catalog from SDSS-DR6 \citep{wen_galaxy_2009}, and the latest catalog of clusters derived from the DESI Legacy Imaging Surveys \citep{wen_catalog_2024}. Detected clusters were preferentially named according to the catalog they were first detected in. Clusters in these catalogs with redshift $0.05 < z < 0.6$ were queried; the lower redshift cut ensures the selected clusters are in the background, and the upper cut reflects difficulties in detecting higher redshift lenses from ground-based observations\footnote{Beyond $z > 0.6$ the small angular size and reduced source density makes it difficult to detect cluster scale ($M \sim 10^{14} M_{\odot}$) lenses.}.

\begin{figure}
    \centering
    \includegraphics[width=0.95\linewidth]{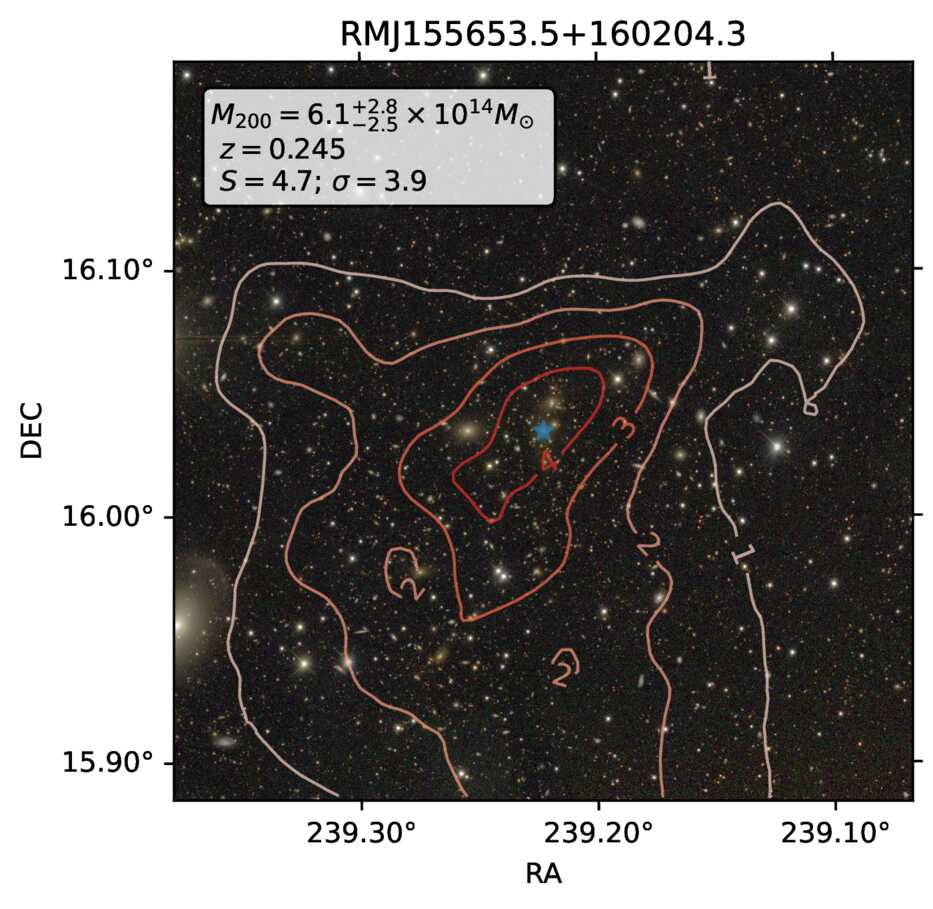}
    \caption{An example cutout which was drawn for each source detected in the mass-map which includes the weak-lensing mass ($M_{200}$), the source redshift ($z$), the peak signal to noise ($S$), and the statistical confidence of the detection ($\sigma$). The boundaries of the cutout were assigned based on the bounding-box of a given footprint, and the location of the peak signal was marked.}
    \label{fig:rm_example}
\end{figure}

\begin{figure*}[th!]
    \centering
    \includegraphics[width=0.48\linewidth]{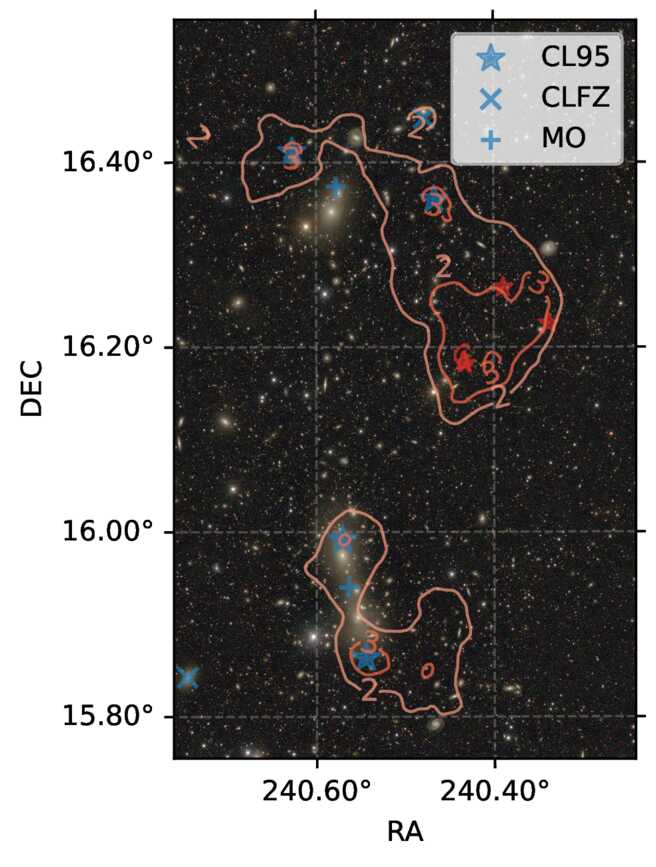}
    \includegraphics[width=0.48\linewidth]{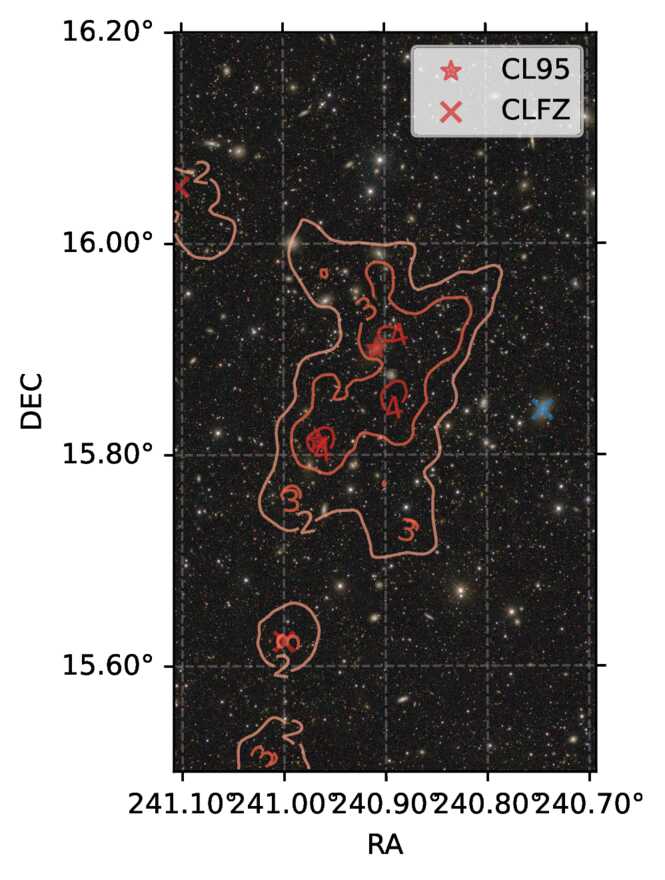}
    \caption{\emph{Left:} Significance contours of the mass map covering A2147; we detect several complexes, including two primary peaks detected in A2147S and a collection of five peaks associated with A2147N. Three of those are associated with background clusters, while two of them are assumed to be in the supercluster itself. \emph{Right:} The background $z \sim 0.1$ complex; though its mass is similar to A2147, the higher redshift results in an enhanced lensing signal. Blue markers are in the foreground, while red markers are in the background; the type of marker specifies the catalog of origin.}
    \label{fig:note_2147}
\end{figure*}

To confirm whether a lensing source is associated with a galaxy cluster, signal-to-noise contours were drawn over the coadds and each peak was manually inspected. If there was an associated galaxy over-density, the optical cluster catalog was queried to search for a nearby cluster and the source was flagged as a match (Fig.~\ref{fig:rm_example}). 

Prior to running mass-fit, catalogs representing four ``scenes'' (collections of foreground deflectors) were built, each of which specify a collection of foreground and background deflectors. Two of these catalogs were derived directly from the mass map: the ``CL95'' catalog contains peaks detected at a strict $95\%$ ($2\sigma$) confidence-level and have an associated over-density of galaxies that appear to belong to a group/cluster; the ``CLFZ'' catalog contains peaks without a strict-cut on the confidence level, but nonetheless have a galaxy over-density and a weak signal in the mass map. The ``MOBG'' and ``\citetalias{monteiro-oliveira_unveiling_2022}'' catalogs were also assembled: the ``MOBG'' catalog carries out a forced fitting of NFW profiles to the substructures identified by \citetalias{monteiro-oliveira_unveiling_2022}, supplemented by a catalog of background clusters from the CL95 catalog; the ``MO'' catalog carries out the same forced fitting but without including the background clusters.

\subsection{Mass Estimates}

For each scene, the procedure outlined in Section \ref{ch:mass_map} is used to compute the reduced shear at the position of each galaxy. Each deflector is assumed to be an NFW, and flat priors are placed over the set of masses $\{ M_i \}$ in a scene; each mass was varied over log-space where $\log (M_{200}) \in (11,16)$. Bootstrapping is used to generate marginalized distributions for each $p(M)$ and to investigate correlations in the mass between nearby peaks; to carry this out, each catalog is resampled with replacement $\sim4000$ times and the fitting is procedure is repeated. Below, the maps and masses of A2147, A2151, A2152, Clump A, and a large background structure are discussed; unless stated otherwise, this discussion concerns the CL95 catalog of sources. Unless specified otherwise, cluster masses are reported out to $R_{200}$.

\subsubsection{Per-Cluster Notes}

\emph{A2147:} A total of five statistically significant peaks are detected (Fig.~\ref{fig:note_2147}); one of these is associated with a complex of rich galaxy clusters at $z\sim0.4$ which are included in the fitting. Two of these are associated with A2147S itself and they appear associated with each of the merging BCGs in the southern complex. In the northern complex, there is not a single peak that aligns with the BCG, rather there are a pair of peaks immediately East and West of it. This curious complex could imply that A2147 is in a more complex state than initially hypothesized from dynamical analyses. Under the CL95 scenario specifically, the mass of the northern and southern complexes are $1.2^{+0.6}_{-0.5} \times 10^{14}~M_{\odot}$ and $2.0^{+0.8}_{-0.7} \times 10^{14}~M_{\odot}$ respectively, while the total mass of the complex is $3.2^{+0.9}_{-0.8} \times 10^{14}~M_{\odot}$. 

The total weak-lensing mass is consistent with the total X-ray mass of A2147, $M_{X200} \sim 3\times10^{14}~M_{\odot}$ \citep{sadibekova_mcxc-ii_2024}, while the dynamical mass of A2147 is $\sim13.5^{+2.1}_{-1.7} \times 10^{14}~M_{\odot}$; a $>5\sigma$ discrepancy from our lensing mass. In their analysis, \citetalias{monteiro-oliveira_unveiling_2022} showed that following periapsis in a 1:3 mass-ratio merger, the dynamical mass of a main-cluster is enhanced by a factor of $\times1.3$ and the dynamical mass of the sub-cluster is enhanced by a factor of $\times 3$. Working from the weak lensing mass of this complex, and assuming A2147S was the ``main cluster'', this implies the enhanced dynamical mass should be $\sim 6.0^{+1.9}_{-1.6} \times 10^{14}~M_{\odot}$. This is consistent with the estimate from \citetalias{monteiro-oliveira_unveiling_2022} at the $2\sigma$ level, implying that the system passed its periapsis in the last $\sim 0.2-0.4~\text{Gyr}$.

\begin{figure*}
    \centering
    \includegraphics[width=0.45\linewidth]{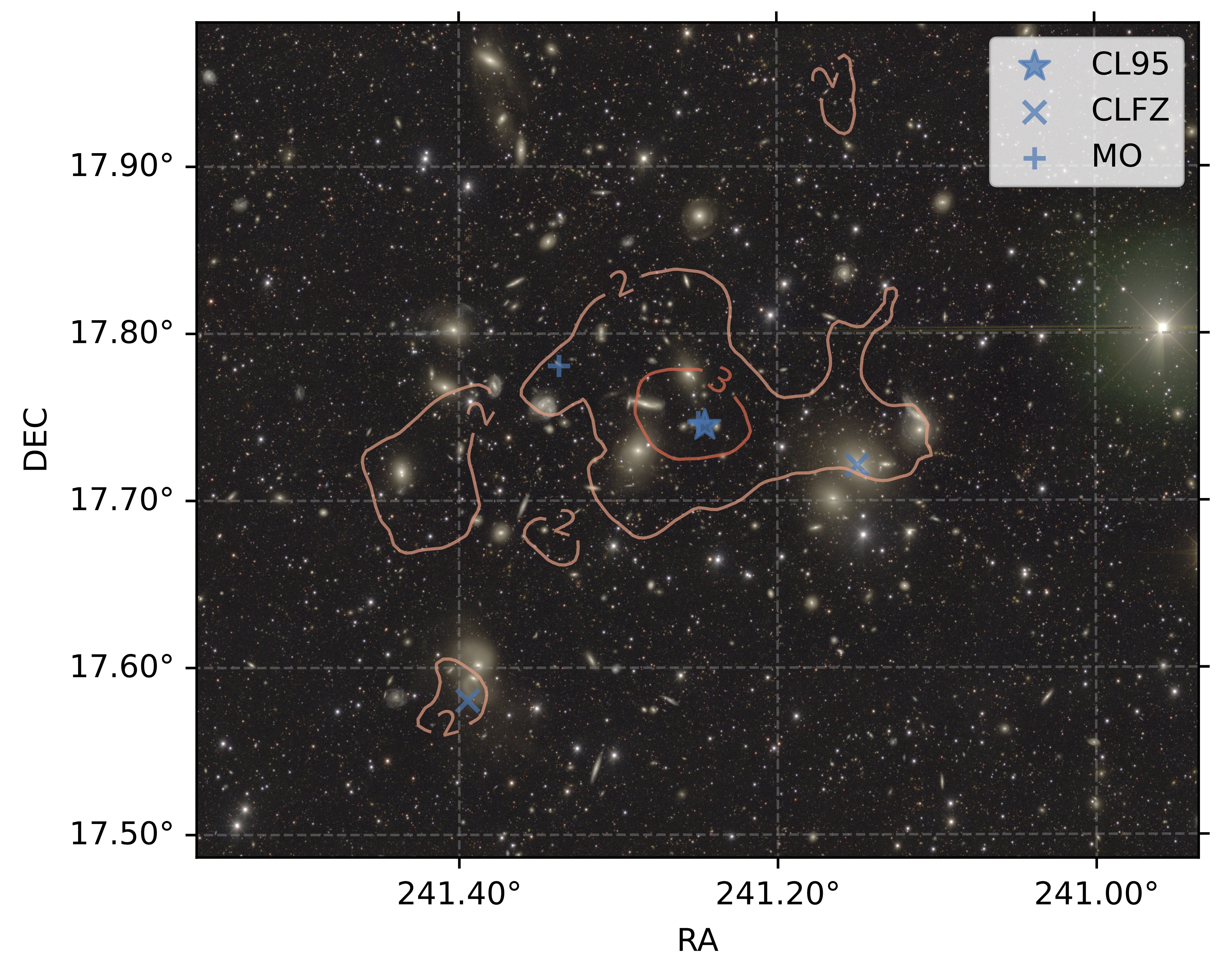}
    \includegraphics[width=0.45\linewidth]{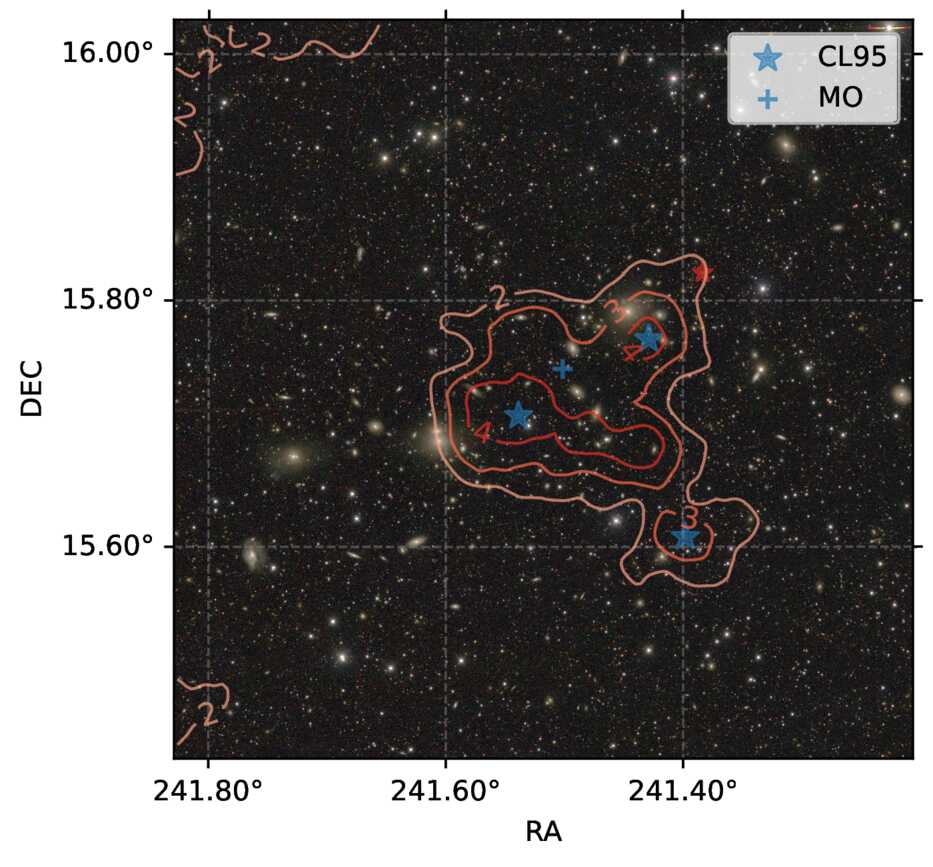}
    \caption{\emph{Left:} A cutout of A2151 showing the footprint which covers A2151C. \emph{Right:} A cutout showing the mass map drawn over Clump A. There are multiple peaks detected, implying that the system is in an early state where multiple groups are merging. The markers are the same as Fig.~\ref{fig:note_2147}.}
    \label{fig:A_2151}
\end{figure*}

\emph{A2151:} The peak lensing signal aligns best with A2151C-F; the associated footprint is extended and consistent with originating from either A2151C-F or A2151C-B (Fig.~\ref{fig:A_2151}). A second, smaller peak which aligns with A2151C-B has been included in the CLFZ catalog, but this detection does not have a high statistical significance. A2151E and an additional peak south of A2151C are also included in the CLFZ catalog. A lensing signal is detected in the northern and southern complexes as well (see Appendix \ref{app:notes}). The total lensing mass of this complex is $2.7^{+1.2}_{-0.8} \times 10^{14}~M_{\odot}$, consistent with \citetalias{monteiro-oliveira_unveiling_2022} and X-ray masses implying that the complex is in the early stages of a merger where its dynamical mass has yet to be enhanced.

\begin{figure}
    \centering
    \includegraphics[width=0.98\linewidth]{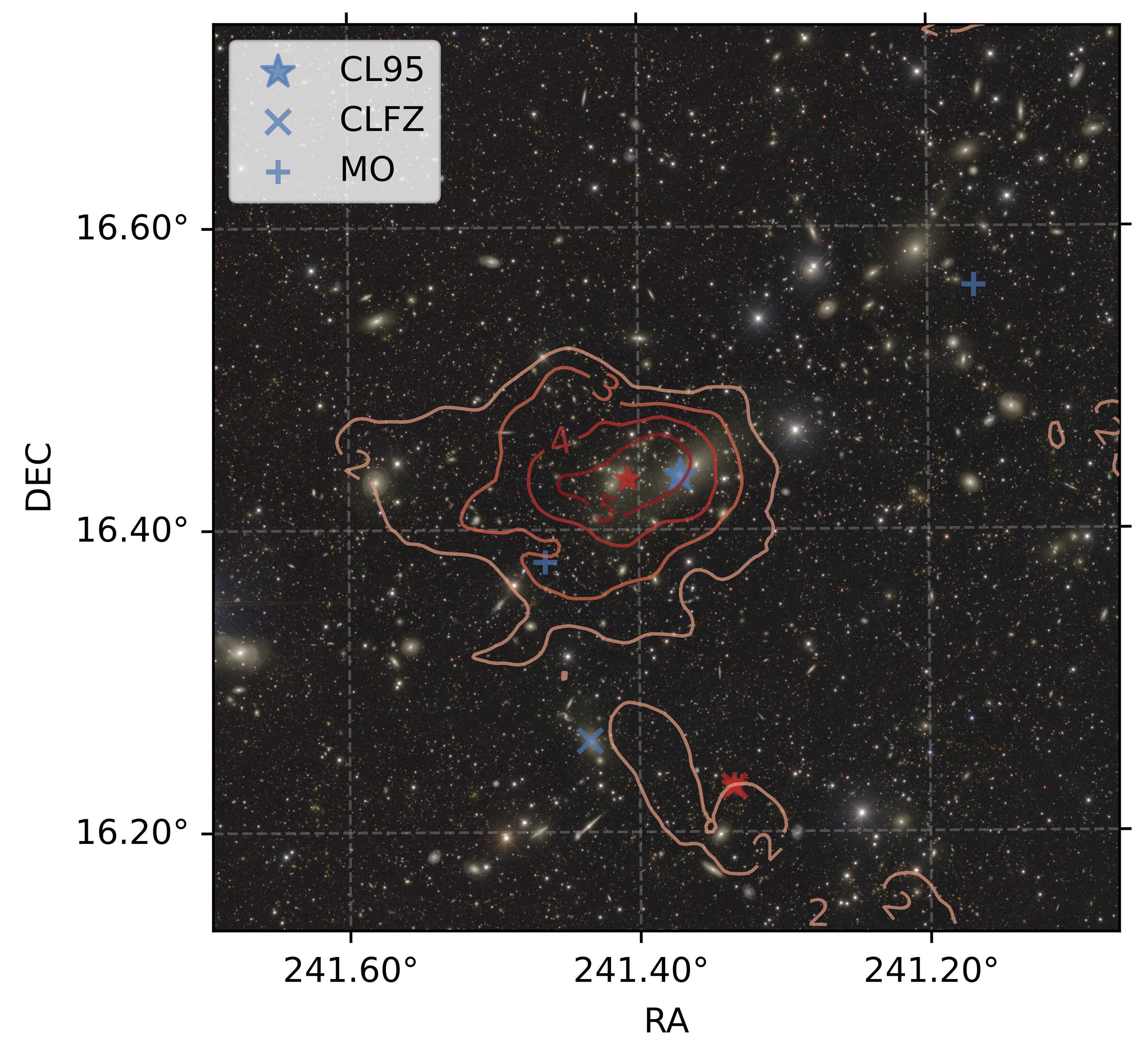}
    \includegraphics[width=0.98\linewidth]{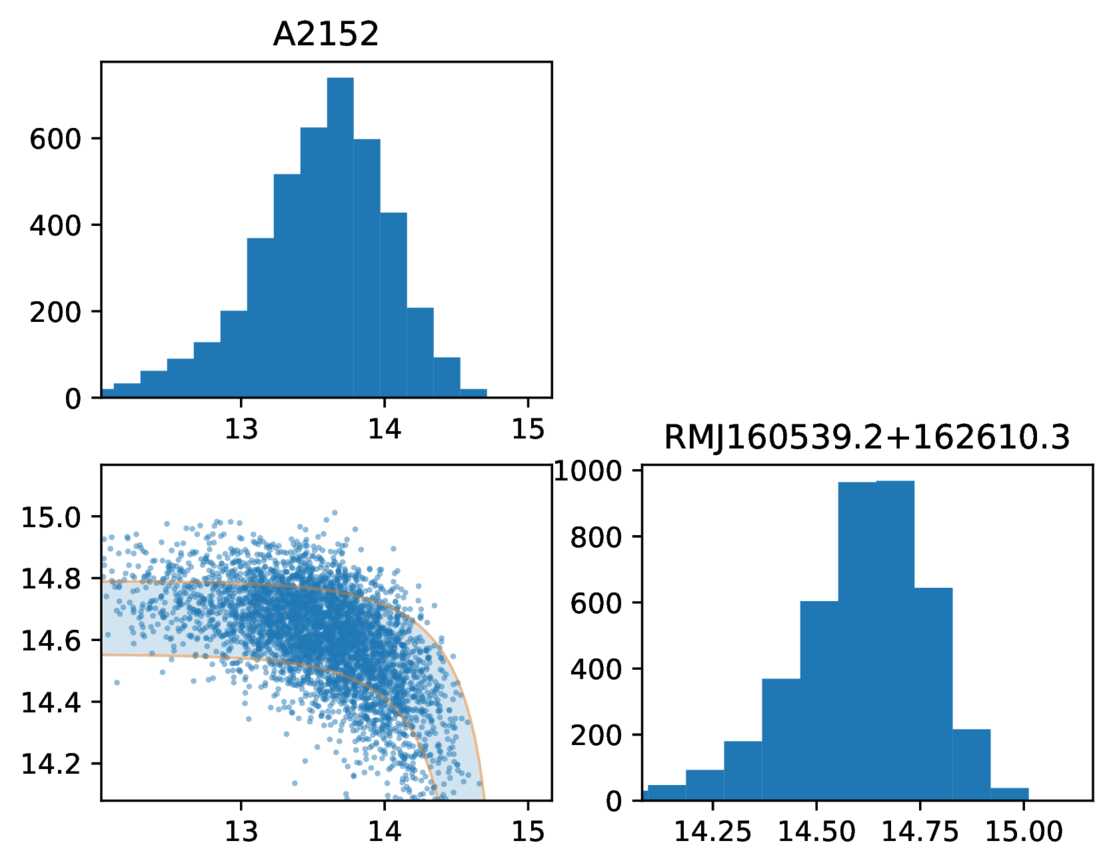}
    \caption{\emph{Top:} A cutout showing the mass map drawn over A2152. The peak lensing signal is noticeably elliptical, similar to the toy-model of a double cluster in Fig.~\ref{fig:mock_dcl}, motivating the assumption that it is composed of two halos at similar redshifts which have blended together in the mass map. \emph{Bottom:} A corner plot comparing $\log(M)$ in the foreground and background deflector; the highlighted region encloses the total mass between the foreground and background within $1\sigma$.}
    \label{fig:2152}
\end{figure}

\emph{A2152:} This cluster is unique due to the presence of a massive background cluster at $z \sim 0.13$, directly behind the apparent BCG of the cluster (Fig.~\ref{fig:2152}). The background cluster, RMJ160539.2, has a large mass and its higher redshift leads to a significantly larger lensing signal compared to the foreground cluster. The footprint in the mass map is noticeably elliptical, similar to the mock double-cluster presented in Fig.~\ref{fig:mock_dcl}; it is plausible that two nearby halos are detected and blended together in the map. Therefore, deflectors were placed at $z\sim 0.1$ for the background cluster and over A2152's BCG, LEDA 57068, at $z\sim 0.04$ for both the CLFZ and CL95 scenarios. The mass estimates for both the foreground and background deflector are non-zero, implying that two halos are detected, but the resulting masses are strongly correlated (Fig.~\ref{fig:2152}). 

We do not detect a weak lensing signal near the A2152N structure identified by \citetalias{monteiro-oliveira_unveiling_2022}, though we do detect weak lensing signal near A2152S that is included in the CLFZ scenario. Under the CL95 scenario, the total mass of A2152 is $0.4^{+0.6}_{-0.3} \times 10^{14}~M_{\odot}$, while under the CLFZ scenario, including A2152S, the total mass is $1.1^{+0.8}_{-0.6} \times 10^{14}~M_{\odot}$. Both values are consistent with the dynamical mass, and the lensing mass cannot be used to confirm if the system has two underlying halos as suggested by \citetalias{monteiro-oliveira_unveiling_2022}.

\emph{Clump A:} There is a rich $z\sim0.4$ redMaPPer cluster overlapping the footprint which is modeled (RMJ160532), though this and the foreground structure's masses are only weakly correlated (Fig.~\ref{fig:A_2151}). Interestingly, this clump has three peaks in the mass map, each of which appears associated with a bright galaxy or a small group of galaxies. Because of its tri-modal nature, the clump could be in the early stages of its formation where the halos of its progenitor clusters have yet to fully merge. For its mass, the lack of X-ray emission is puzzling; this could be because underlying ICM could still be bound to each of the individual $\lesssim 10^{14}~M_{\odot}$ halos, over which ROSAT observations are generally incomplete \citep{voges_rosat_1993}; a deeper followup is needed to confirm whether the ICM is absent from this structure. Clump A does not have a dedicated X-ray mass estimate due to its absence from ROSAT. From \citetalias{monteiro-oliveira_unveiling_2022}, the dynamical mass is $3.1^{+0.6}_{-0.6} \times 10^{14}~M_{\odot}$; this is consistent with the lensing mass, which is $2.2^{+0.6}_{-0.6} \times 10^{14}~M_{\odot}$.

\emph{Background clusters:} In X-ray, A2147 has an anomaly east of A2147S that does not appear associated with an over-density of cluster members (Fig.~\ref{fig:note_2147}). However, this X-ray peak does align with a collection of multiple clusters in the background located at $z\sim 0.1$. The structure is composed of a pair of merging clusters (J160338 \& SS1; Appendix \ref{app:notes}) with a total mass on the order of $\sim 5\times 10^{14}~M_{\odot}$. Despite having a lensing mass comparable to A2147, the higher redshift of this structure enhances its overall lensing signal. Two other background clusters, RMJ160532 \& RMJ160731, have background sources that appear strongly lensed.

\emph{Shear-selected Structures:} Three sources were associated with galaxy overdensities, but lacked a cluster in the optical catalog. These candidate ``shear-selected'' clusters are named SS1, SS2, and SS3; they happen to be located in the background at $z\sim 0.1$.

\emph{CLFZ in the foreground:} A handful of CLFZ sources appear associated with overdensities of galaxies in the foreground. This includes a weak detection which could be associated with A2152S, detections which could be from groups in A2147S (A2147S3) and groups in A2157N (A2147N3), a new group south of A2151C (A2151Sb), and a detection which could be associated with A2151N (A2151N1).

\subsubsection{Total Mass of Hercules}

\begin{figure}[htp]
    \centering
    \includegraphics[width=0.98\linewidth]{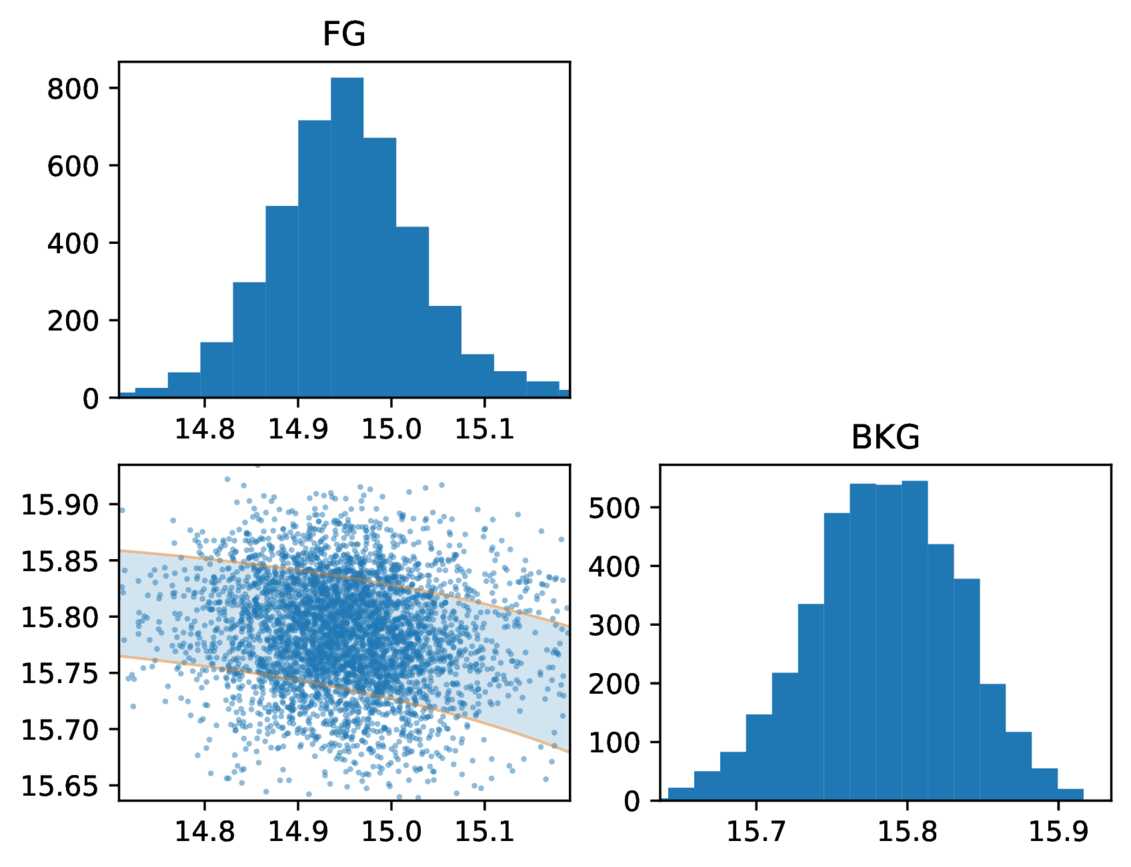}
    \includegraphics[width=0.98\linewidth]{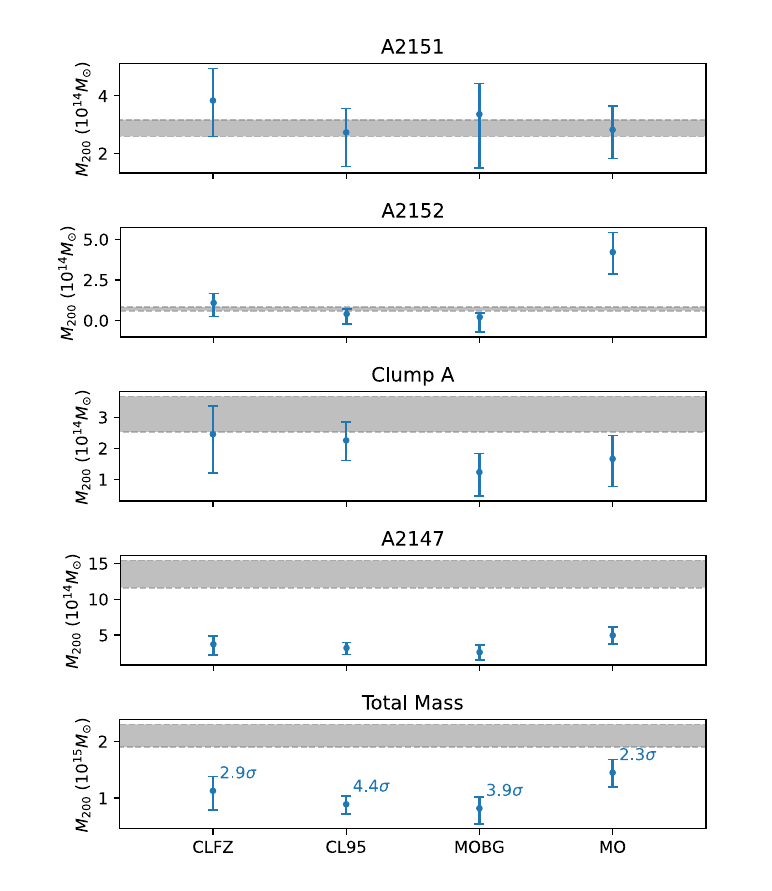}
    \caption{\emph{Top:} A corner plot comparing the $\log(M)$ in the foreground and background of our analysis under the CL95 scenario. \emph{Bottom:} A comparison of the weak lensing (blue) and dynamical mass estimates (gray) for each complex of Hercules across the four scenarios.}
    \label{fig:total_mass}
\end{figure}

The mass of A2147, A2151, A2152, Clump A, and the total mass of the Hercules supercluster are presented in Fig.~\ref{fig:total_mass}; the total lensing mass of the complex is $8.9^{+1.7}_{-1.4} \times 10^{14}~M_{\odot}$. Generally, the dynamical and weak lensing masses of each structure are consistent, with the exception of A2147. Working under the CL95 scenario, the disagreement between the total lensing and dynamical mass of the supercluster is detected with a $>4\sigma$ confidence; this is primarily driven by the large disagreement in the dynamical and lensing masses of A2147. The significance of this result shifts between \citetalias{monteiro-oliveira_unveiling_2022} and MOBG, specifically the omission of background clusters weakens the result by $\sim 1.8\sigma$. This is primarily driven by the increased mass of A2152 and A2147 under the \citetalias{monteiro-oliveira_unveiling_2022} scenario, both of which are biased due to the presence of multiple massive background clusters. Although the background and foreground masses overall are uncorrelated (Fig.~\ref{fig:total_mass}), this shows that the mass of individual clusters in the foreground can be biased by nearby background clusters. This has been mitigated by including background clusters explicitly and marginalizing over their mass when computing the final mass of a foreground cluster.

\section{Conclusion}

In this paper, we present the new Gen3 pipeline that will support future data processing for the LoVoCCS collaboration, and demonstrated its utility by carrying out a weak lensing analysis of the Hercules Supercluster. The new pipeline addresses several key systematics that were left unresolved in the Gen2 pipeline, including correcting for the BFE, using \texttt{metadetect} for shear calibration rather than the HSC-Y1 calibration, and implementing a less aggressive background model enabling a subset of low surface brightness (LSB) science. We have also explicitly computed the photometric/astrometric uncertainty and shown that we are nearing LSST-Y1 standards, but have fallen short due to the empirical nature of joint-calibration.

We also introduce a new suite of tools to support cluster weak lensing. The biased mass estimates of A2147 and A2152 demonstrate the importance of including, and marginalizing over, massive deflectors which are also present along the line-of-sight. Our estimators for the mass map have been optimized and we have begun automatically detecting and assigning a statistical significance to peaks in the map. In upcoming papers, these techniques will be applied to the complete sample of LoVoCCS clusters and, eventually, LSST observations. However, based on the current timeline for operations, LoVoCCS will remain the deepest and most homogeneous view of clusters in the southern sky until at least LSST-Y2 data products are released, giving LoVoCCS observations a strong legacy value until at least 2030.\footnote{LSST operates on a rolling cadence, leaving fields covered in-homogeneously until a few years of survey operations have passed \citep{leistedt_uniform_2026}.}

The deep observations presented here enable studies of the galaxy population in Hercules down to at least $M_r \sim -12$ and can support LSB science. These avenues will be explored in a followup paper, which will study the galaxy population and its similarities to the mass maps presented in this work. In particular, the intracluster light and the blue galaxy populations provide independent ways to confirm the dynamical state of the complex \citep{englert_intracluster_2025,kong_merger-induced_2026}.

Concerning the Hercules Supercluster, our weak lensing maps have led to the identification of several candidate substructures which comprise A2147 and Clump A. Moreover, we have confirmed that the dynamical and weak lensing masses of A2147 disagree, consistent with the hypothesis that its dynamical state is biasing the dynamical mass estimate. Follow-up studies will soon push the magnitude completeness fainter by utilizing DESI observations or targeted followups with the Subaru Prime Focus Spectrograph, and provide a more complete picture of the dynamical state of A2147 and the supercluster overall. By combining future dynamical analyses with broadband studies of the stellar mass (in the form of galaxies and intracluster light) and weak lensing studies of the dark matter, a complete history of the supercluster can be assembled.

\begin{acknowledgments}

This work was initially completed and presented as a portion of A.M.E.'s doctoral dissertation; A.M.E. would like to thank Jonathan Pober and Savvas Koushiappas for their comments during this stage of the writing.


We thank the members of the Observational Cosmology, Gravitational Lensing and Astrophysics Research Group at Brown University for contributing to the testing of \texttt{lovoccs\_pipe} on different galaxy clusters, especially previous/current students and group members: 

A.M.E., Z.E., S.H., and J.N. acknowledge support from the NASA Rhode Island Space Grant Consortium.
A.M.E., Z.E., S.H., and I.D. acknowledges support from the Department of Energy grant \#DE-SC0010010

I.D. and D.C. are thankful for support from the National Science Foundation (No. AST-2108287; Collaborative Research; LoVoCCS: The Local Volume Complete Cluster Survey). 
G.W. gratefully acknowledges support from the National Science Foundation through grant AST-2347348. 
M.D., D.T., and A.E. are grateful for support from the National Aeronautics and Space Administration Astrophysics Data Analysis Program (NASA-80NSSC22K0476).
P.N.~acknowledges support from Department of Energy grant \#DE-SC0017660. E.P. is supported by NASA grant 80NSSC23K0747.

K.U. acknowledges support from the National Science and Technology Council of Taiwan (grants NSTC 112-2112-M-001-027-MY3 and NSTC 115-2112-M-001-027-) and from the Academia Sinica Investigator Award (AS-IA-112-M04). 

J.S. was supported by the National Research Foundation of Korea (NRF) grant funded by the Korea government (MSIT) (No. RS-2023-00210597).

H.M. was supported by JSPS KAKENHI Grant Numbers 22K21349, 23H00108, and 24KK0065, JST FOREST Program Grant Number JPMJFR2137, and Tokai Pathways to Global Excellence (T-GEx), part of MEXT Strategic Professional Development Program for Young Researchers.

M.M. acknowledges support from grant RYC2022-036949-I financed by the MICIU/AEI/10.13039/501100011033 and by ESF+, grant PID2024-158845NB-I00  financed by MICIU/AEI/10.13039/501100011033 and by ERDF, EU, and program Unidad de Excelencia Mar\'{i}a de Maeztu CEX2020-001058-M.

This research was conducted using computational resources and services at the Center for Computation and Visualization, Brown University. 

This work has made use of data from the European Space Agency (ESA) mission {\it Gaia} (\url{https://www.cosmos.esa.int/gaia}), processed by the {\it Gaia} Data Processing and Analysis Consortium (DPAC,\url{https://www.cosmos.esa.int/web/gaia/dpac/consortium}). Funding for the DPAC has been provided by national institutions, in particular the institutions participating in the {\it Gaia} Multilateral Agreement. 

The Pan-STARRS1 Surveys (PS1) and the PS1 public science archive have been made possible through contributions by the Institute for Astronomy, the University of Hawaii, the Pan-STARRS Project Office, the Max-Planck Society and its participating institutes, the Max Planck Institute for Astronomy, Heidelberg and the Max Planck Institute for Extraterrestrial Physics, Garching, The Johns Hopkins University, Durham University, the University of Edinburgh, the Queen's University Belfast, the Harvard-Smithsonian Center for Astrophysics, the Las Cumbres Observatory Global Telescope Network Incorporated, the National Central University of Taiwan, the Space Telescope Science Institute, the National Aeronautics and Space Administration under Grant No. NNX08AR22G issued through the Planetary Science Division of the NASA Science Mission Directorate, the National Science Foundation Grant No. AST–1238877, the University of Maryland, Eotvos Lorand University (ELTE), the Los Alamos National Laboratory, and the Gordon and Betty Moore Foundation.

The national facility capability for SkyMapper has been funded through ARC LIEF grant LE130100104 from the Australian Research Council, awarded to the University of Sydney, the Australian National University, Swinburne University of Technology, the University of Queensland, the University of Western Australia, the University of Melbourne, Curtin University of Technology, Monash University and the Australian Astronomical Observatory. SkyMapper is owned and operated by The Australian National University's Research School of Astronomy and Astrophysics. The survey data were processed and provided by the SkyMapper Team at ANU. The SkyMapper node of the All-Sky Virtual Observatory (ASVO) is hosted at the National Computational Infrastructure (NCI). Development and support of the SkyMapper node of the ASVO has been funded in part by Astronomy Australia Limited (AAL) and the Australian Government through the Commonwealth's Education Investment Fund (EIF) and National Collaborative Research Infrastructure Strategy (NCRIS), particularly the National eResearch Collaboration Tools and Resources (NeCTAR) and the Australian National Data Service Projects (ANDS).

Funding for the Sloan Digital Sky Survey (SDSS) has been provided by the Alfred P. Sloan Foundation, the Participating Institutions, the National Aeronautics and Space Administration, the National Science Foundation, the U.S. Department of Energy, the Japanese Monbukagakusho, and the Max Planck Society. The SDSS Web site is \url{https://www.sdss.org}.
The SDSS is managed by the Astrophysical Research Consortium (ARC) for the Participating Institutions. The Participating Institutions are The University of Chicago, Fermilab, the Institute for Advanced Study, the Japan Participation Group, The Johns Hopkins University, Los Alamos National Laboratory, the Max-Planck-Institute for Astronomy (MPIA), the Max-Planck-Institute for Astrophysics (MPA), New Mexico State University, University of Pittsburgh, Princeton University, the United States Naval Observatory, and the University of Washington. 
This paper makes use of software developed for the Large Synoptic Survey Telescope. We thank the LSST Project for making their code available as free software at  \url{https://www.lsst.org/about/dm}. 
This research has made use of the Spanish Virtual Observatory (\url{https://svo.cab.inta-csic.es}) project funded by MCIN/AEI/10.13039/501100011033/ through grant PID2020-112949GB-I00. 

This research has made use of the NASA/IPAC Extragalactic Database (NED), which is funded by the National Aeronautics and Space Administration and operated by the California Institute of Technology.

This research has made use of the SIMBAD database, operated at CDS, Strasbourg, France.

This research has made use of data, software and/or web tools obtained from the High Energy Astrophysics Science Archive Research Center (HEASARC), a service of the Astrophysics Science Division at NASA/GSFC and of the Smithsonian Astrophysical Observatory's High Energy Astrophysics Division.

This project used data obtained with the Dark Energy Camera (DECam), which was constructed by the Dark Energy Survey (DES) collaboration. This work is based on observations at Cerro Tololo Inter-American Observatory, NSF's NOIRLab (NOIRLab Prop. ID 2019A-0308; PI: I. Dell'Antonio). 
This project used public archival data from the Dark Energy Survey (DES). Funding for the DES Projects has been provided by the U.S. Department of Energy, the U.S. National Science Foundation, the Ministry of Science and Education of Spain, the Science and Technology Facilities Council of the United Kingdom, the Higher Education Funding Council for England, the National Center for Supercomputing Applications at the University of Illinois at Urbana-Champaign, the Kavli Institute of Cosmological Physics at the University of Chicago, the Center for Cosmology and Astro-Particle Physics at the Ohio State University, the Mitchell Institute for Fundamental Physics and Astronomy at Texas A\&M University, Financiadora de Estudos e Projetos, Funda{\c c}{\~a}o Carlos Chagas Filho de Amparo {\`a} Pesquisa do Estado do Rio de Janeiro, Conselho Nacional de Desenvolvimento Cient{\'i}fico e Tecnol{\'o}gico and the Minist{\'e}rio da Ci{\^e}ncia, Tecnologia e Inova{\c c}{\~a}o, the Deutsche Forschungsgemeinschaft, and the Collaborating Institutions in the Dark Energy Survey.
The Collaborating Institutions are Argonne National Laboratory, the University of California at Santa Cruz, the University of Cambridge, Centro de Investigaciones Energ{\'e}ticas, Medioambientales y Tecnol{\'o}gicas-Madrid, the University of Chicago, University College London, the DES-Brazil Consortium, the University of Edinburgh, the Eidgen{\"o}ssische Technische Hochschule (ETH) Z{\"u}rich,  Fermi National Accelerator Laboratory, the University of Illinois at Urbana-Champaign, the Institut de Ci{\`e}ncies de l'Espai (IEEC/CSIC), the Institut de F{\'i}sica d'Altes Energies, Lawrence Berkeley National Laboratory, the Ludwig-Maximilians Universit{\"a}t M{\"u}nchen and the associated Excellence Cluster Universe, the University of Michigan, the National Optical Astronomy Observatory, the University of Nottingham, The Ohio State University, the OzDES Membership Consortium, the University of Pennsylvania, the University of Portsmouth, SLAC National Accelerator Laboratory, Stanford University, the University of Sussex, and Texas A\&M University.
Based on observations at Cerro Tololo Inter-American Observatory, a programme of NOIRLab (NOIRLab Prop. 2012B-0001; PI J. Frieman).

The Legacy Surveys consist of three individual and complementary projects: the Dark Energy Camera Legacy Survey (DECaLS; Proposal ID \#2014B-0404; PIs: David Schlegel and Arjun Dey), the Beijing-Arizona Sky Survey (BASS; NOAO Prop. ID \#2015A-0801; PIs: Zhou Xu and Xiaohui Fan), and the Mayall z-band Legacy Survey (MzLS; Prop. ID \#2016A-0453; PI: Arjun Dey). DECaLS, BASS and MzLS together include data obtained, respectively, at the Blanco telescope, Cerro Tololo Inter-American Observatory, NSF's NOIRLab; the Bok telescope, Steward Observatory, University of Arizona; and the Mayall telescope, Kitt Peak National Observatory, NOIRLab. The Legacy Surveys project is honored to be permitted to conduct astronomical research on Iolkam Du'ag (Kitt Peak), a mountain with particular significance to the Tohono O'odham Nation.
BASS is a key project of the Telescope Access Programme (TAP), which has been funded by the National Astronomical Observatories of China, the Chinese Academy of Sciences (the Strategic Priority Research Programme `The Emergence of Cosmological Structures' Grant \# XDB09000000), and the Special Fund for Astronomy from the Ministry of Finance. The BASS is also supported by the External Cooperation Programme of Chinese Academy of Sciences (Grant \# 114A11KYSB20160057), and Chinese National Natural Science Foundation (Grant \# 11433005).
The Legacy Survey team makes use of data products from the Near-Earth Object Wide-field Infrared Survey Explorer (\textit{NEOWISE}), which is a project of the Jet Propulsion Laboratory/California Institute of Technology. \textit{NEOWISE} is funded by the National Aeronautics and Space Administration.
The Legacy Surveys imaging of the DESI footprint is supported by the Director, Office of Science, Office of High Energy Physics of the U.S. Department of Energy under Contract No. DE-AC02-05CH1123, by the National Energy Research Scientific Computing Center, a DOE Office of Science User Facility under the same contract; and by the U.S. National Science Foundation, Division of Astronomical Sciences under Contract No. AST-0950945 to NOAO.

This research is based on data obtained from the Astro Data Archive at NSF's NOIRLab. These data are associated with the observing programs listed in Table \ref{tab:props}.

\end{acknowledgments}

\facilities{Blanco (DECam), Astro Data Archive}
\software{LSST Science Pipelines \citep{bosch_hyper_2018,bosch_overview_2019},
          \texttt{astropy} \citep{collaboration_astropy_2022},
          \texttt{photutils} \citep{bradley_photutils_2016},
          \texttt{numpy} \citep{harris_array_2020},
          \texttt{scipy} \citep{virtanen_scipy_2020},
          \texttt{matplotlib} \citep{hunter_matplotlib_2007},
          Source Extractor \citep{bertin_sextractor_1996},
}

\clearpage

\appendix

%
%

\section{LoVoCCS South \& Extensions}

\startlongtable
\begin{deluxetable}{ccccccc}
    \tablecaption{The extended Southern sky LoVoCCS catalog \label{tab:lv}}
    \tablehead{
    \colhead{Name} & \colhead{RA [deg]} & \colhead{DEC [deg]} & \colhead{$z$} & \colhead{$L_{500}$ [$10^{44} \text{ erg}\text{ s}^{-1}$]} & \colhead{Source} & \colhead{Observations}
    }
    \startdata
A2029            & 227.73 & 5.72   & 0.077 & 8.75  & LV     & C         \\
A401             & 44.74  & 13.58  & 0.073 & 6.10  & LV     & C         \\
A85              & 10.46  & -9.30  & 0.056 & 5.09  & LV     & C         \\
A3667            & 303.13 & -56.83 & 0.055 & 4.86  & LV     & C         \\
A3827            & 330.48 & -59.95 & 0.098 & 4.23  & LV     & C         \\
A3266            & 67.85  & -61.44 & 0.058 & 3.96  & LV     & C         \\
A3628            & 248.18 & -75.13 & 0.149 & 3.87  & AR     & P         \\
A3571            & 206.87 & -32.85 & 0.039 & 3.85  & LV     & C         \\
A1651            & 194.84 & -4.19  & 0.085 & 3.84  & LV     & C         \\
A754             & 137.28 & -9.67  & 0.055 & 3.84  & LV     & C         \\
A3112            & 49.49  & -44.24 & 0.075 & 3.82  & LV     & C         \\
RXC J1958.2-3011  & 299.56 & -30.19 & 0.117 & 3.63  & AR     & C         \\
A399             & 44.46  & 13.05  & 0.072 & 3.60  & LV     & C         \\
A2597            & 351.33 & -12.13 & 0.083 & 3.55  & LV     & C         \\
A1650            & 194.67 & -1.76  & 0.084 & 3.46  & LV     & C         \\
A2837            & 13.19  & -80.27 & 0.129 & 3.42  & LV     & C         \\
RXC J1558.3-1410 & 239.60 & -14.17 & 0.095 & 3.20  & LV     & C         \\
A3558            & 201.99 & -31.50 & 0.047 & 3.14  & LV     & C         \\
A3695            & 308.70 & -35.81 & 0.089 & 2.99  & LV     & C         \\
A3921            & 342.49 & -64.43 & 0.093 & 2.86  & LV     & C         \\
RXC J1524.2-3154  & 231.05 & -31.90 & 0.103 & 2.83  & AR     & Cu        \\
A2426            & 333.64 & -10.37 & 0.093 & 2.79  & LV     & C         \\
A3158            & 55.72  & -53.64 & 0.059 & 2.76  & LV     & C         \\
RXC J1217.6+0339 & 184.42 & 3.66   & 0.077 & 2.74  & LV     & C         \\
A2811            & 10.54  & -28.54 & 0.107 & 2.73  & LV     & C         \\
A780             & 139.53 & -12.09 & 0.052 & 2.71  & LV     & C         \\
A2420            & 332.58 & -12.18 & 0.084 & 2.66  & LV     & C         \\
A1285            & 172.58 & -14.58 & 0.101 & 2.60  & LV     & C         \\
A3911            & 341.58 & -52.73 & 0.096 & 2.57  & LV     & C         \\
    \enddata 
    \tablecomments{The coordinates, redshift, and $L_{500}$ were extracted from MCXC-II \citep{sadibekova_mcxc-ii_2024}. This catalog includes the original LoVoCCS sources (LV), the SZ-extension (LV-SZ), and the archival extension (AR) as discussed in Sec. \ref{sec:lv}. The observing status specifies whether a cluster's observations are complete (C), complete except for $u$-band (Cu), partially observed with data suitable for lensing (P), or incomplete (IC). This table is published in its entirety in the machine-readable format. A portion is shown here for guidance regarding its form and content.}
    \digitalasset

\end{deluxetable}

\section{Individual Peaks}\label{app:notes}

\begin{figure*}
    \centering
    \includegraphics[width=0.3\linewidth]{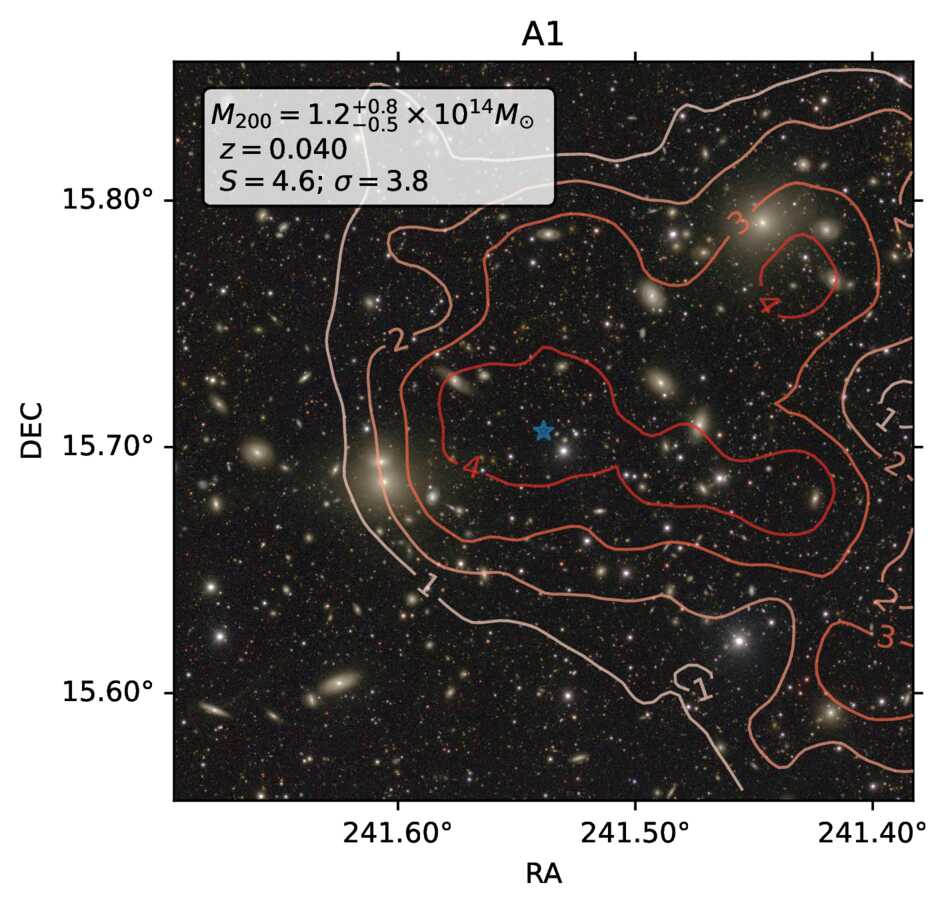}
    \includegraphics[width=0.3\linewidth]{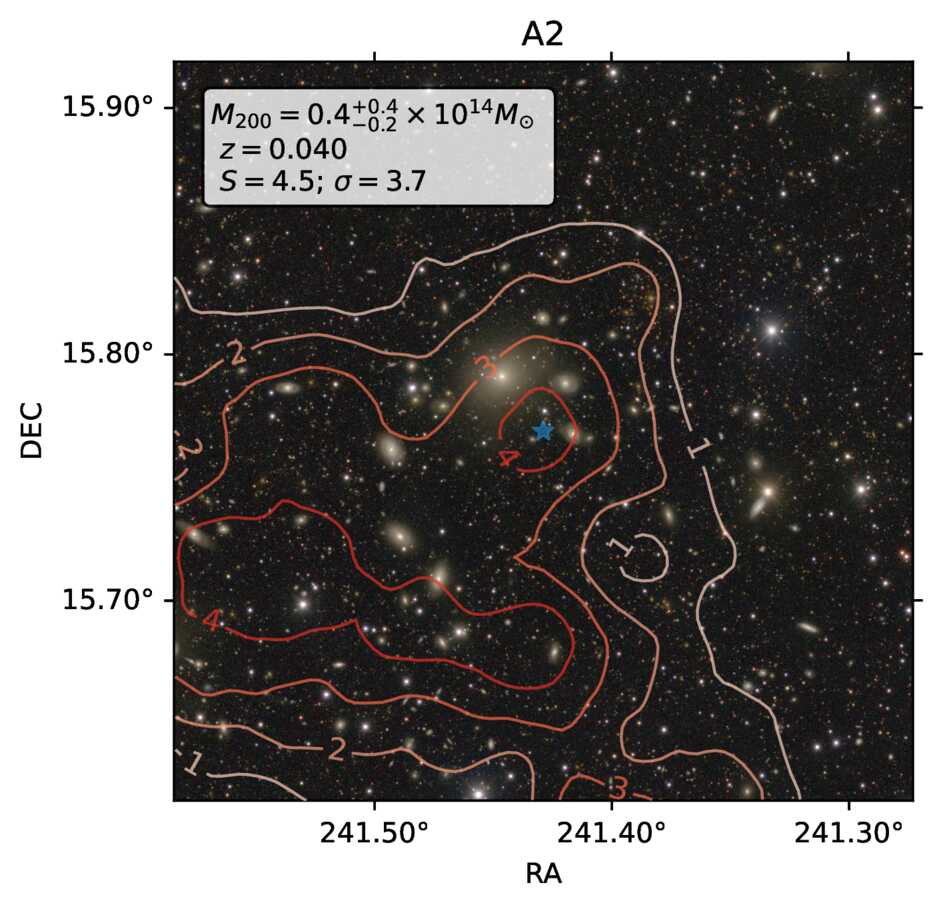}
    \includegraphics[width=0.3\linewidth]{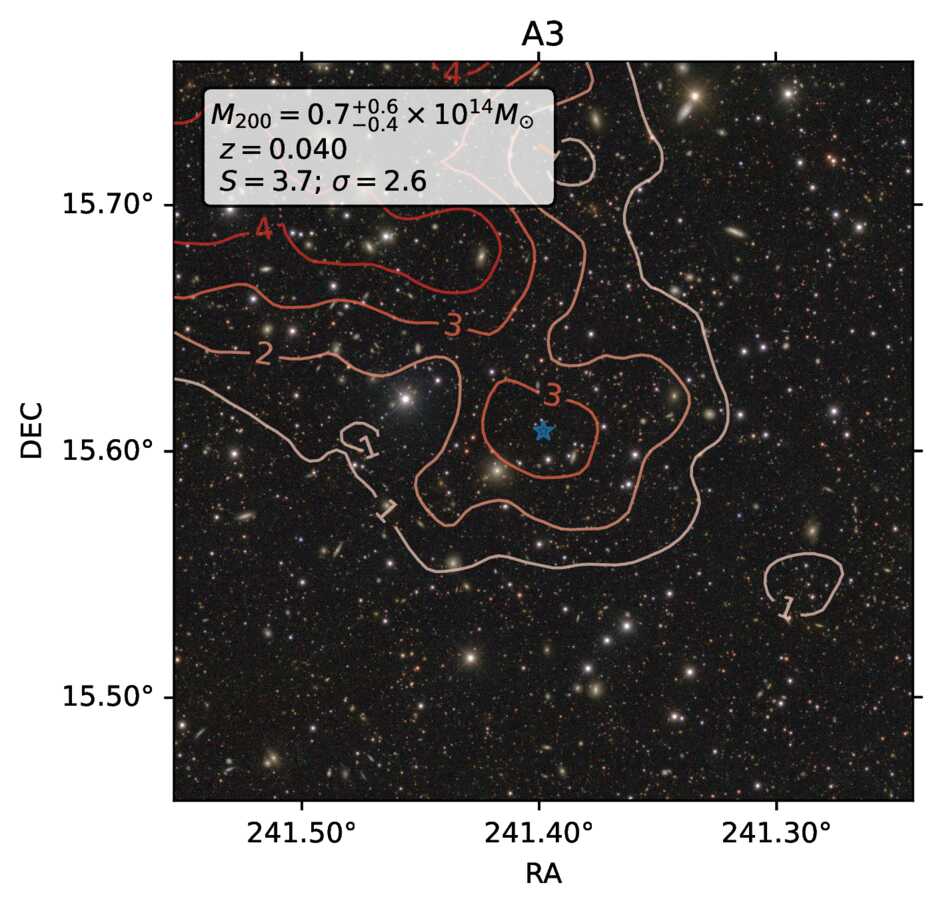}
    \includegraphics[width=0.3\linewidth]{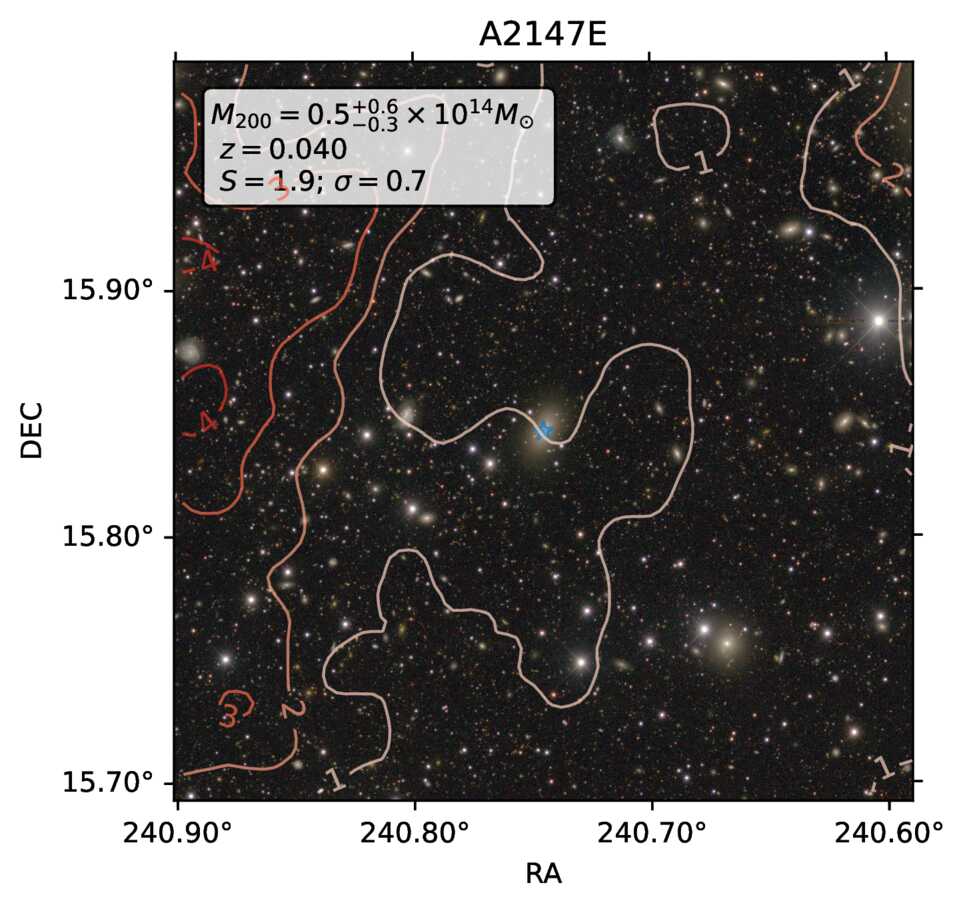}
    \includegraphics[width=0.3\linewidth]{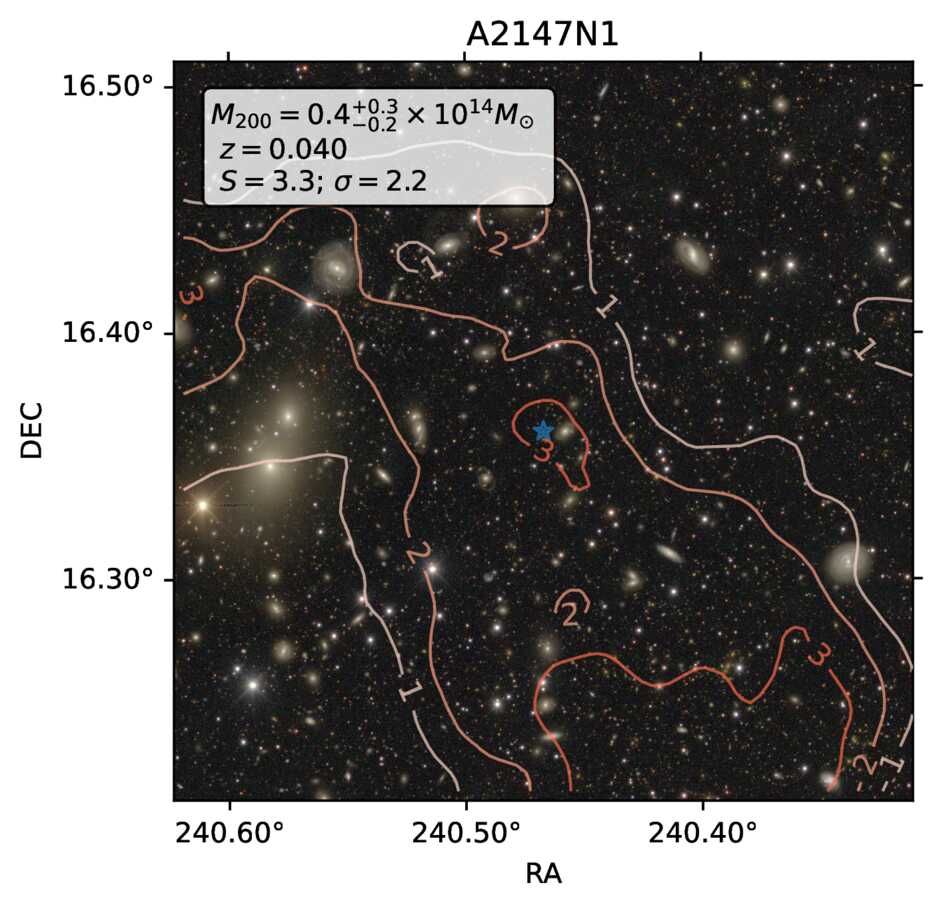}
    \includegraphics[width=0.3\linewidth]{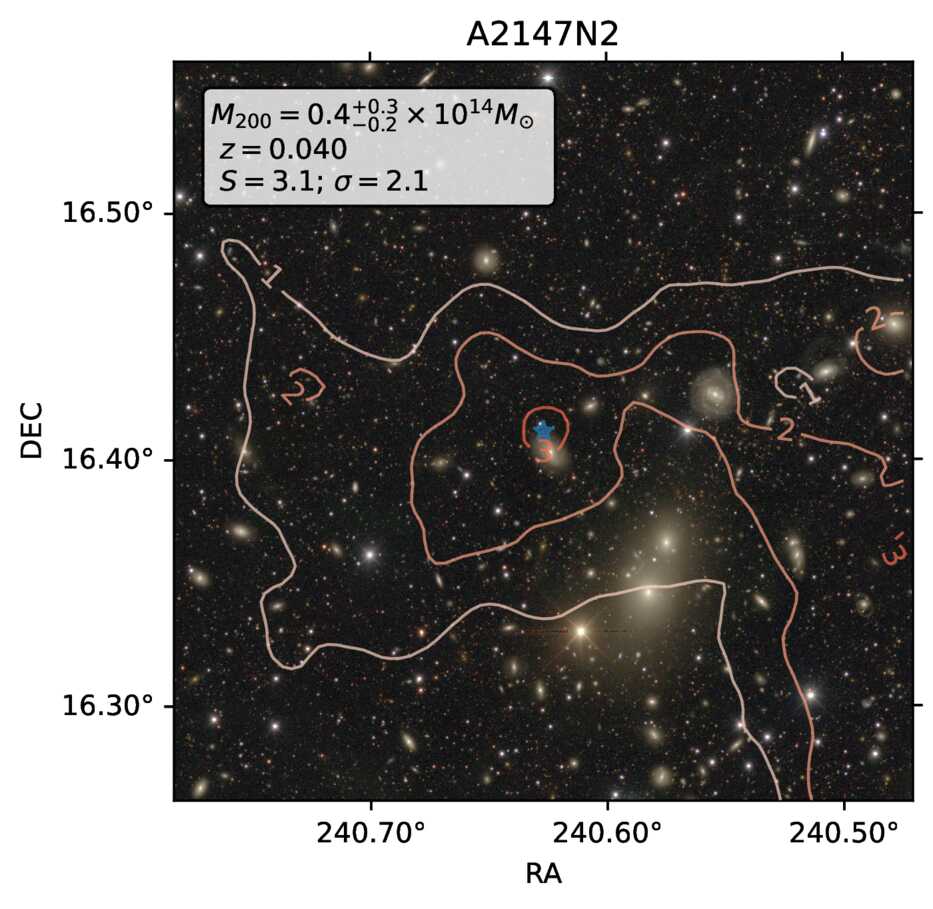}
    \includegraphics[width=0.3\linewidth]{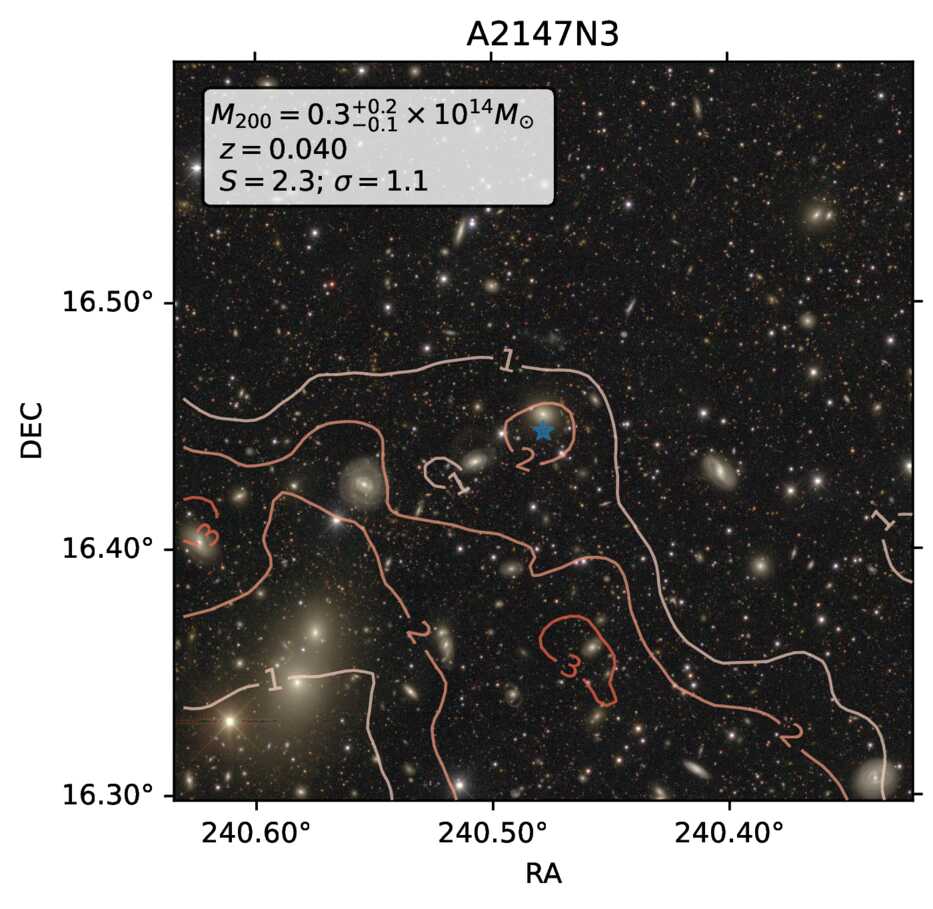}
    \includegraphics[width=0.3\linewidth]{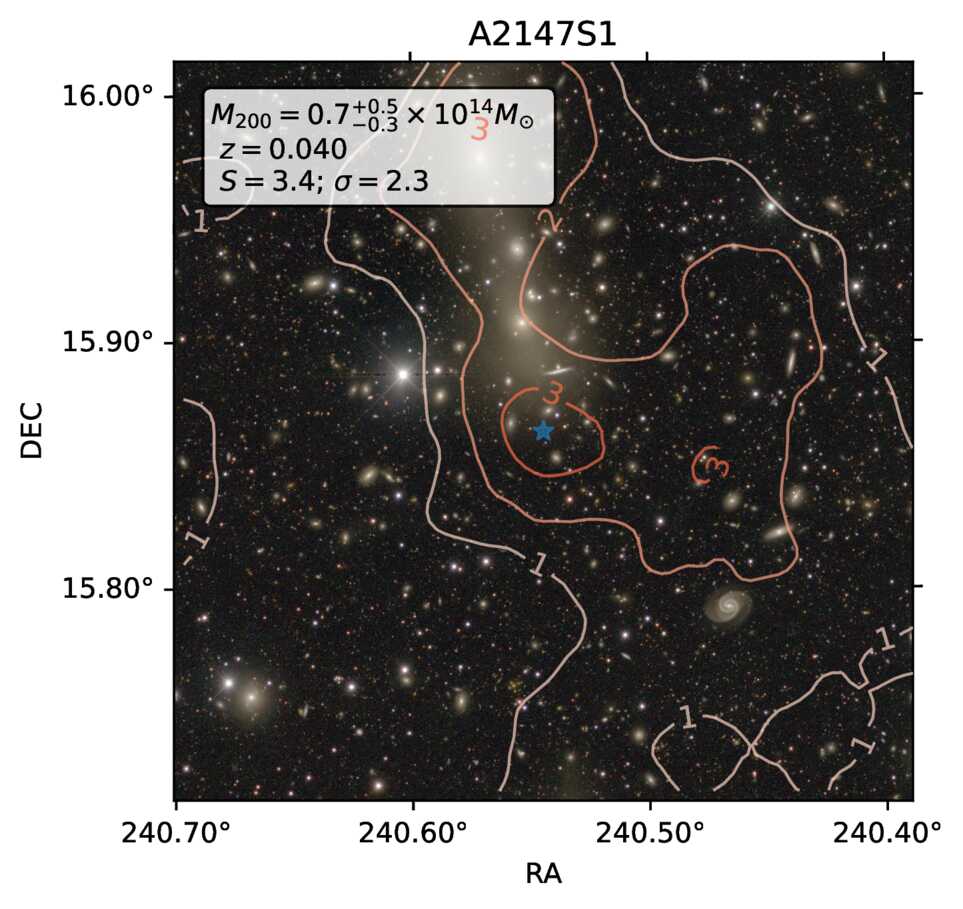}
    \includegraphics[width=0.3\linewidth]{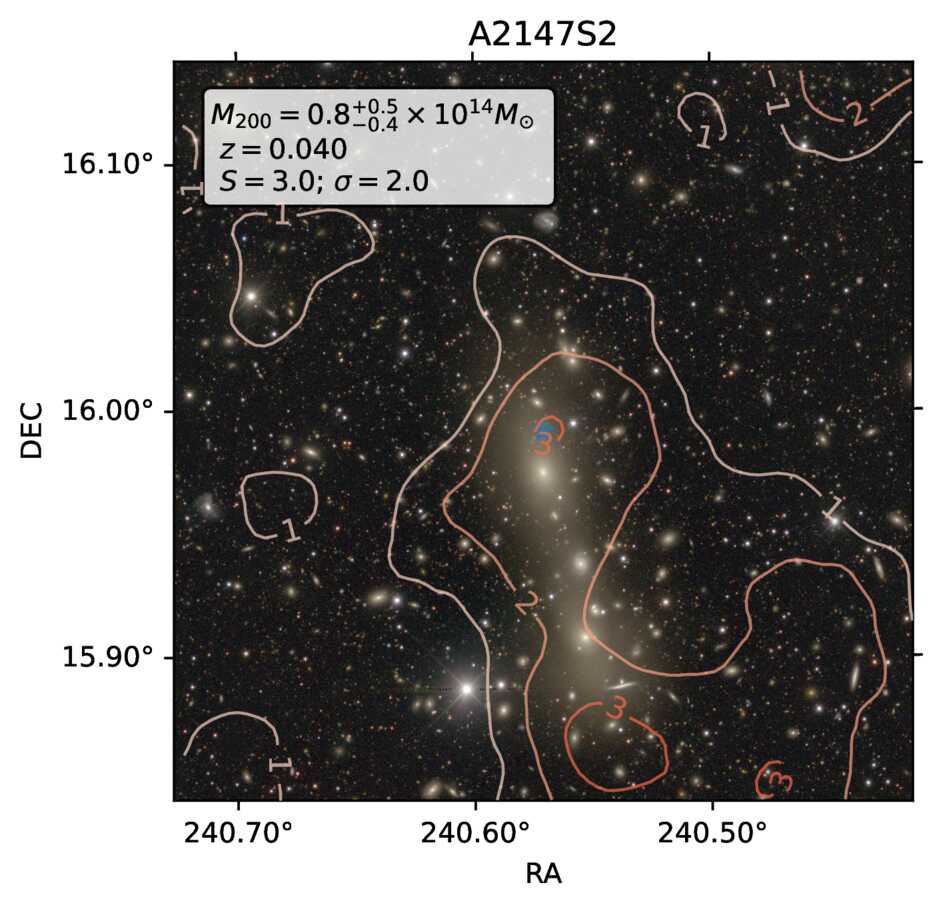}
    \includegraphics[width=0.3\linewidth]{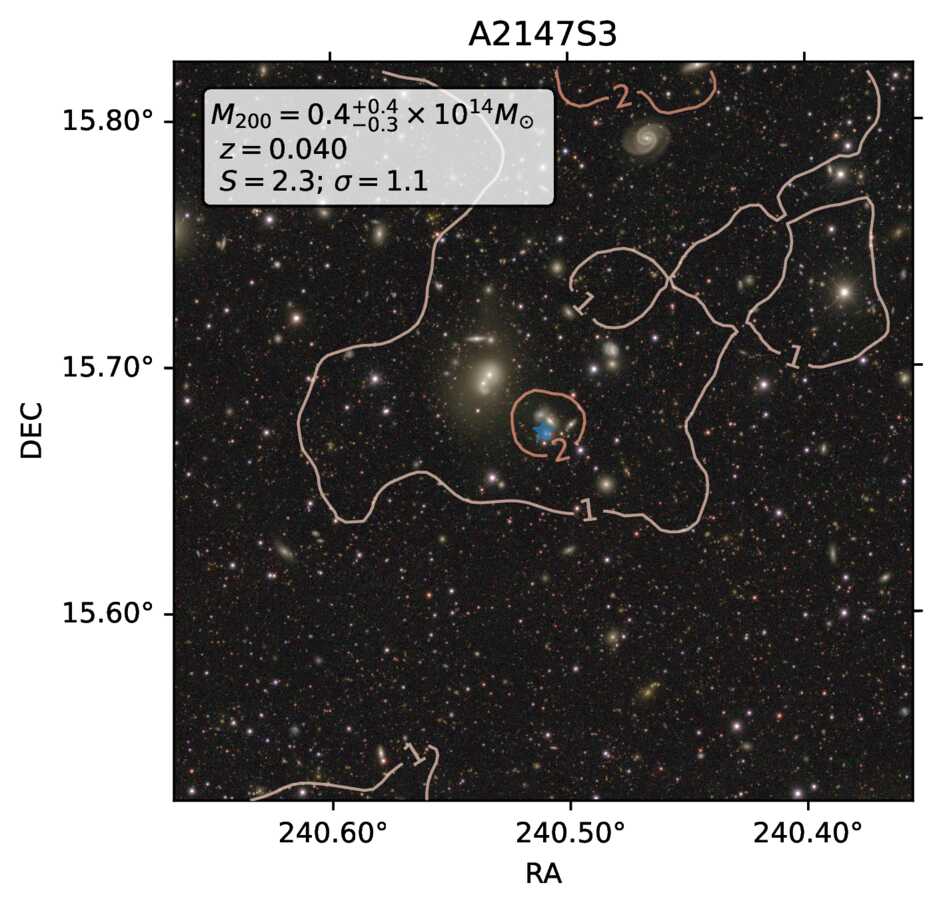}
    \includegraphics[width=0.3\linewidth]{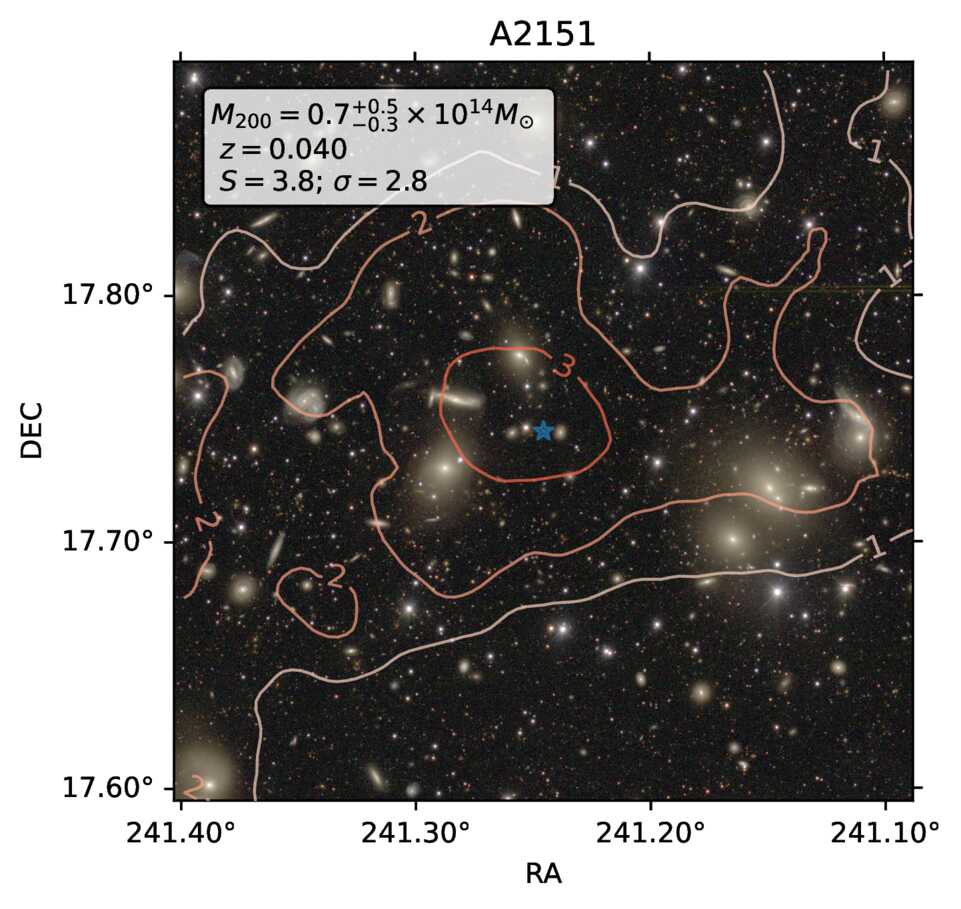}
    \includegraphics[width=0.3\linewidth]{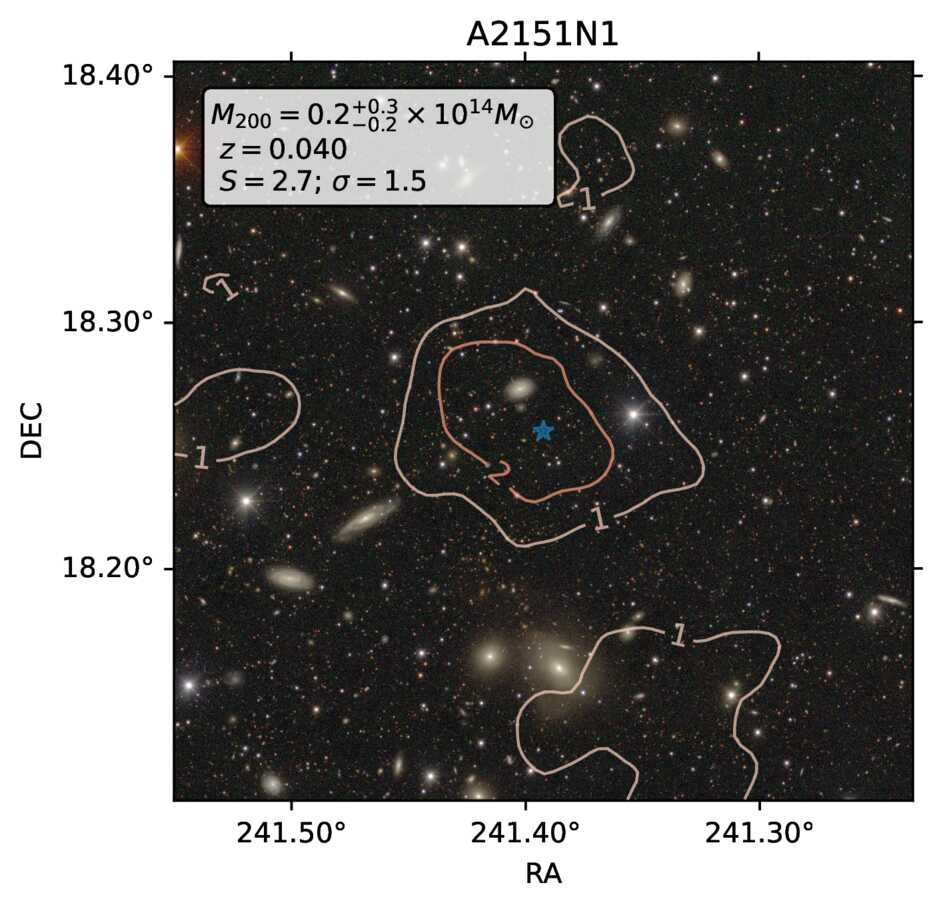}
    \caption{Postage-stamps of individual sources; $irg$ cutouts of lensing sources detected in the complete CLFZ catalog with the mass maps overlaid. The coordinates, masses, and redshifts of this figure is available as the Data behind the Figure.}
    \label{fig:placeholder}
    \digitalasset
\end{figure*}

\begin{figure*}
    \centering
    \includegraphics[width=0.3\linewidth]{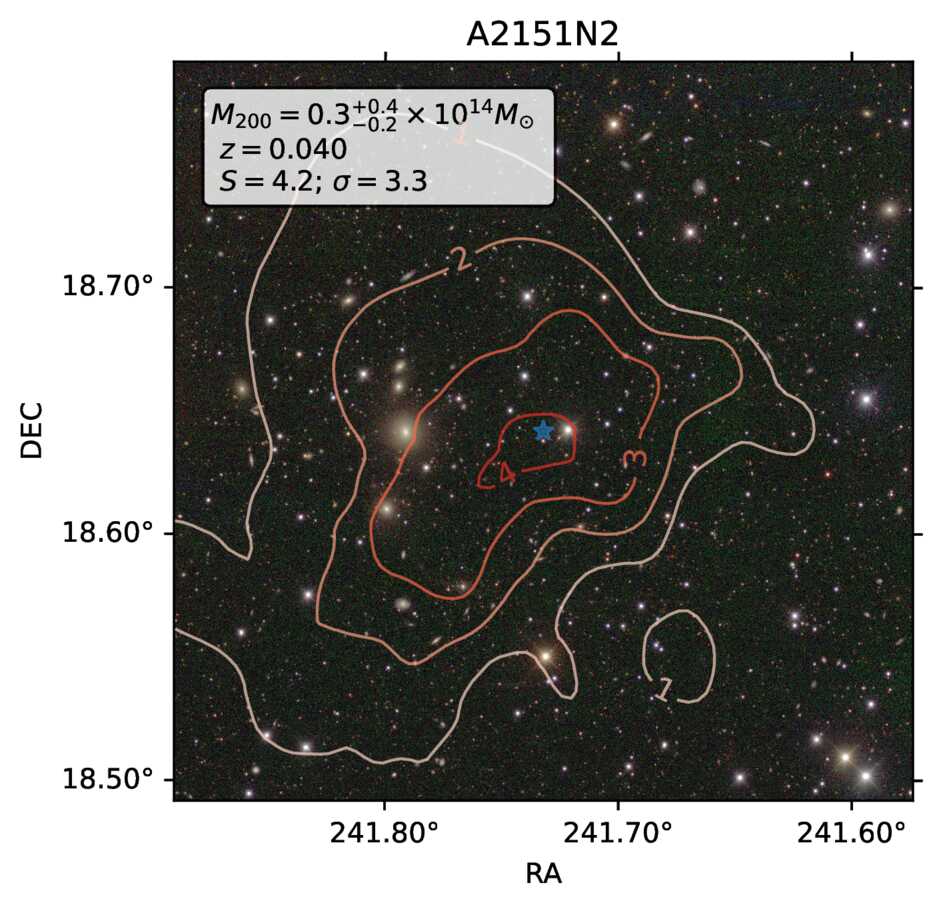}
    \includegraphics[width=0.3\linewidth]{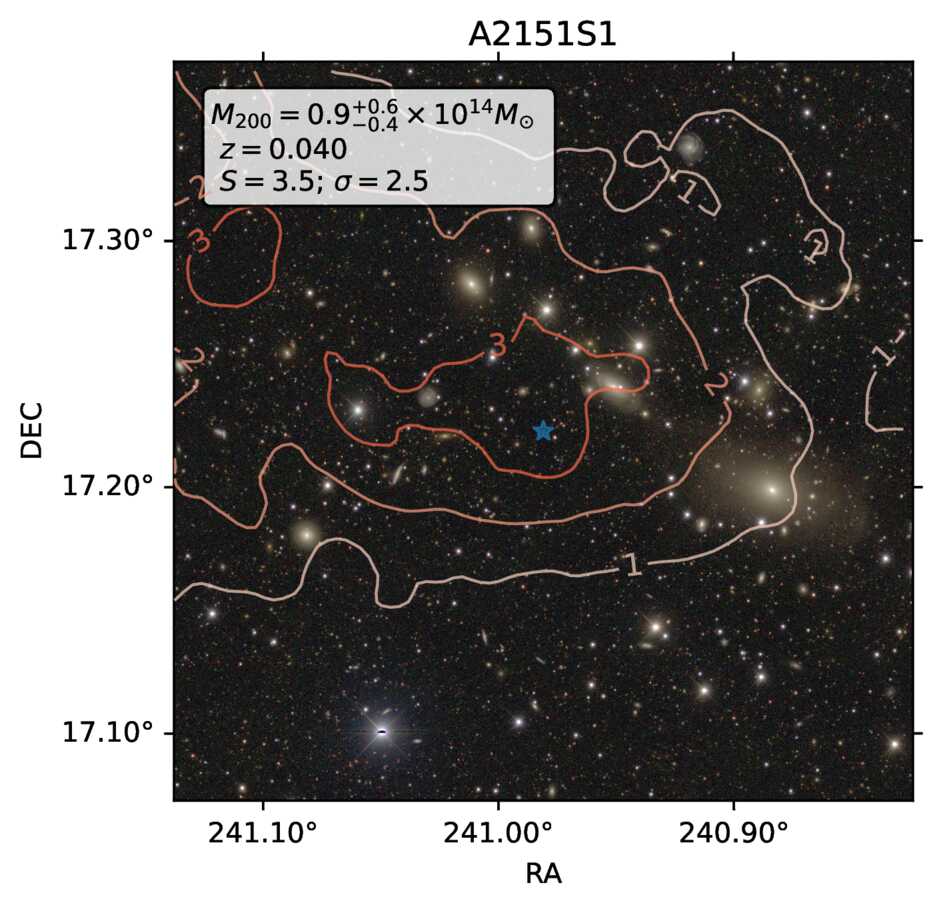}
    \includegraphics[width=0.3\linewidth]{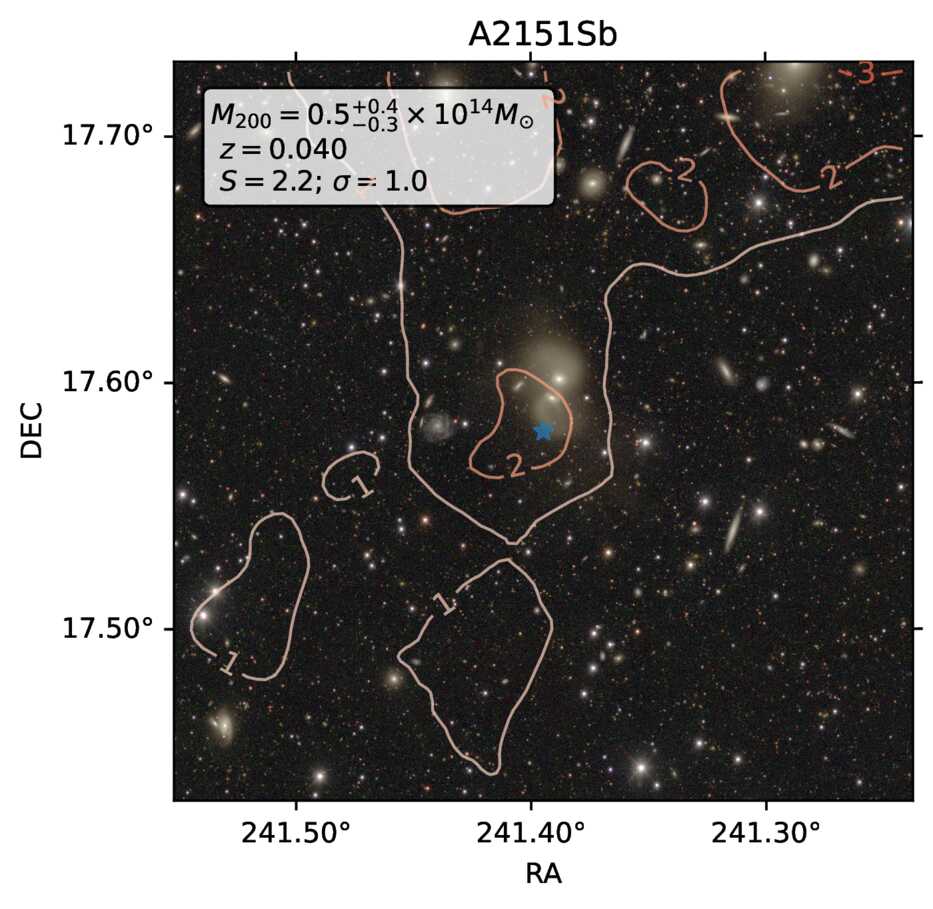}
    \includegraphics[width=0.3\linewidth]{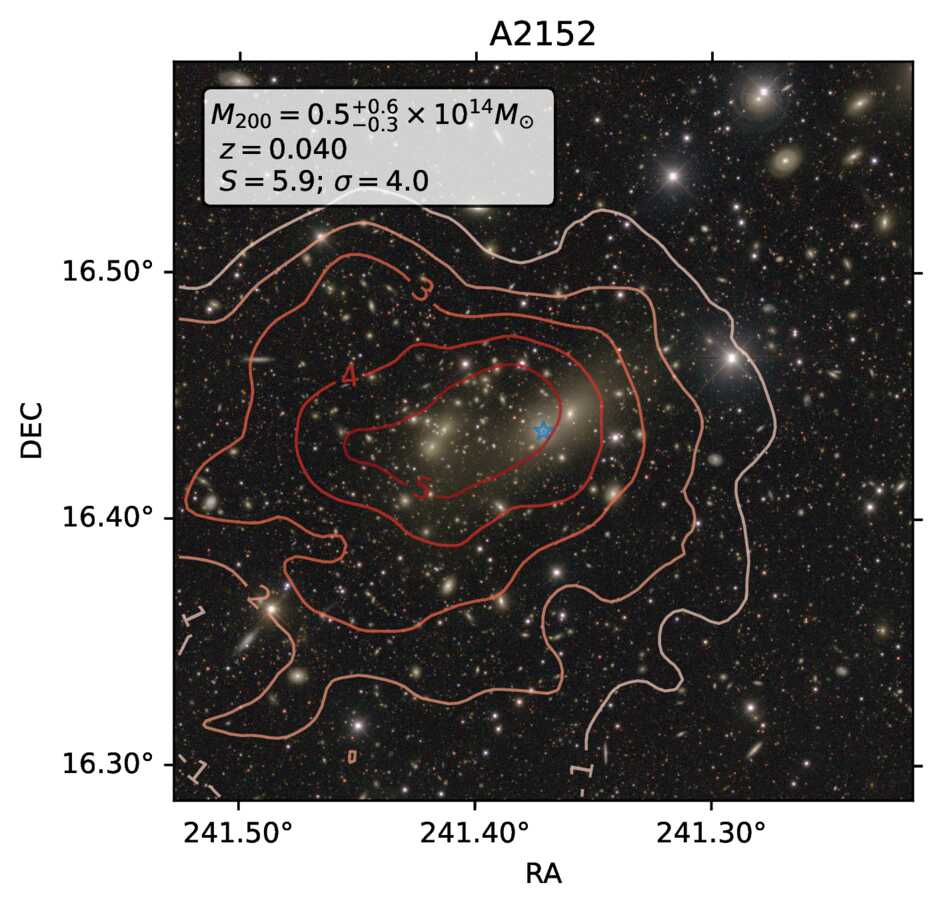}
    \includegraphics[width=0.3\linewidth]{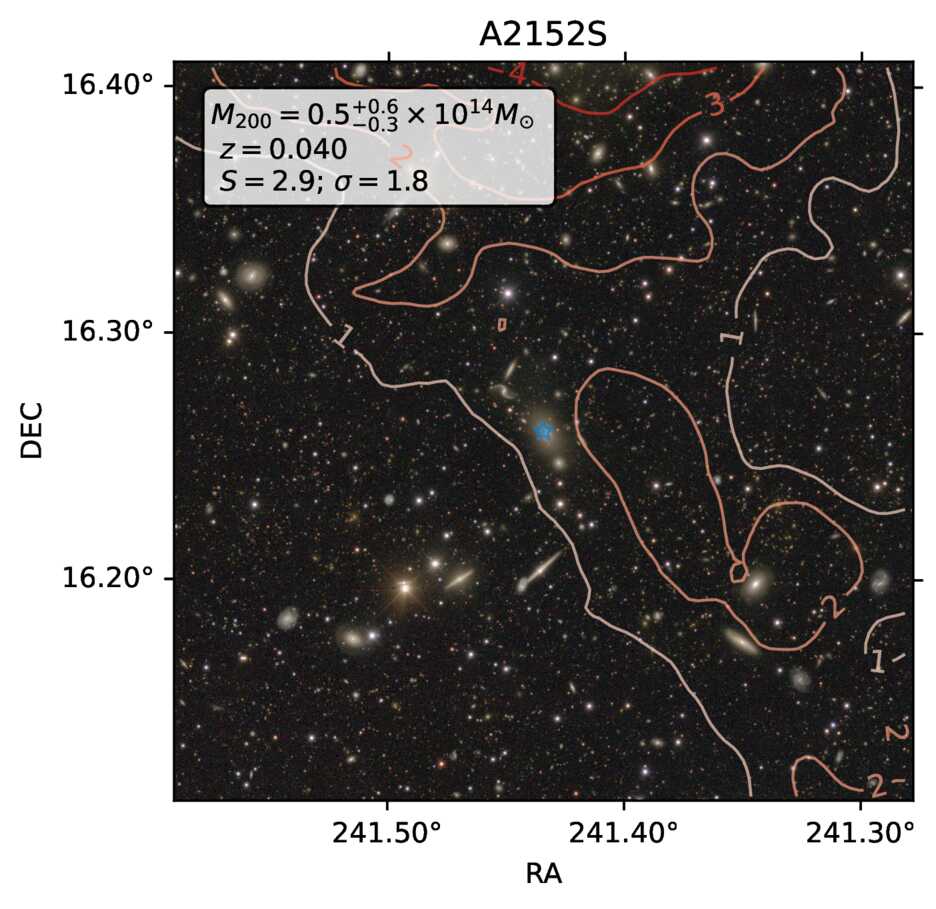}
    \includegraphics[width=0.3\linewidth]{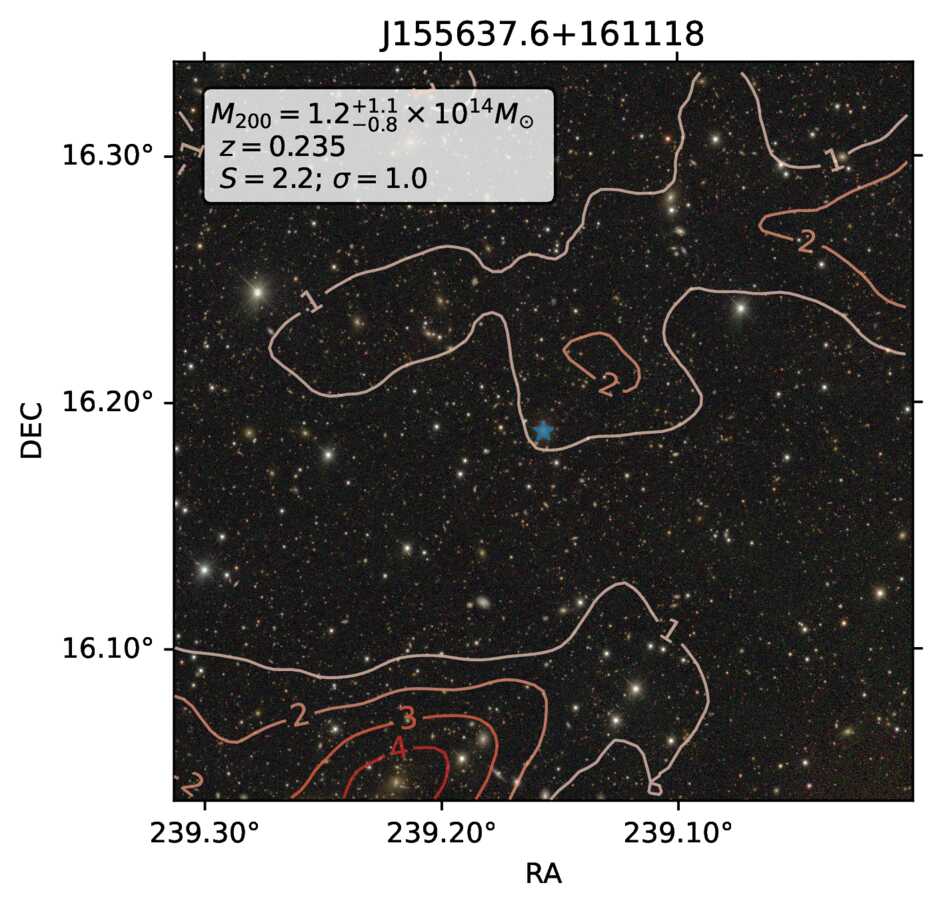}
    \includegraphics[width=0.3\linewidth]{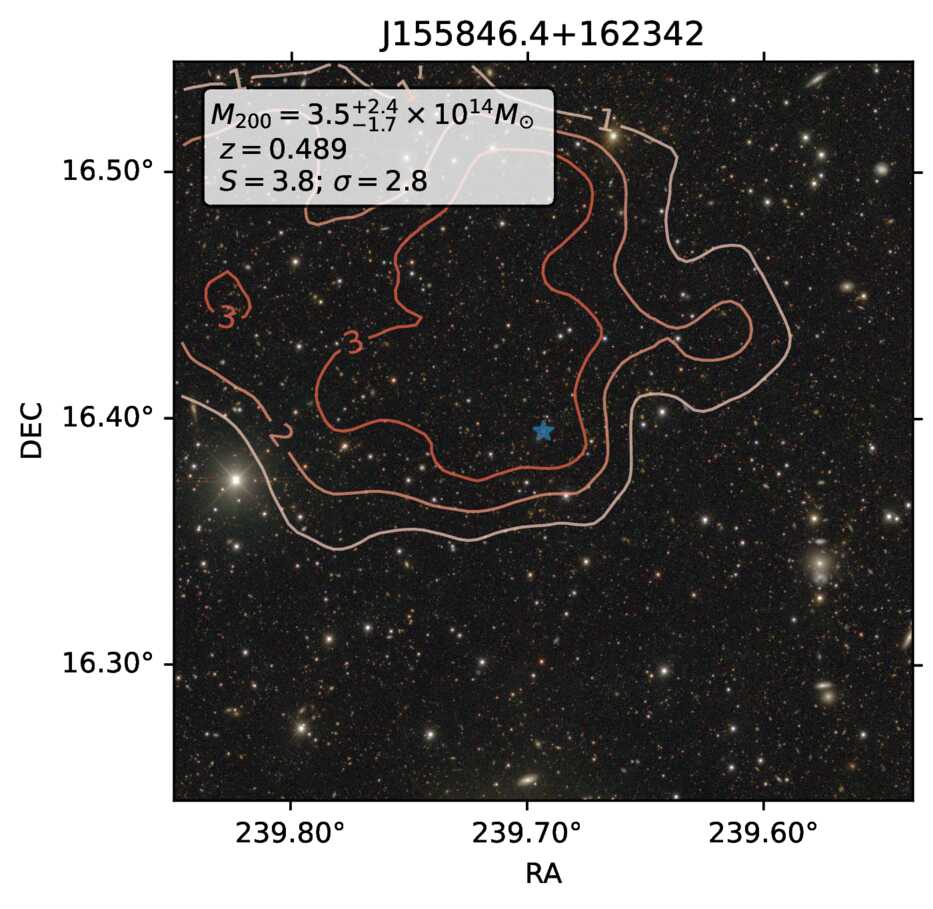}
    \includegraphics[width=0.3\linewidth]{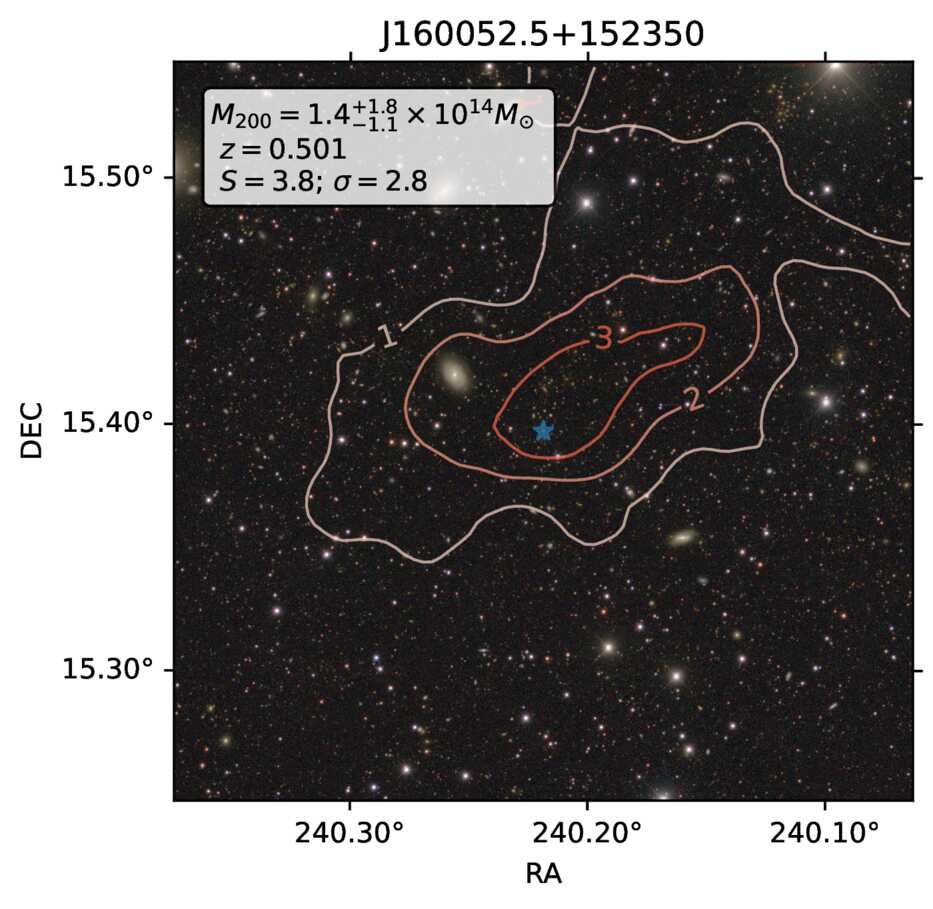}
    \includegraphics[width=0.3\linewidth]{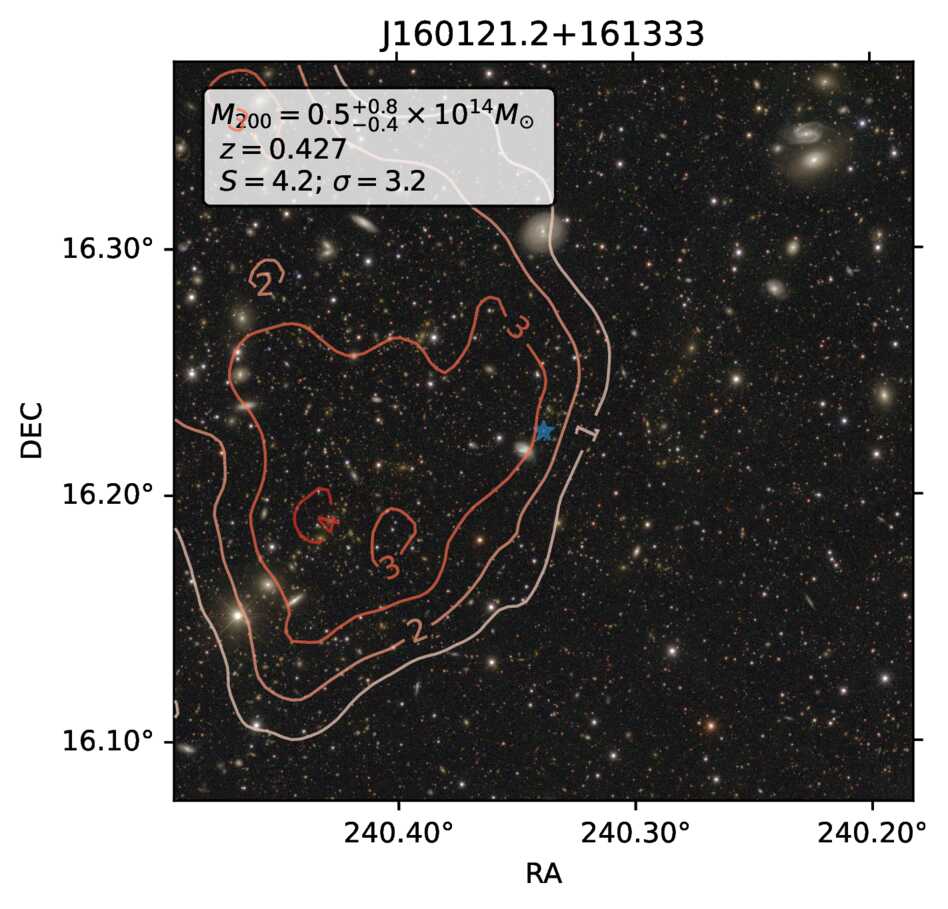}
    \includegraphics[width=0.3\linewidth]{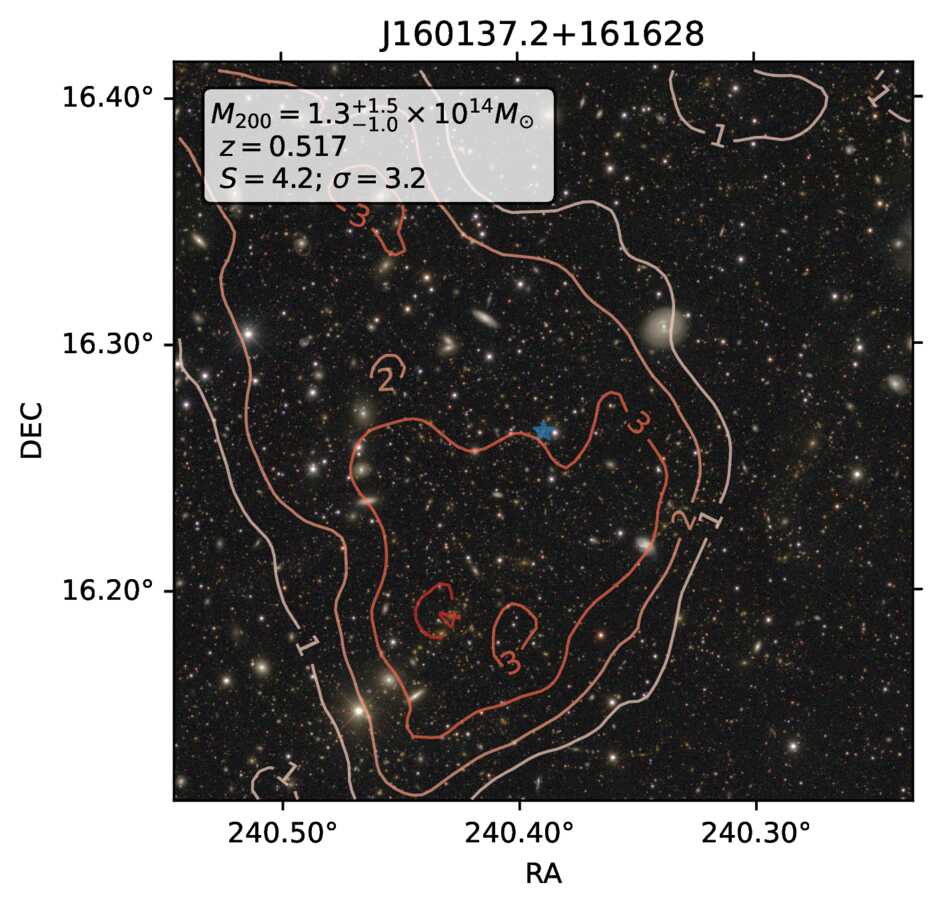}
    \includegraphics[width=0.3\linewidth]{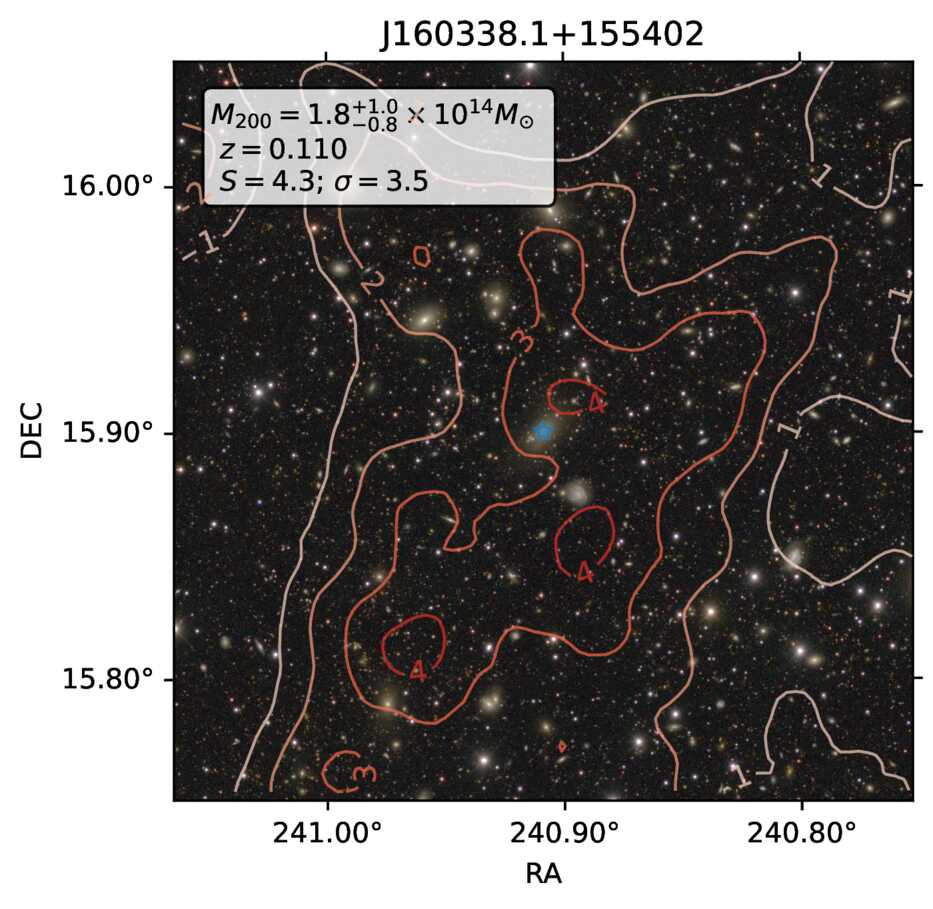}
    \includegraphics[width=0.3\linewidth]{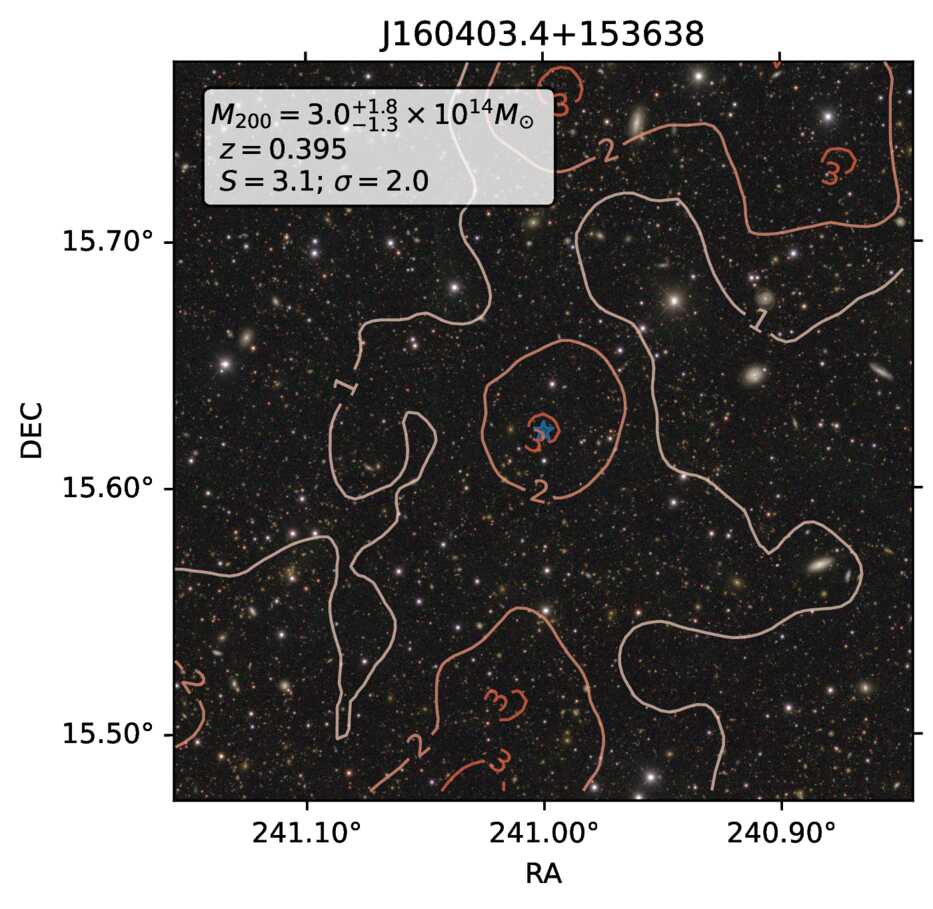}
    \caption{Postage-stamps of individual sources (II)}
    \label{fig:placeholder}
\end{figure*}

\begin{figure*}
    \centering
    \includegraphics[width=0.3\linewidth]{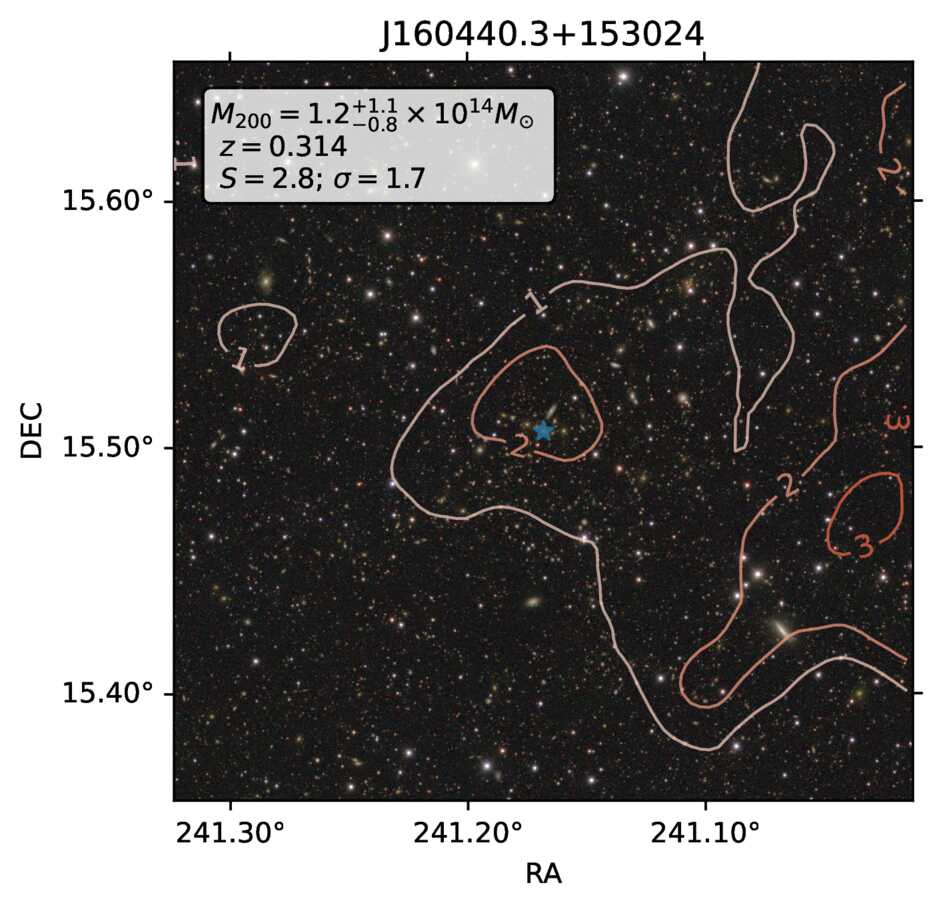}
    \includegraphics[width=0.3\linewidth]{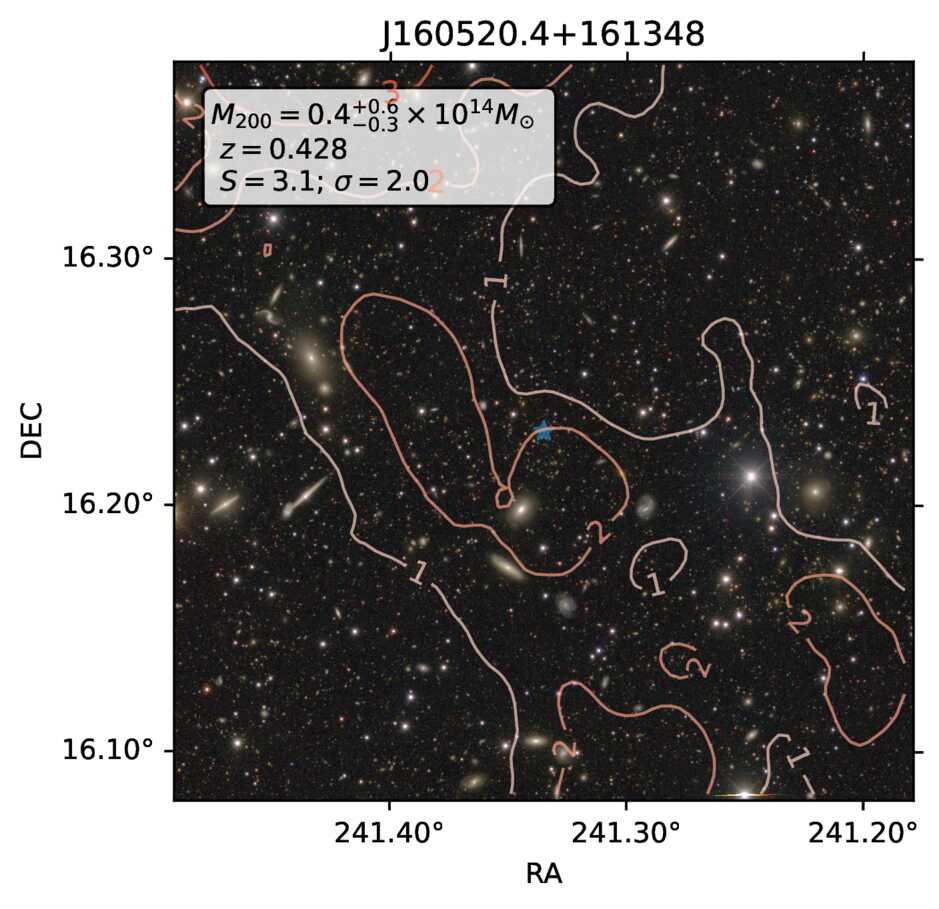}
    \includegraphics[width=0.3\linewidth]{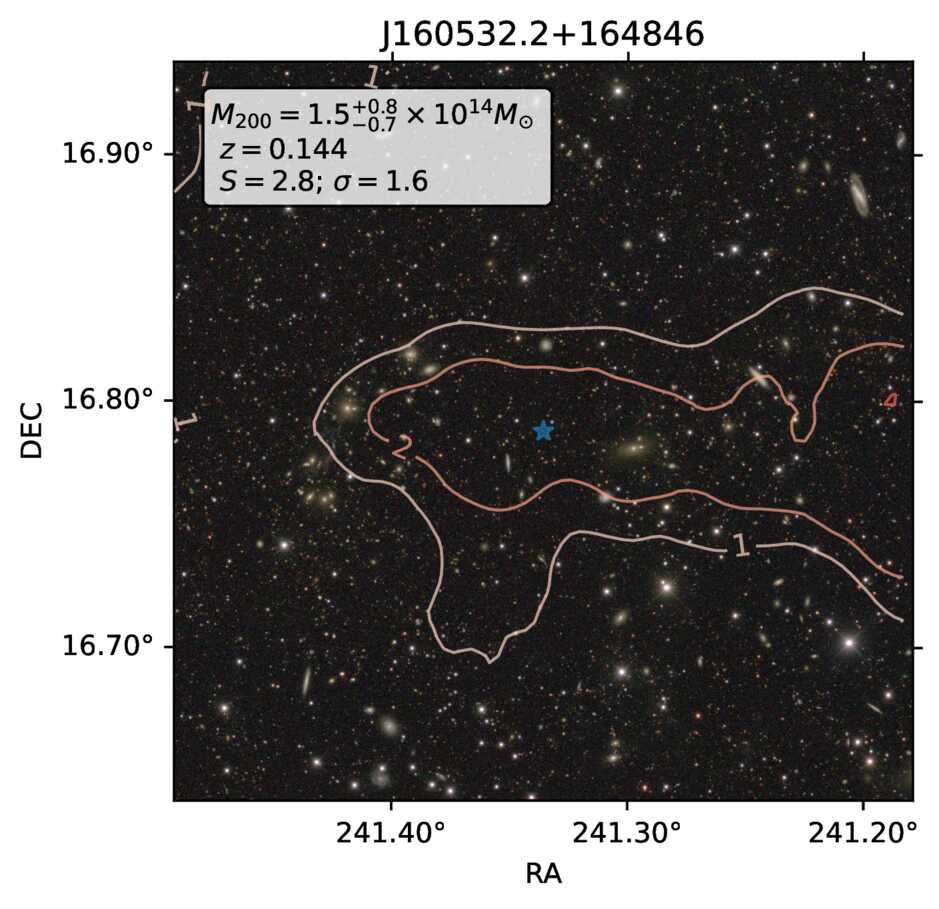}
    \includegraphics[width=0.3\linewidth]{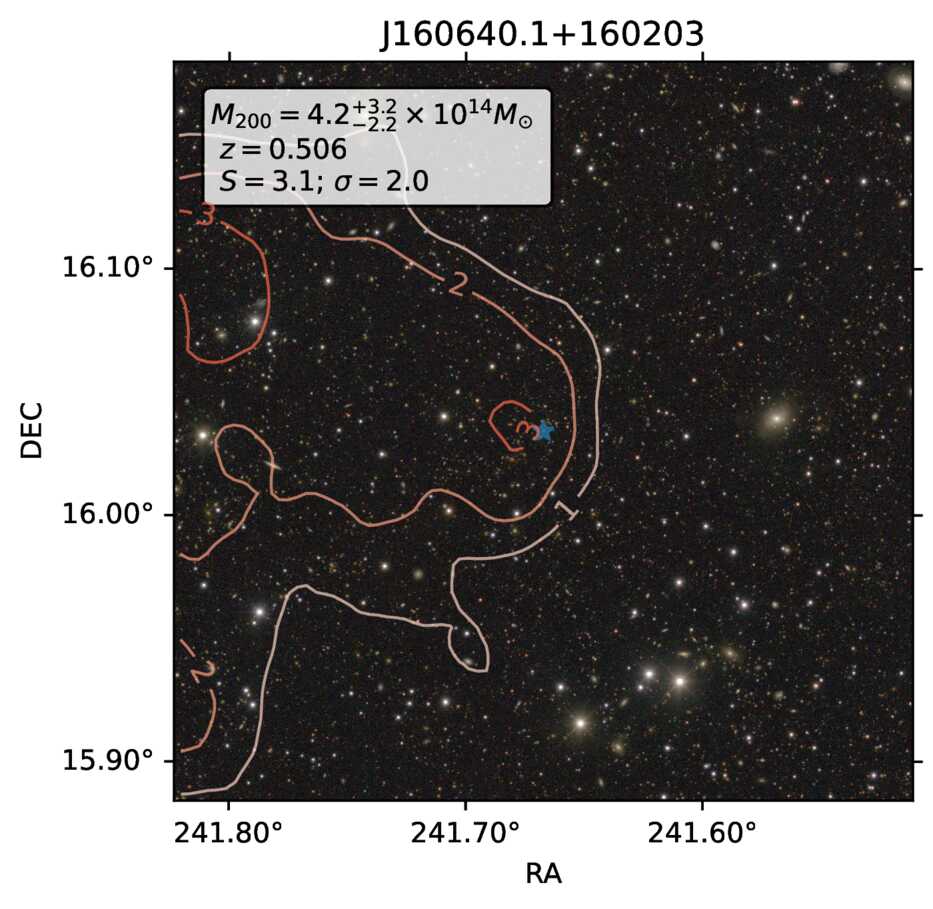}
    \includegraphics[width=0.3\linewidth]{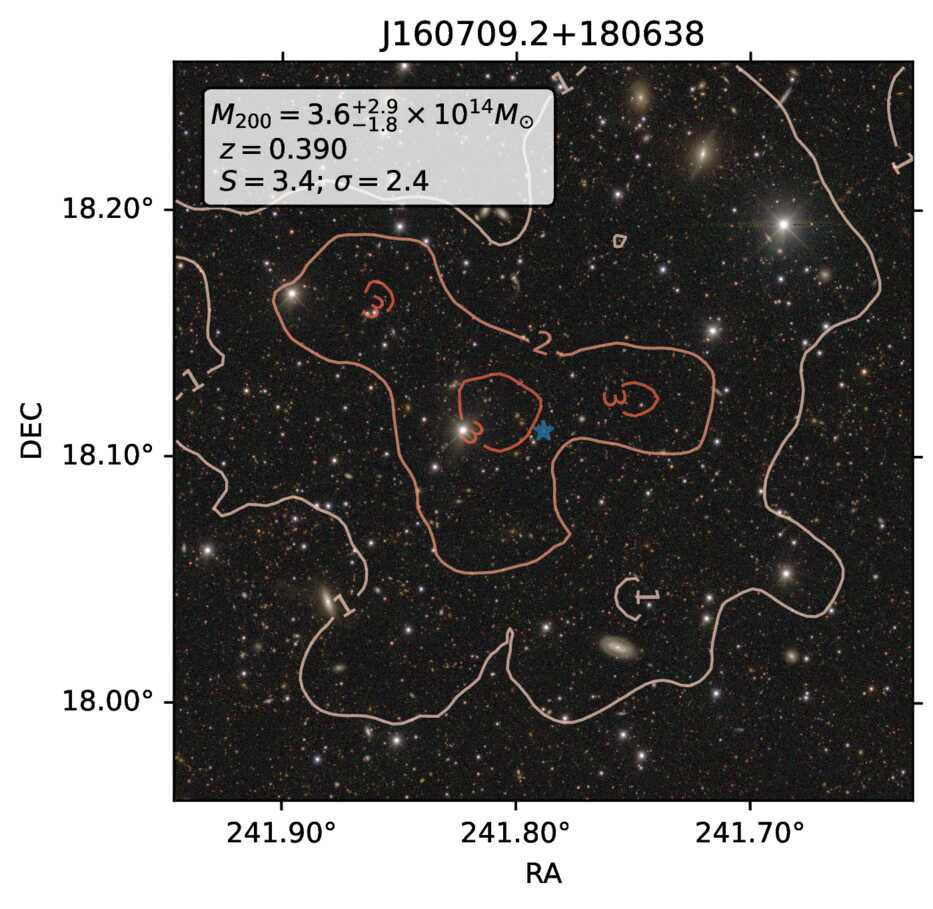}
    \includegraphics[width=0.3\linewidth]{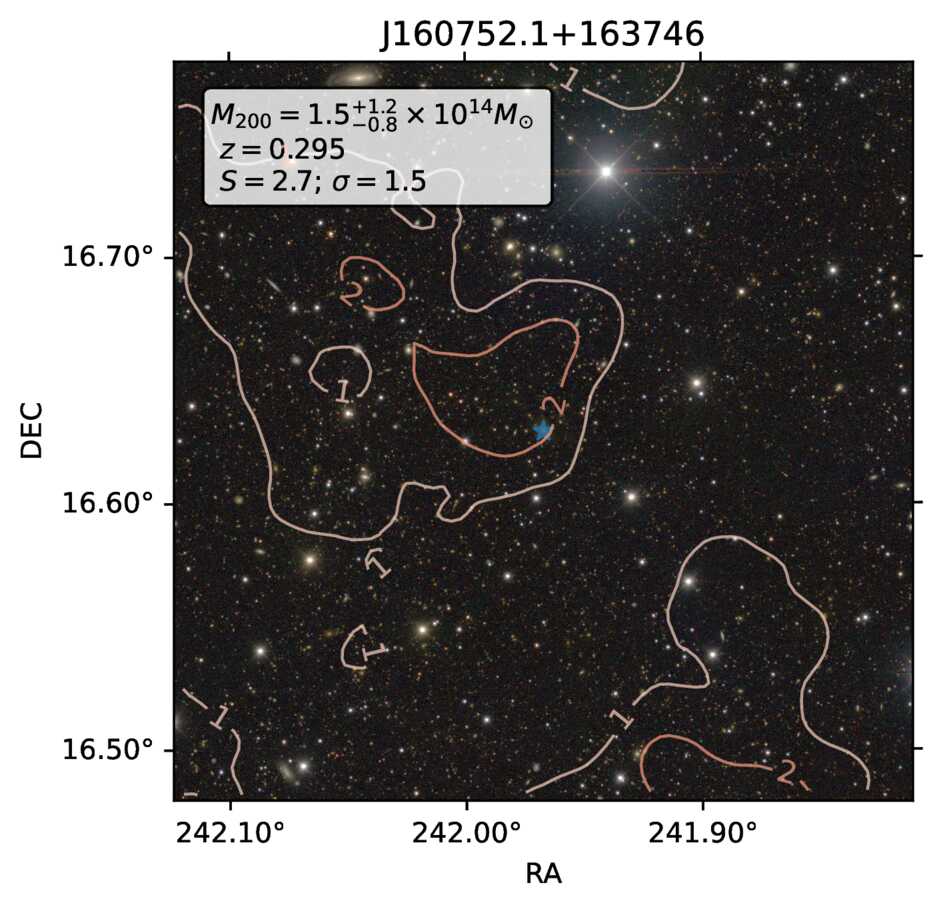}
    \includegraphics[width=0.3\linewidth]{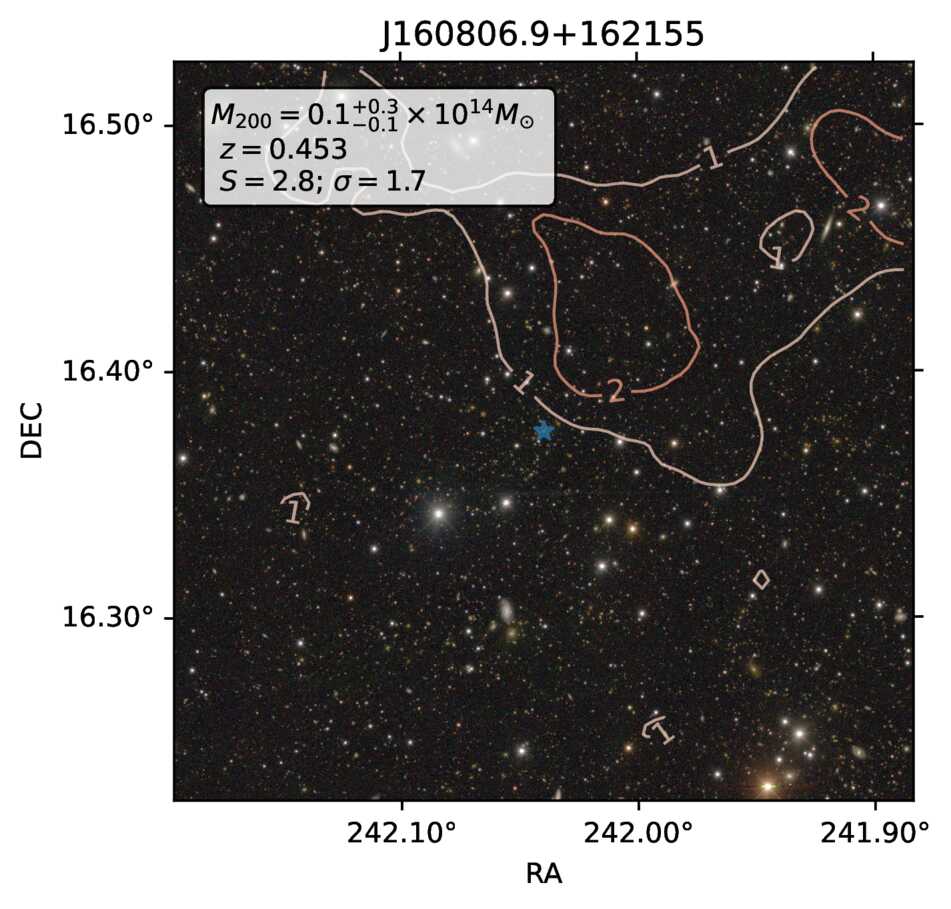}
    \includegraphics[width=0.3\linewidth]{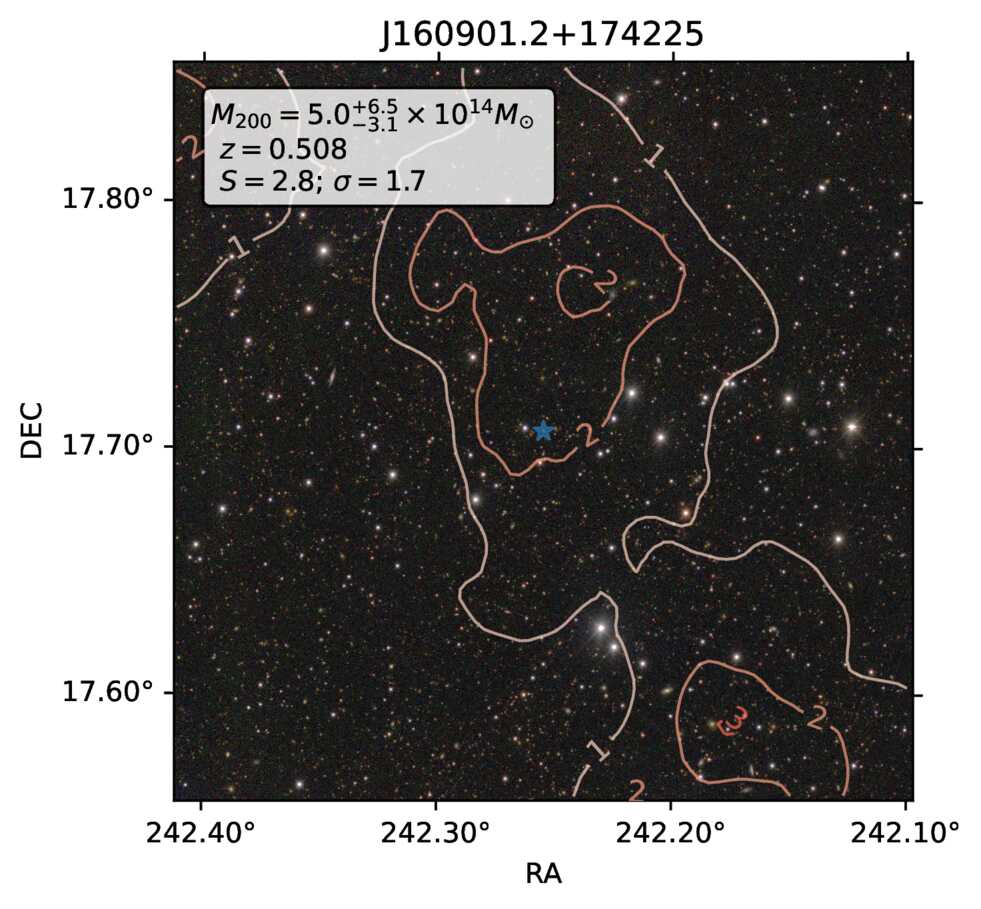}
    \includegraphics[width=0.3\linewidth]{RMJ155653.jpg}
    \includegraphics[width=0.3\linewidth]{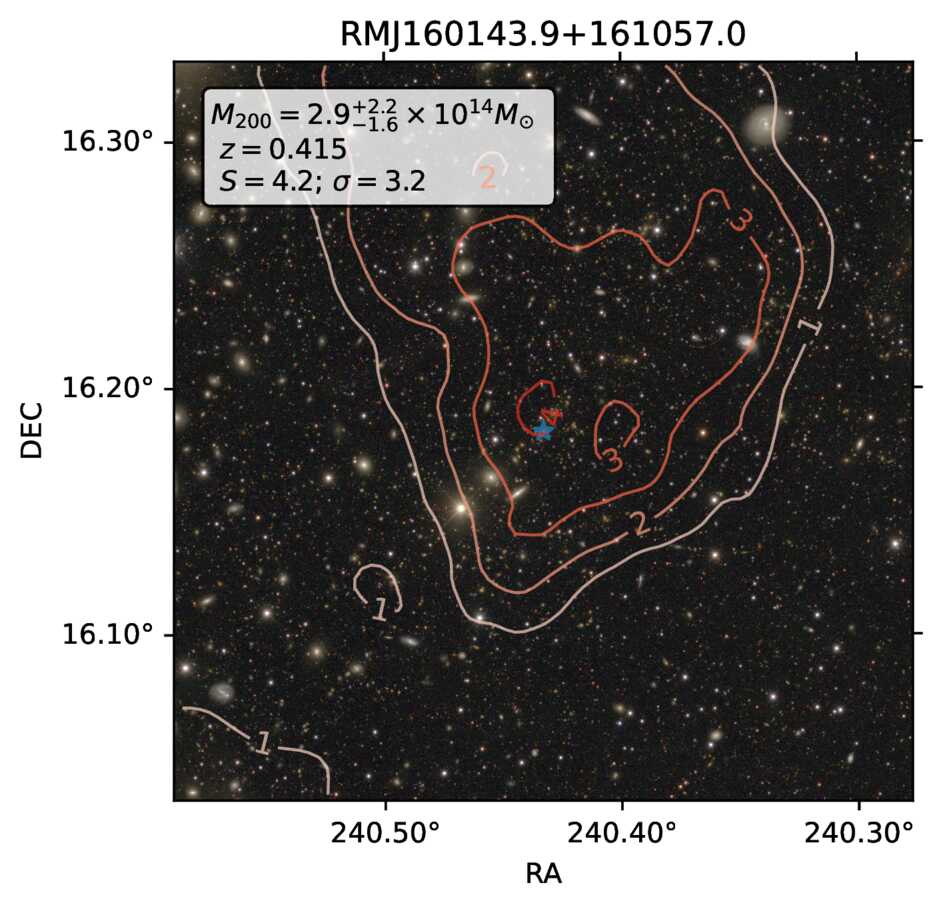}
    \includegraphics[width=0.3\linewidth]{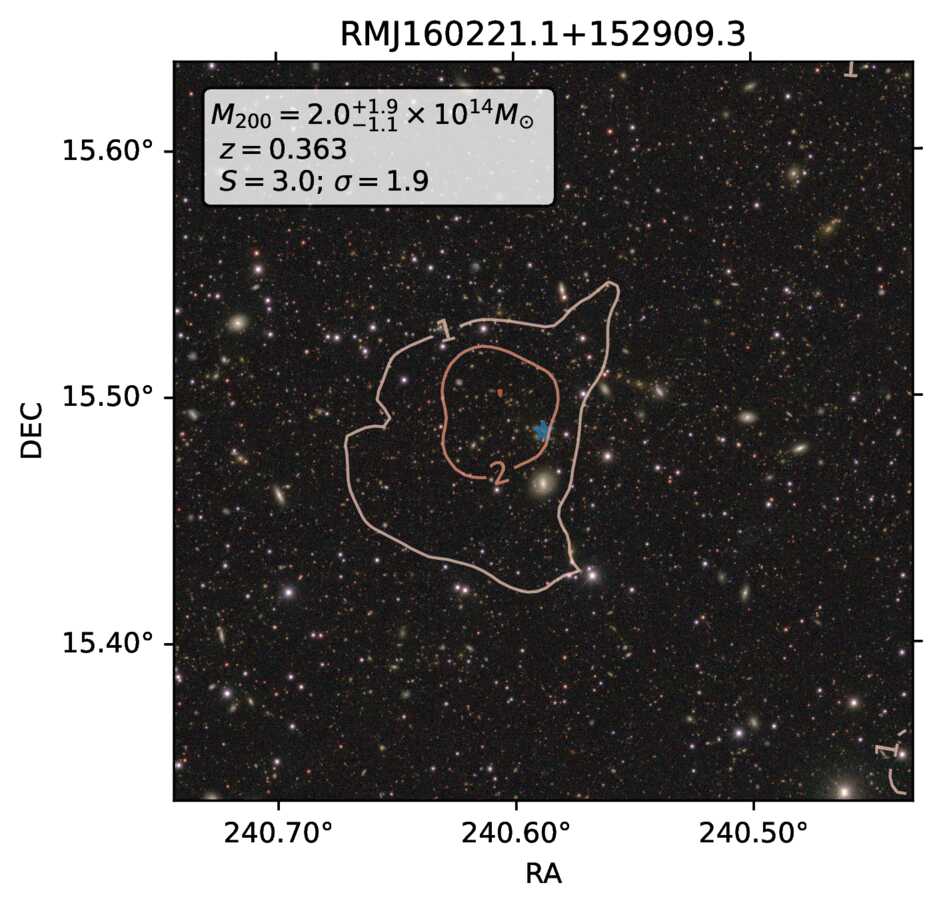}
    \includegraphics[width=0.3\linewidth]{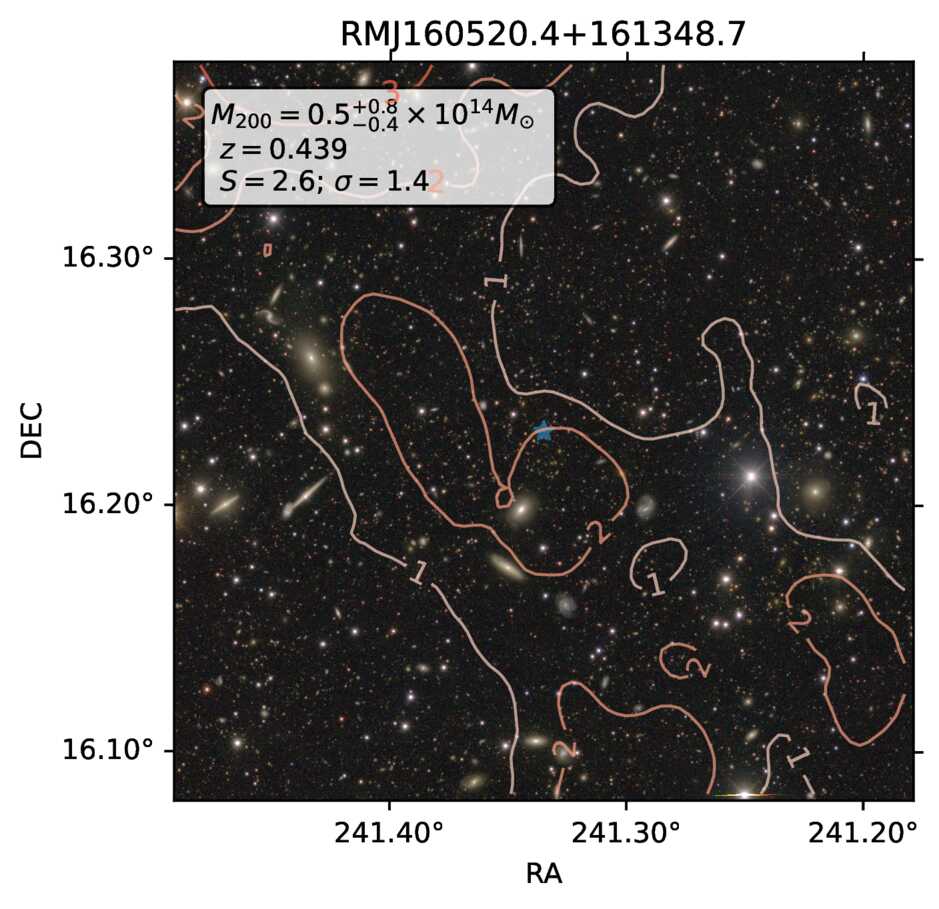}
    \caption{Postage-stamps of individual sources (III)}
    \label{fig:placeholder}
\end{figure*}

\begin{figure*}
    \centering
    \includegraphics[width=0.3\linewidth]{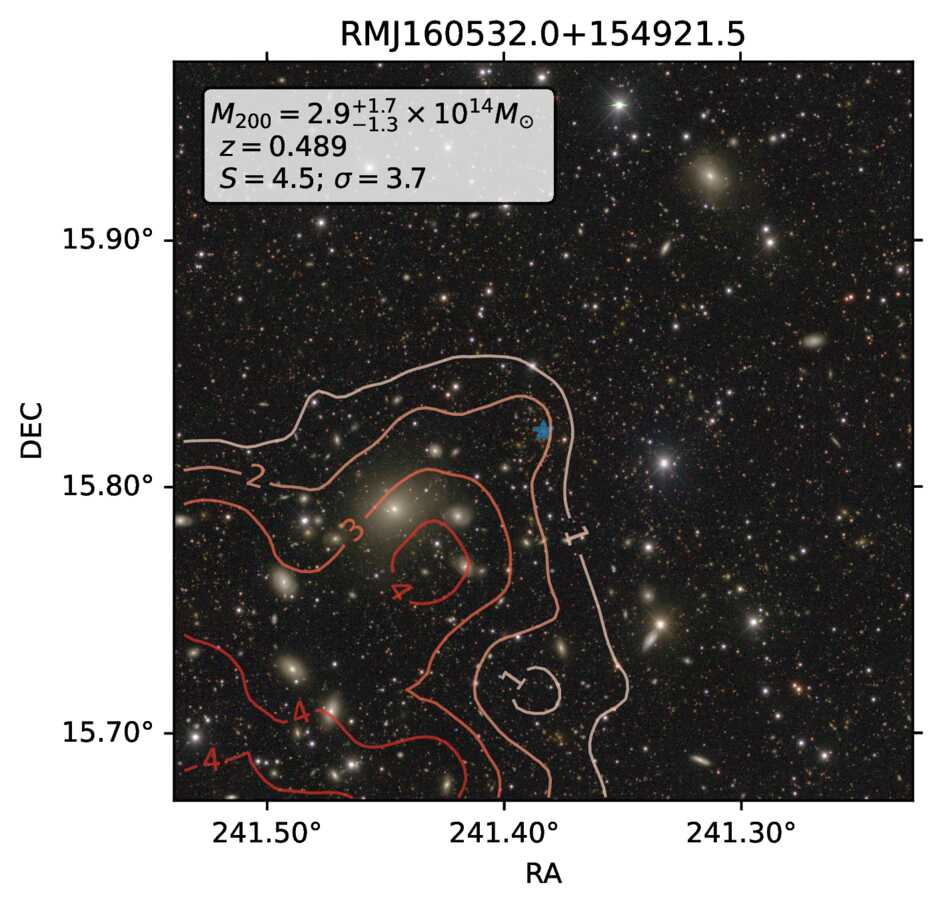}
    \includegraphics[width=0.3\linewidth]{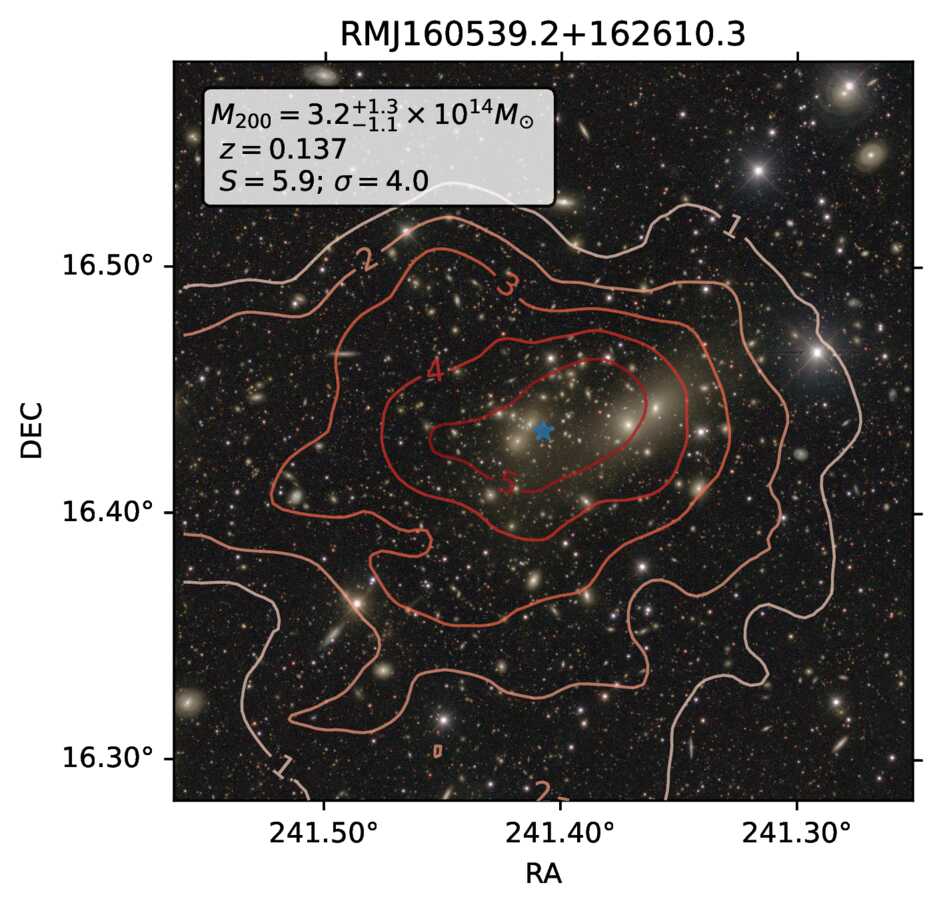}
    \includegraphics[width=0.3\linewidth]{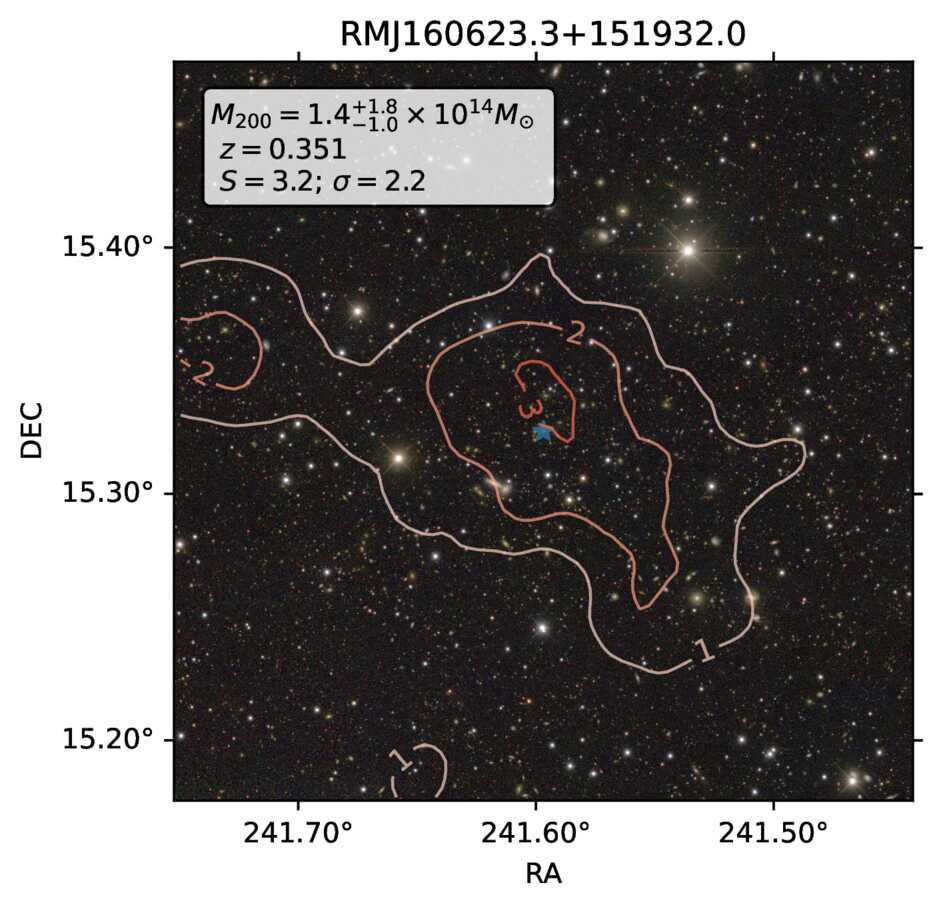}
    \includegraphics[width=0.3\linewidth]{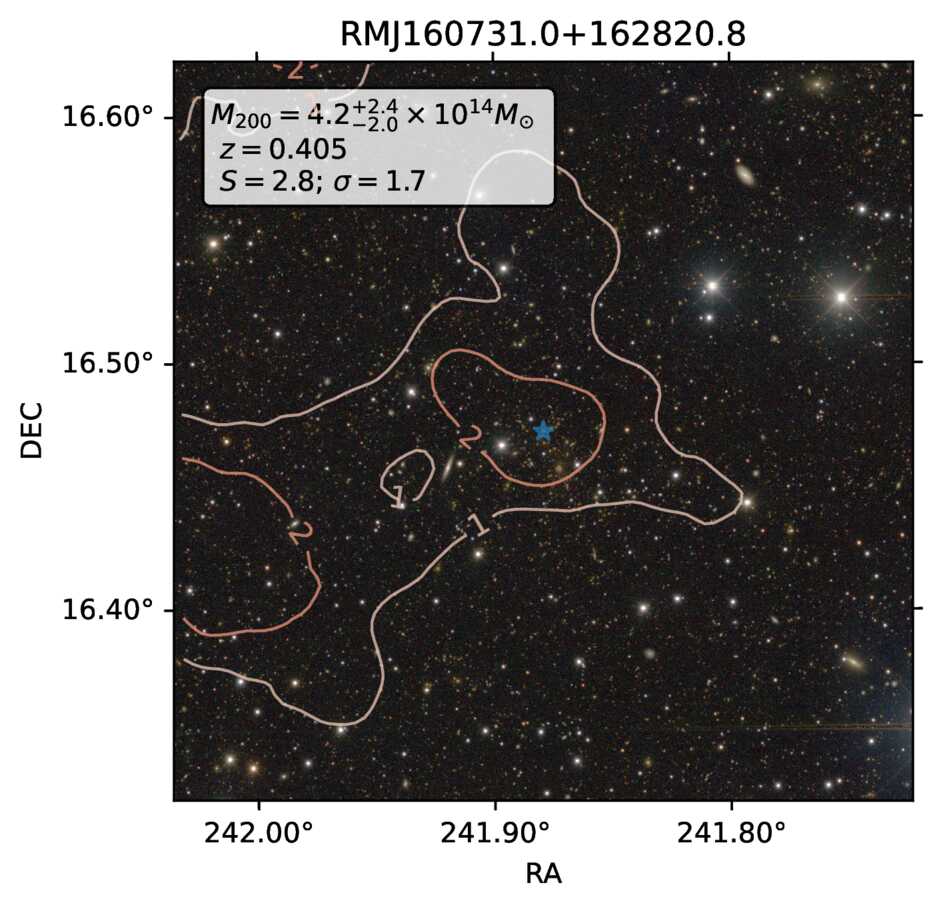}
    \includegraphics[width=0.3\linewidth]{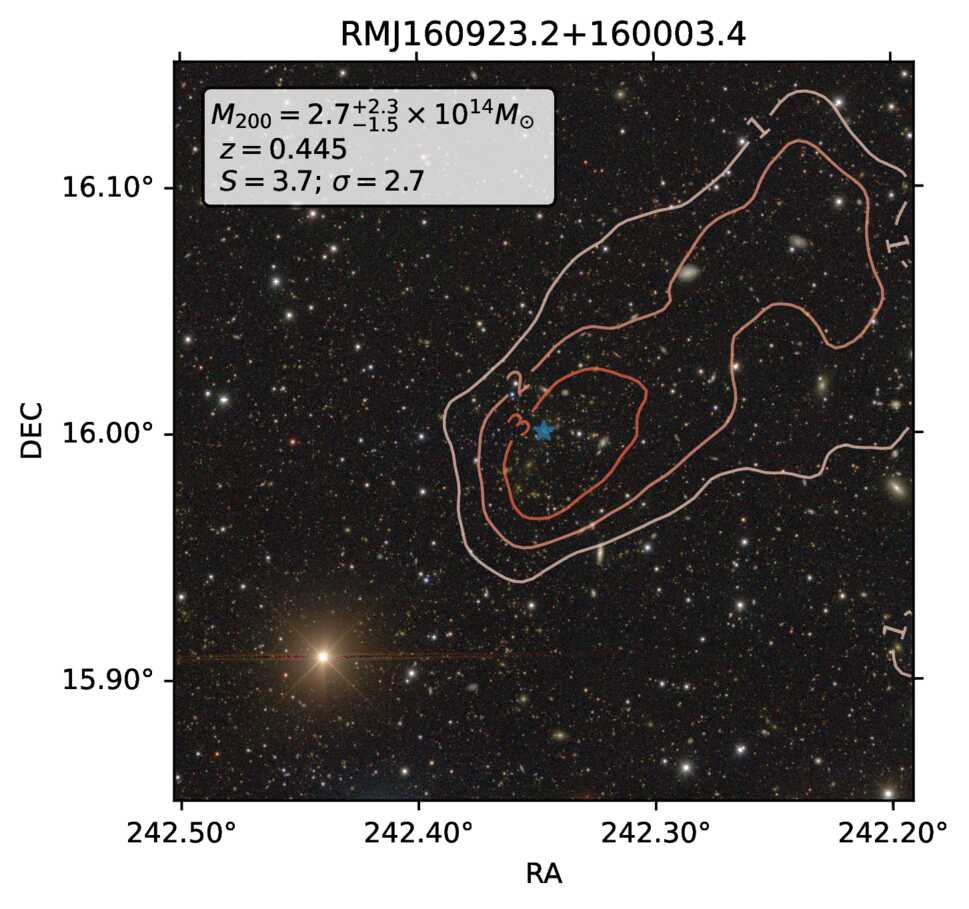}
    \includegraphics[width=0.3\linewidth]{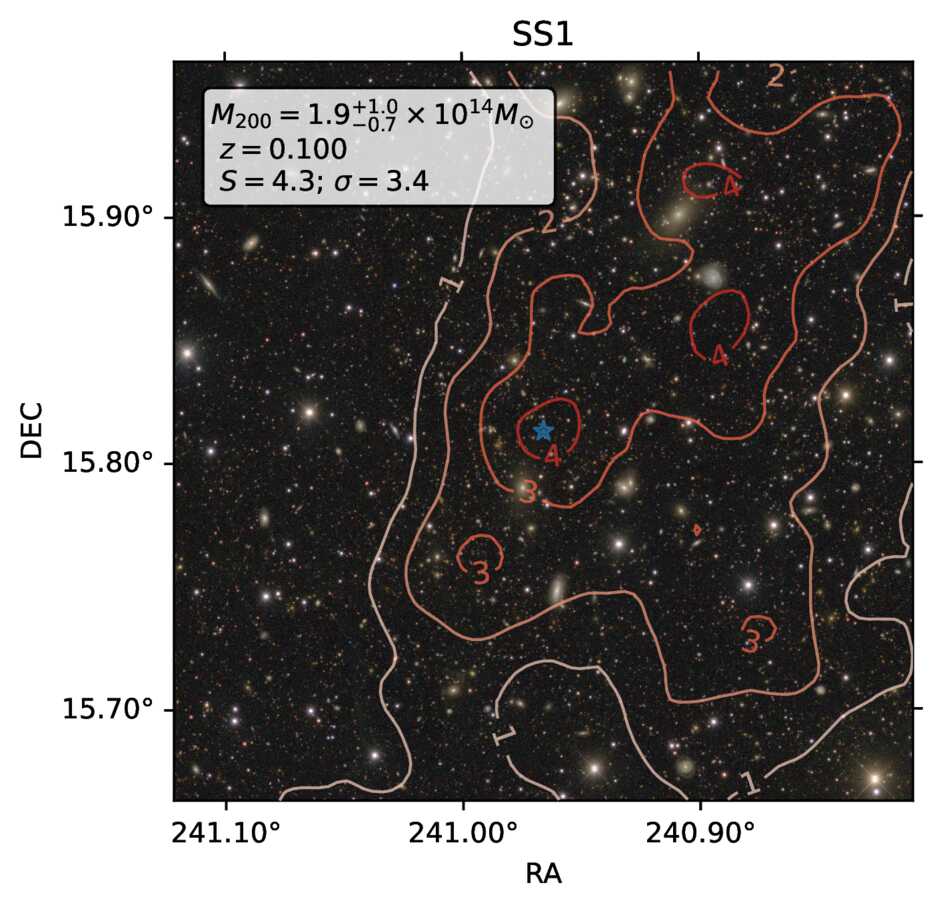}
    \includegraphics[width=0.3\linewidth]{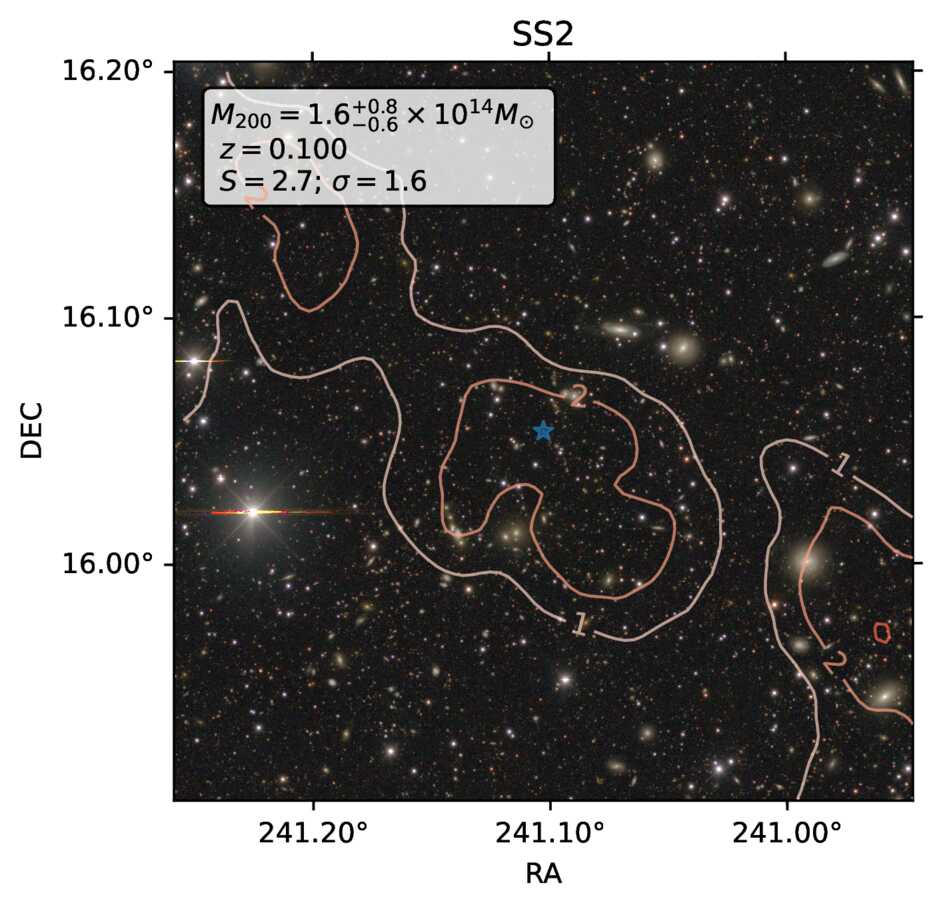}
    \includegraphics[width=0.3\linewidth]{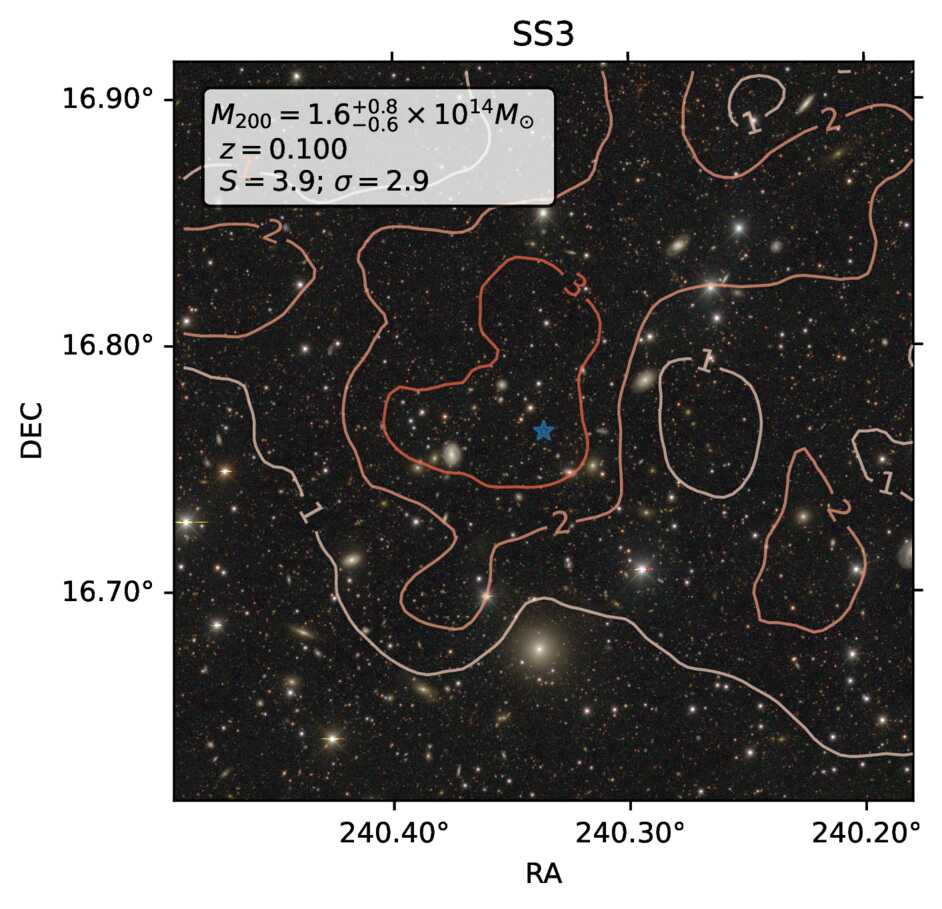}
    \caption{Postage-stamps of individual sources (IV)}
    \label{fig:placeholder}
\end{figure*}

\clearpage

\section{Table of Exposures}

\begin{deluxetable*}{ccccccc}[htbp!]
\tablecaption{Table of Proposals}\label{tab:props}
\tablehead{ \colhead{Proposal ID} & \colhead{PI} & \colhead{u} & \colhead{g} & \colhead{r} & \colhead{i} & \colhead{z} }
\startdata
2013A-0613   & Munoz         & -     & 1800  & -     & 1800  & -     \\
2013A-9999   & Walker        & -     & -     & 30    & 25    & -     \\
2014B-0404   & Schlegel      & -     & 6202  & 4142  & -     & 7158  \\
2015A-0130   & Crnojevic     & -     & 2520  & 1260  & -     & -     \\
2019A-0305   & Drlica-Wagner & -     & 180   & 450   & 4140  & 360   \\
2019A-0308   & Dell'Antonio  & 6579  & 2300  & 4490  & 3900  & 6000  \\
2020B-0053   & Brout         & -     & 170   & 170   & 145   & 145   \\
2023A-237157 & Narayan       & -     & 415   & 505   & 465   & 280   \\
2024B-763968 & Narayan       & -     & 110   & 60    & 60    & 60    \\
2025A-632499 & Englert       & 18450 & 13110 & 18130 & 15900 & 7595  \\
\hline
             & {\bf Total}         & 25029 & 26807 & 29237 & 26435 & 21598 \\
\enddata
\tablecomments{Exposure times are listed in seconds.}
\end{deluxetable*}

\clearpage



\bibliography{references,references-2,sample701}{}
\bibliographystyle{aasjournalv7}



\end{document}